\documentclass[traditabstract]{aa}  
\usepackage{graphicx}
\usepackage{tabularx}
\usepackage{rotating, booktabs}
\usepackage{txfonts}
\usepackage{hyperref}
\hypersetup{colorlinks = true, linkcolor = blue, urlcolor=blue, citecolor=blue} 

\usepackage{url}
\usepackage{epsfig}

\def\ltsima{$\; \buildrel < \over \sim \;$}
\def\simlt{\lower.5ex\hbox{\ltsima}}
\def\gtsima{$\; \buildrel > \over \sim \;$}
\def\simgt{\lower.5ex\hbox{\gtsima}}

\def\ergs{{erg s$^{-1}$}}
\def\cm2{{cm$^{-2}$}}

\def\fhx{{$F_{\rm 2-10}$}}

\def\lum{{$L_{\rm 2-10}$}}

\def\p1{{Paper I}}

\def\xmm{{\em XMM--Newton}}
\def\chandra{{\em Chandra}}

\def\chandra{{\em Chandra}}

\def\xmm{{\em XMM--Newton}}

\def\f14{{10$^{-14}$}}
\def\f13{{10$^{-13}$}}
\def\f12{{10$^{-12}$}}
\def\f11{{10$^{-11}$}}

\def\4u{{4U~1344$-$60}}

\def\lbol{{$L_{\rm Bol}$}}

\def\lx6{{$L_x\!-\!L_{\rm 6\mu m}$}}
\def\aox{$\alpha_{OX} \,$}
\def\nh{$N_{\rm H} \,$}

\begin{document}

   \title{The WISSH Quasars Project}

   \subtitle{III. X-ray properties of hyper-luminous quasars}

   \author{S. Martocchia\inst{1,2}, E. Piconcelli\inst{2}, L. Zappacosta\inst{2}, F. Duras\inst{2,3}, G. Vietri\inst{2,4}, C. Vignali\inst{5,6}, S. Bianchi\inst{3}, M. Bischetti\inst{2,7}, A. Bongiorno\inst{2}, M. Brusa\inst{5,6}, G. Lanzuisi\inst{6}, A. Marconi\inst{8,9}, S. Mathur\inst{10}, G. Miniutti\inst{11}, F. Nicastro\inst{2}, G. Bruni\inst{12,13}, and F. Fiore\inst{2}}
\titlerunning{X-ray properties of hyper-luminous quasars}
   \authorrunning{S. Martocchia et al.}
   \institute{Astrophysics Research Institute, Liverpool John Moores University, 146 Brownlow Hill, Liverpool L3 5RF, UK
   \and INAF - Osservatorio Astronomico di Roma, via Frascati 33, 00078 Monteporzio Catone, Italy      
   	 \and Dipartimento di Matematica e Fisica, Universit\`a degli Studi Roma Tre, via della Vasca Navale 84, I--00146, Roma, Italy         
   	          \and Universit\`a degli Studi di Roma "La Sapienza", Piazzale Aldo Moro 5, I--00185 Roma, Italy
   	    \and Dipartimento di Fisica e Astronomia, Universit\`a di Bologna, viale Berti Pichat 6/2, I--40127 Bologna, Italy
   \and INAF - Osservatorio Astronomico di Bologna, via Ranzani 1, I--40127 Bologna, Italy      
    \and Universit\`a degli Studi di Roma "Tor Vergata", Via Orazio Raimondo 18, I--00173 Roma, Italy
    \and Dipartimento di Fisica e Astronomia, Universit\`a di Firenze, Via G. Sansone 1, I--50019, Sesto Fiorentino (Firenze), Italy
   \and INAF - Osservatorio Astrofisico di Arcetri, Largo E. Fermi 5, I--50125, Firenze, Italy	 
    \and Ohio State University, 140 West 18th Avenue, Columbus, OH 43210, USA 
   	 \and Centro de Astrobiolog\'{i}a (CSIC--INTA), Depto. de Astrof\'{i}sica, ESAC campus, Camino Bajo del Castillo s/n, E-28692 Villanueva de la Ca\~{n}ada, Spain         
   	    \and INAF - Istituto di Astrofisica e Planetologia Spaziali, via Fosso del Cavaliere 100, I-00133 Rome, Italy
   	    \and  Max Planck Institute for Radio Astronomy, Auf dem Hugel 69, D-53121 Bonn, Germany}

  \date{}

 
  \abstract{We perform a survey of the X-ray properties of 41 objects from the WISE/SDSS selected Hyper-luminous (WISSH) quasars sample, which includes 86 broad-line quasars with bolometric luminosity $L_{\rm Bol} \simgt 2 \times 10^{47} \, {\rm erg\, s^{-1}}$ shining at $z$ $\sim$ 2-4. We  use both proprietary and archival {\it Chandra} and {\it XMM-Newton} observations.
 Twenty-one quasars have sufficient quality data to perform a spectroscopic analysis, while for the remaining sources, X-ray properties are derived through hardness-ratio analysis (apart for six sources which result to be undetected). The bulk ($\sim$ 70\%) of the detected WISSH quasars exhibit \nh\ $<$ 5 $\times$ 10$^{22}$ \cm2, in agreement with their optical Type 1 AGN classification. All but three quasars show  unabsorbed 2-10 keV luminosities \lum\ $\geq$  10$^{45}$ \ergs. 
Thanks to their extreme radiative output across the Mid-IR-to-X-ray
range, WISSH quasars therefore offer the opportunity to significantly extend and validate the  existing relations involving \lum. Specifically, we study the X-ray luminosity as a function of (i)  X-ray-to-Optical (X/O) flux ratio,  (ii) mid-IR luminosity ($L_{MIR}$),  (iii) $L_{\rm Bol}$ as well as (iv) \aox\ versus the 2500\AA\ luminosity.
We find that the WISSH quasars show (i) unreported very low  X/O  ($<0.1$)  compared to typical AGN values;
(ii)  \lum/$L_{MIR}$ ratios significantly smaller than those derived for AGN with lower luminosity; (iii) a
large X-ray bolometric correction  $k_{\rm Bol,X}$ $\approx$ 100-1000; and (iv) steep  -2 \simgt \aox\ \simgt\ -1.7.
These results lead to a scenario in which the X-ray emission of hyper-luminous quasars is relatively weaker compared to lower-luminosity AGN.
 Models predict that such an X-ray weakness can be relevant for the acceleration of powerful high-ionization emission line-driven winds, commonly detected in the UV spectra of WISSH quasars, which can in turn perturb the X-ray corona and  weaken its emission. 
Accordingly,  hyper-luminous quasars represent the ideal laboratory
to study the link between the AGN energy output and wind acceleration.
Additionally, WISSH quasars exhibit very large SMBH masses ($\log[M_{\rm BH}/M_{\odot}]$ \simgt\ 9.5). This
 enables  a more robust modeling  of the  $\Gamma$-$M_{\rm BH}$  relation by increasing the statistics
 at high masses. We derive a flatter $\Gamma$ dependence than previously found over the broad range  5 \simlt\ $\log(M_{\rm BH}/M_{\odot})$ \simlt\ 11.
Finally, we estimate that only 300 ks observation of X-IFU on board  {\it Athena} will offer a detailed view of 
the properties of  absorption features associated to powerful X-ray SMBH winds for a representative sample of WISSH quasars.
}


   \maketitle
%

\section{Introduction}

X-ray observations have been demonstrated to be a key tool to probe the nature of the innermost region of
active galactic nuclei (AGN). 
The power-law like spectrum of the X-ray continuum emission is interpreted as the result of Compton up-scattering of thermal UV photons produced from the optically-thick accretion disk  surrounding the SMBH inside an optically-thin corona, with an electron temperature of $kT_e$ $\sim$ 50--100 keV (\citealt{hm91}; \citeyear{hm93}; \citealt{haardt94, petrucci00, reis13}).
This two-phase, disk-corona  model has been supported by broad-band UV-to-X-rays  and ultra-hard X-ray observations (e.g., \citealt{zdz96};  \citealt{nandra00}; \citealt{petrucci13}), although the exact geometry and  size of the corona are still largely unconstrained.
Furthermore, the X-ray radiation emitted by the corona impinging onto the underlying accretion disk and the circumnuclear gas produces the so-called {\it reflection} emission, whose bell shape peaked at $\sim$ 30 keV is the result of photo-electric absorption  al low energies and Compton scattering at high energies \citep{ferland88, matt91}.
The X-ray spectral continuum in AGN can be also modified by the presence of cold and warm absorbers, located at different distances ($\sim$ 0.1 pc -- 100 pc) from the SMBH along our line of sight \citep{bianchi12}.
Absorption occurring in highly-ionized,  high-velocity ($\sim$ a few  10$^{4}$  ${\rm km \, s^{-1}}$) material very close to the accretion disk ($\sim$ 100 gravitational radii) has been also revealed (\citealt{tombesi15}; \citealt{kingpounds15} and references therein).

Thanks to their high X-ray fluxes, local Seyfert-like AGN (\lbol\ $\sim$ 10$^{44-45}$ \ergs) and moderately-luminous quasars (\lbol\ $\sim$ 10$^{45-46}$ \ergs) have been typically targeted by X-ray facilities, 
and  our knowledge on the origin of the X-ray continuum  and the physical and spectral properties of the X-ray emitting/absorbing regions in AGN have been basically derived by observations of these classes of sources \citep{reynolds97, nandra97, picoPG05, turnermiller09}.
On the contrary, the properties of the X-ray emission and absorption in quasars at the tip of the luminosity scale (and hence, rare), which typically shine at $z >$2, have remained less investigated since they need time-consuming observations and because of their low number density.

A fundamental improvement in the study of the luminous (\lbol\ $\sim$ 10$^{46-47}$ \ergs) and hyper-luminous (\lbol\ $>$ 10$^{47}$ \ergs) quasars in the $\sim$0.3--10 keV band have been provided by \xmm\ and \chandra\ thanks to their  high sensitivity and angular resolution. Both survey programs and targeted observations have been successful in  providing unprecedented constraints on X-ray spectral and evolutionary properties of these powerful AGN, both unobscured and obscured ones (see \citealt{vignali03, page04, lafranca05, just07, bianchi07, shemmer08}; \citealt{young09, reeves09, vignali10, pico15}). 

It has emerged that the slope of the X-ray continuum does not show strong dependence on redshift or luminosity 
(\citealt{nanni17} and references therein) .
On the contrary, it has been found that the intensity of the reflection features in the spectra of luminous quasars  is weak compared to AGN at lower luminosities (e.g., \citealt{reevesturner00}; \citealt{jim05}; \citealt{bianchi07}; Zappacosta et al. 2017 submitted). 
In addition,  many studies report that the UV-to-X-ray spectral energy distribution (typically described by the relationship between the 2500 \AA\ and 2 keV monochromatic luminosities, i.e. \aox) depends primarily upon the UV luminosity  \citep{avni82, vignali03, steffen06, lusso10, lussorisaliti16}, while the redshift dependence is weak.
These results lend support to a scenario whereby the mechanism responsible for the primary continuum emission in AGN remains almost identical from $z \approx 0$ to $z \approx 6$, while the more the AGN luminosity increases, the less the energy in the X-ray band is emitted relative to that in the UV/optical range.
More recently, some works have reported that the X-ray to mid-infrared (MIR) luminosity relation shows a different
behavior from low to high  MIR luminosity AGN. Specifically, the ratio between X-ray and MIR luminosity  is smaller for 
MIR luminous ($\geq$ 10$^{46}$ \ergs) objects \citep{lanzuisi09, stern15}.

A comprehensive investigation of nuclear properties of hyper-luminous quasars is also important for our understanding of the outflows launching mechanism and, hence, AGN-galaxy self-regulated growth.  
Many models indicate that an intense radiation field can be able to
accelerate winds out from the immediate vicinity of the AGN accretion disk
\citep{murray95, proga05, kingpounds03} via line-driving and radiation pressure.
 A low X-ray illumination
of the outflowing gas is necessary to avoid the
suppression of UV line driving by over-ionization, and explains the  large blueshifts ($>$ 2000 ${\rm km \, s^{-1}}$)  of the CIV emission line observed in the most luminous quasars (i.e. the so-called "wind-dominated" quasar population, e.g. \citealt{richards11}).
Furthermore, the relation between AGN luminosity and wind terminal velocity observed for broad (FWHM $\sim$ 10$^3$ ${\rm km \, s^{-1}}$) absorption line (BAL) quasars can be indeed easily accounted for by a radiatively–driven outflow \citep{laor02}.
\citet{brandt00} found that the equivalent width of the CIV BALs anti-correlates with the \aox,
indicating a strong link between powerful winds and soft X-ray weakness in quasars.
It has been also reported that 
objects typically showing strong outflows (i.e. BAL quasars and AGN-dominated ultra-luminous IR galaxies, ULIRGs; \citealt{sturm11,cicone14}) are indeed X-ray weak compared to normal quasars \citep{imanishi04, sabra01}.
The X-ray spectrum of BAL quasars and ULIRGs typically shows large obscuration and the X-ray absorbing medium has been considered the main cause for a reduced X-ray emission and very steep \aox\ values (e.g. \citealt{greenmathur96}; \citealt{mathur00}; \citealt{gallagher02,picoPG05}). The X-ray absorber can also  shield the BAL clouds from the ionizing continuum  \citep{kaastra14}.
Remarkably, even taking into account the measured X-ray absorption, a large fraction of BAL quasars still remain X-ray
weak \citep{gibson08, luo13},
i.e. they are unable to produce strong X-ray emission. The prototype of such a class of intrinsically weak quasar is  Mrk 231 for which
a recent NuSTAR observation has confirmed the intrinsic X-ray weakness \citep{teng14}.

The latter is also a typical feature observed in the so-called {\it weak emission line quasars} (i.e. objects showing Ly$\alpha$ and CIV emission with an equivalent width EW $\leq$ 10 \AA) and quasars with highly blueshifted ($>$ 2000 km s$^{-1}$) CIV emission lines (e.g., \citealt{wu12}).

In this paper, we present the X-ray spectral properties of the WISE/SDSS selected hyper-luminous  (WISSH) quasars sample and report on the correlations between the X-ray and multiwavelength (Optical, UV and MIR) properties. The WISSH quasars project consists of a multi-band (from millimeter wavelengths up to hard X-rays) investigation of 86 hyper-luminous ($L_{\rm Bol} \geq 2 \times 10^{47} \, {\rm erg\, s^{-1}}$), broad-line  quasars
at $z \sim 1.8-4.8$. This sample has been obtained by cross-correlating the {\it WISE} All-Sky source catalog (for 22$\mu$m flux density $S_\nu$(22$\mu$m) $>$ 3 mJy)
and  the SDSS DR7 quasar catalog  (for 1.5 $<z<$ 5). We refer to \citet{bischetti16} for a detailed description of the WISSH quasar sample and the main goals of this multi-band  project aimed at performing a systematic study of the nuclear, outflows and host galaxy properties of the most powerful quasars.
This paper has been organized as follows. Details about X-ray observations of the sources in the  WISSH sample, data reduction and source detection are presented in Sect. \ref{sec:xobs}. In Sect. \ref{sec:Xresults} we present the results of the X-ray data analysis and we discuss the X-ray properties of the WISSH sample. In Sect. \ref{sec:Xmulti} we report on the correlation of X-ray versus Optical and MIR properties, while in Sect. \ref{sec:bolo} results on X-ray bolometric corrections  and  relations involving SMBH mass and Eddington ratio have been outlined. 
In Sect. \ref{subsec:xweakness} we summarise our results and discuss the relative  X-ray weakness of WISSH quasars compared to AGN at lower luminosities. We outline and conclude about possible future perspectives for the WISSH project with $Athena$ X-ray observatory in Sect. \ref{subsec:athena}.

A $\Lambda$CDM cosmology with $\Omega_\Lambda$ = 0.73,   $\Omega_M$ = 0.27 and $H_0$ = 70 ${\rm km \, s^{-1}}$  is assumed throughout.
Hereafter, errors correspond to 1$\sigma$ while upper limits are reported at 90\% confidence level, unless specified otherwise.


\begin{table*}
    \caption{The X-ray WISSH sample and log of the X-ray observations. Columns give the following information:(1) SDSS ID, (2) redshift, (3) SDSS AB magnitudes in the $i$ band, (4) X-ray observatory, (5) observation ID, (6) date of the X-ray observation, (7) net exposure time in ks, (8) net counts in the 0.5-8.0 keV band, (9) Galactic absorption by \citet{kalberla05} (in units of $10^{20} \, {\rm cm^{-2}}$), (10) rest-frame 6 $\mu m$ luminosities from Duras et al. (2017, in prep.)  (in units of Log[$L/{\rm erg \, s^{-1}}$]).}      
    \label{tab:xdata}
    \centering          
    \begin{tabular}{c c c c c c c c c c}     
        \toprule\toprule
        SDSS & z & $i$  & X-ray Obs. & Obs. ID & Obs. Date & Exp (ks) & Net Counts & $N_{\rm H}^{\rm Gal}$ & Log$\lambda L_{6 \mu m}$\\
        (1)   &(2)& (3) & (4)        & (5)     & (6)       & (7)        & (8) & (9) & (10)\\
        \midrule
         J0045+1438$^B$        & 1.992$^a$  & 16.99	   & Chandra & 6889   & 2006-07-24   & 11.5 & 10.9 $\pm$  3.6	   & 5.27			&  46.96  \\
 J0209-0005  & 2.856$^a$  & 16.99		   & Chandra$^\dagger$ & 17078  & 2014-11-18	   & 29.7 & 146.2 $\pm$ 12.3		& 2.32  &  47.09   \\
 J0735+2659  & 1.982$^a$  & 16.14		   & Chandra$^\dagger$ & 17077  & 2015-10-02	   & 24.7 & 187.3 $\pm$ 13.8		& 4.87  &  47.07   \\
 J0745+4734	      & 3.225$^b$  & 16.29	   & Chandra & 1330   & 2012-01-01	   & 1.5  & 81.4 $\pm$ 9.1		& 5.76  	&  47.31   \\
 J0747+2739	      & 4.11$^a$   & 17.91	   & Chandra & 3561   & 2002-12-03	   & 5.0  & 21.5 $\pm$ 4.7	 & 3.66 		&  46.83   \\
 J0801+5210  & 3.263$^b$  & 16.76		   & Chandra$^\dagger$ & 17081  & 2014-12-11	   & 43.5 & 173.5 $\pm$ 13.4		& 4.32  &  47.20      \\
 J0900+4215	      & 3.294$^b$  & 16.69	   & Chandra & 6810   & 2006-02-09	   & 3.9  & 109.6 $\pm$  10.5	 & 1.23 		&  47.26     \\
 J0904+1309  & 2.974$^a$  & 17.04		   & XMM$^\dagger$     & 0745010301 & 2014-11-24 & 24.6 & 1330.6 $\pm$ 38.7		& 2.83  &  47.32     \\
 J0947+1421	& 3.04$^a$   & 17.01		  & Chandra & 13325  & 2012-05-28   & 1.6  & $4.8 \pm 3.0$		& 3.05  		&  47.05     \\
 J1014+4300	      & 3.126$^a$  & 16.38	  & Chandra & 6809   & 2006-06-14   & 4.1  & 33.5 $\pm$ 5.8	  & 1.34			&  47.16     \\
 J1027+3543	      & 3.112$^a$  & 16.59	  & Chandra & 13312  & 2012-04-02   & 1.6  & 26.8 $\pm$ 5.2	       & 0.99			&  47.40     \\
 J1057+4555	      & 4.14$^a$   & 17.29	  & Chandra & 878	   & 2000-06-14   & 2.9  & 27.5 $\pm$ 5.3	       & 0.90		&  47.24	\\
 J1106+6400	      & 2.22$^b$   & 15.98	  & Chandra & 6811   & 2006-07-16   & 3.65 & 123.5 $\pm$ 11.1	  & 1.05			&  47.08	\\
 J1110+4831	      & 2.957$^a$  & 16.55	  & XMM     & 0104861001 & 2002-06-01 & 26.8 & 425.5 $\pm$ 22.0 	       & 1.73		&  47.27	\\
 J1111+1336  & 3.492$^b$  & 17.18		  & Chandra$^\dagger$ & 17082  & 2015-01-26   & 43.1 & 180.7 $\pm$ 13.6 	& 1.48  	&  47.07	\\
 J1159+1337	     & 3.984$^a$  & 17.56	  & Chandra & 13323  & 2012-06-29   & 1.6  & $2.9 \pm 1.7$	       & 2.26			&  47.12    \\
 J1200+3126	      & 2.993$^a$  & 16.36	  & Chandra & 13309  & 2012-03-18	  & 1.5  & 15.8 $\pm$ 4.0	       & 1.64		&  47.15     \\
 J1201+0116	     & 3.247$^b$  & 17.32	  & Chandra & 13345  & 2012-02-10	  & 1.6  & $\leq$ 5.4		       & 1.76		&  47.07      \\
 J1201+1206	      & 3.512$^b$  & 17.31	  & Chandra & 13324  & 2012-06-11	  & 1.6  & 15.8 $\pm$ 4.0	       & 1.81		&  47.10     \\
 J1210+1741$^B$     & 3.64$^a$   & 17.73	  & Chandra & 13366  & 2012-07-09   & 1.6  & $\leq 2.5$ 	       & 2.83			&  47.16      \\
 J1215-0034$^B$       & 2.707$^a$  & 17.13	  & Chandra & 4201   & 2003-11-14   & 44.5 & 137.7 $\pm$ 11.8	  & 1.95			&  47.08       \\
 J1236+6554	      & 3.424$^b$  & 17.19	  & Chandra & 6817   & 2006-08-29   & 4.1  & 20.5  $\pm$  4.8	  & 1.71			&  47.03	\\
 J1245+0105$^B$       & 2.798$^a$  & 18.12	  & Chandra & 2974   & 2002-05-03   & 6.7  & 10.4  $\pm$  3.3	  & 1.52			&  46.88	\\
 J1249-0159	      & 3.638$^a$  & 17.73	  & Chandra & 13335  & 2011-12-03   & 1.6  & 3.8 $\pm$ 2.0	       & 1.61			&  47.01	\\
 J1250+2631	      & 2.044$^a$  & 15.37	  & XMM     & 0143150201 & 2003-06-18 & 18.6 & 3099.6 $\pm$  56.8	       & 0.86		&  47.22	\\
 J1328+5818$^B$      & 3.133$^a$  & 18.57	  & XMM     & 0405690501 & 2006-11-25 & 43.1 & 34.3  $\pm$  8.9        & 1.58			&  46.83	 \\
 J1333+1649	      & 2.089$^a$  & 15.99	  & Chandra & 867	       & 2000-04-03 & 3.0  & 159.6 $\pm$  12.6         & 1.66		&  47.06	  \\
 J1421+4633	  & 3.454$^b$  & 17.22     & Chandra & 12859	    & 2011-06-20 & 23.6 & 51.2  $\pm$ 7.3		& 1.12  		&  47.04	  \\
 J1422+4417	     & 3.647$^b$  & 17.57	  & Chandra & 13360	 & 2011-11-11 & 1.5  & $\leq 2.5$	       & 1.01			&  47.38	  \\
 J1426+6025	      & 3.189$^a$  & 16.23	  & XMM     & 0402070101 & 2006-11-12 & 5.8  & 181.9 $\pm$  14.4	       & 1.74		&  47.43	  \\
 J1433+0227	    & 4.62$^a$   & 18.33	  &  Chandra & 3959	  & 2003-04-20 & 3.5  & $\leq$ 3.8		       & 2.59		&  47.02     \\
 J1441+0454	      & 2.059$^a$  & 17.08	  & Chandra & 12860	& 2012-02-28 & 21.5 & 78.2 $\pm$  8.9	       & 2.75			&  46.80     \\
 J1506+5220$^B$ 	 & 4.068$^a$  & 18.29	   & Chandra & 4071    & 2012-10-24   & 4.9  & $\leq$ 5.1			& 1.85  	&  47.09      \\
 J1513+0855$^B$ & 2.897$^a$  & 17.09		    & Chandra$^\dagger$ & 17079  & 2016-04-06	& 29.7 & 298.0 $\pm$ 17.4	  & 2.84	&  47.30     \\
 J1521+5202	  & 2.218$^b$  & 15.44         & Chandra & 15334 & 2013-10-22	 & 37.4 & 87.5  $\pm$  9.5		& 1.58  		&  47.22     \\
 J1538+0855   & 3.564$^b$  & 17.00 	  & Chandra & 13314   & 2012-05-02   & 1.6 & $\leq 2.5$ 		& 3.06  		&  47.16       \\
 J1549+1245$^B$  & 2.365$^b$  & 17.38		   & XMM$^\dagger$     & 0763160201 & 2016-02-04 & 30.4 & 520.6 $\pm$	   32.4  & 3.47        &   47.14	\\
 J1621-0042	      & 3.71$^a$   & 17.26	   & Chandra & 2184	 & 2001-09-05  & 1.6  & 27.6  $\pm$ 5.3      & 6.59		       &   47.08	\\
 J1639+2824$^B$       & 3.801$^a$  & 17.21	   & Chandra & 13315	 & 2011-11-24	 & 1.5  & 5.8 $\pm$ 2.4 		& 3.24         &   47.52	\\
 J1701+6412	      & 2.737$^a$  & 15.84	   & Chandra & 9756	 & 2007-11-14	  & 32.3 & 214.4 $\pm$ 14.7	& 2.28  	       &   47.32	\\
 J2123-0050  & 2.283$^b$  & 16.34		   & XMM$^\dagger$     & 0745010401 & 2014-11-14 & 21.9 & 785.2 $\pm$  30.4	   & 3.84      &   47.05	 \\
        \bottomrule\bottomrule
    \end{tabular}
    \tablefoot{ $B$ : BAL Quasars according to the modified absorption index (AI$_{1000} $>$ 100$, \citealt{bruni12}), a more conservative version of the \cite{hall02} definition. The detailed analysis of the BAL properties of WISSH quasars will be presented in a forthcoming paper (Bruni et al. in prep.). $^a$: redshifts from the SDSS DR10 catalog.  $^b$: redshifts from LBT-LUCI near-IR spectroscopy (Vietri et al. in preparation).  $^\dagger$ : Proprietary data observations.}
\end{table*}

\section{X-ray Observations}
\label{sec:xobs}

\subsection{X-ray Coverage of the WISSH sample}

The X-ray sample of WISSH quasars (X-WISSH, hereafter) consists of 41 objects observed with {\it Chandra} or {\it XMM-Newton}, corresponding to 48\% of X-ray coverage for the WISSH sample. Table \ref{tab:xdata} provides information about sources in the X-WISSH sample and  X-ray observations (34 observations by {\it Chandra} and 7 by {\it XMM-Newton}). All sources but J1328+5818 (\xmm\ Obs. ID. 0405690501) were the principal targets of the observations. We used 33 archival and 8 proprietary observations. 
The latter represent 20\% of the sample (namely, J0209-0005, J0735+2659, J0801+5210, J1111+1336, J1513+0855 with \emph{Chandra ACIS-S} and J0904+1309, J1549+1245, J2123-0050 with \xmm). These sources are indicated with a cross in column 4 in Table \ref{tab:xdata}.

\subsection{Data Reduction and Source Detection}
\label{subsec:dr}

{\it Chandra} data were reduced and analysed with the \emph{Chandra Interactive Analysis of Observations}  CIAO 4.7 package\footnote{See \url{http://cxc.harvard.edu/ciao}} with CALDB 4.7.0 by standard procedures following the CIAO {\it Science Threads}. For each source, we created an image in the 0.5-8 keV energy band from the event file by using the \texttt{dmcopy} tool.
In order to extract source and background counts, we used a circular region centered on the SDSS source position, with radii between $2''$ and $6''$. For the background regions, we usually considered an annulus centered on the source position with inner radius ranging from $2''$ to $7''$ and outer radius $\geq 30''$. However, in some cases a circular aperture (radius $\geq 20-30''$) close to the source position was preferred in order to avoid contamination from nearby X-ray sources. Net counts were extracted using the \texttt{dmextract} task. Errors at 1$\sigma$ confidence level were estimated either according to the Gaussian statistics for large number of counts ($N \gtrsim 20$) or according to the Gehrels approximation for lower number of counts ($1\sigma=1+\sqrt{N+0.75}$, \citealt{gehrels}). However, the corresponding lower limit ($1\sigma=\sqrt{N-0.25}$) is smaller, and \texttt{dmextract} adopts the larger error to be conservative.
Using Poisson statistics, we then calculated the false-positive probability in order to establish the robustness of the X-ray detection. We found that 6 sources (i.e., $\sim 15\%$ of X-WISSH) resulted undetected, according to a probability threshold of $10^{-5}$. A 90\% upper limit on count rate was assumed for these sources.

For the quasars observed with \xmm, the data reduction was carried out with \emph{Science Analysis System} SAS\footnote{See \url{http://xmm.esac.esa.int/sas/}} software package, 14.0.0 version. The EPIC $pn$ Observation Data Files (ODFs) have been processed with the \texttt{epproc} task in order to generate the event files. These files were filtered in order to remove flaring background periods. We applied the standard methods described in the SAS threads. Accordingly, we extracted  lightcurves in the 10-12 keV  range in order to exclude intervals contaminated by high-energy particles background. Good Time Interval (GTI) tables were then generated by setting rates less than 0.4 cts/s ($RATE \leq 0.4$). The cleaned event file was created for energies extending from 0.3 keV up to 10 keV. 
Counts were extracted  by using a circular aperture with radii of $\sim 20-30''$( $\sim 70-100''$)
for  the source(background) region.
Table \ref{tab:xdata} reports net counts in the 0.5-8.0 keV band with corresponding uncertainties.

\section{Results of X-ray Analysis}
\label{sec:Xresults}
\subsection{Hardness Ratio Analysis}
\label{subsec:HR}
As a first step,  we performed the Hardness Ratio (HR hereafter) analysis for all the detected sources, i.e. 35 objects, in order to make inferences about the basic X-ray emission properties of these quasars. 

We created images in the soft (0.5-2 keV) and hard (2-8 keV) bands from each source event file and we extracted net counts with relative uncertainties in each band as in Sect. \ref{subsec:dr}.
The HR compares the number of counts obtained in two or more energy bands. We specifically used this definition of Hardness Ratio:
\begin{equation}
	HR=\frac{H-S}{H+S},
\end{equation}
where $H$ is the number of counts in the hard band and $S$ is the number of counts in the soft band.

A power law model with photon index $\Gamma = 1.8$  modified by intrinsic absorption with  5 $\times$ $10^{21}$ $\leq$ \nh\  $\leq$   $10^{24}$ \cm2\ was assumed to simulate the spectrum of our sources with WebPIMMS\footnote{See \url{https://heasarc.gsfc.nasa.gov/cgi-bin/Tools/w3pimms/w3pimms.pl}}.
The lower value of \nh\ =  5 $\times$ $10^{21}$ \cm2\ was set by the $z$ \simgt\ 2 of our sources, for which the photoelectric cut-off falls outside the 
energy range of our observations.
Specifically, we used Chandra WebPIMMS\footnote{See \url{http://cxc.harvard.edu/toolkit/pimms.jsp}}, provided by the Chandra Data Center, as it allows to specify the observation cycle for Chandra data, taking into account the decreasing quality of the ACIS detector caused by piled-up dust over time.

This analysis revealed that the majority of the sources ($\sim 90$\% of the X-WISSH) have negative HR values, suggesting little absorption in the X-rays. Indeed, by comparing the HR values with the simulated absorbed power-law models, we were able to constrain the \nh values and we found that the bulk of the objects are consistent with \nh $\leq 5 \times 10^{22} \,{\rm cm^{-2}}$. In Table \ref{tab:data} we report the HR-based \nh\ values except for the sources with best available data for which X-ray spectroscopic analysis was performed (see  Sect. \ref{subsec:xspec}). 
In case of a value of HR corresponding to a \nh\ consistent within uncertainties with   5 $\times$ $10^{21}$ \cm2, we list the \nh\ value as 90\% upper limit.
To estimate X-ray fluxes, we used WebPIMMS by assuming  an absorbed  power law  with $\Gamma=1.8$ and \nh\ corresponding to the estimated HR value (see Table \ref{tab:data}).
For J1157$+$1337 and  J1236$+$6554 which had  upper limits on HR, we used  5 $\times$ $10^{21}$ \cm2\
to derive their fluxes and luminosities.
Table \ref{tab:data} also lists the unabsorbed 2-10 keV luminosity ($L_{2-10}$ hereafter). 

Finally, for the 6 undetected sources (see Sect. \ref{subsec:dr}), we assumed a \nh\ equal to the maximum value of the sample, i.e. \nh$= 4 \times 10^{23} \, {\rm cm^{-2}}$, in order to derive  conservative values for X-ray fluxes and luminosities.

\subsection{X-ray Spectroscopy}
\label{subsec:xspec}

We performed  X-ray spectroscopy for the 21 sources  in the X-WISSH sample  detected with $\geq 40$ net counts.
We created the redistribution matrix file (RMF) and auxiliary response file (ARF), using \texttt{mkacisrmf} and \texttt{mkarf} CIAO tools for \emph{Chandra} sources and \texttt{rmfgen} and \texttt{arfgen} SAS tools for \emph{XMM} sources. We extracted the spectra with the same source and background regions used for the counts estimation (Sect. \ref{subsec:dr}).
In fitting the spectra using  \texttt{XSPEC v.12.8.2}, we consistently applied the Cash statistics to all spectra (\citealt{cash79}), being more appropriate for low-count spectra.
Accordingly, we rebinned the source plus background spectra in order to ensure that
at least one count is included in each spectral bin.
Spectra collected with \emph{Chandra}(\xmm) were fitted in the  0.3-8(0.3-10) keV band.

A first characterization of the continuum shape of these sources was obtained by fitting each spectrum with a simple model consisting of a power law plus Galactic absorption (denominated  PL model hereafter). 
The majority of the sources exhibit photon index close to $\Gamma \sim 1.8$ which is the typical value for a broad line quasar (\citealt{picoPG05}). However, eight sources show a flat photon index ($\Gamma \leq 1.5$),  likely due to the presence of intrinsic \nh not accounted for in the PL model. We then fitted each spectrum by adding an extra absorption component to the PL model (named APL hereafter).
We found  that for 13 out of 21 sources, i.e. $\sim 60\%$ of the X-ray spectral sample, the PL model provides a reasonable fit to the spectra, while
the presence of an extra  absorption component is required by the data for seven WISSH quasars.
For two sources, namely J1333+1649 and J1421+4633, fitting with the  APL model  yielded an upper limit on \nh and the resulting $\Gamma$ remained quite flat ($\sim 1.5$) and, therefore,  we  fixed the photon index to the canonical AGN value of $\Gamma = 1.8$.
From the spectral analysis, it  emerged that the majority of the sources show little or no intrinsic absorption, with 14 out of 21 quasars  exhibiting a value of  \nh  $ < \, {\rm few} \, 10^{22} \, {\rm cm^{-2}}$.
Finally,  the \nh\ values  inferred from the spectral analysis are consistent with those based on HR.

\begin{figure}
	\centering
	\includegraphics[trim={2.5cm 0.5cm 0 5cm}, clip=true, scale=0.35, angle=-90]{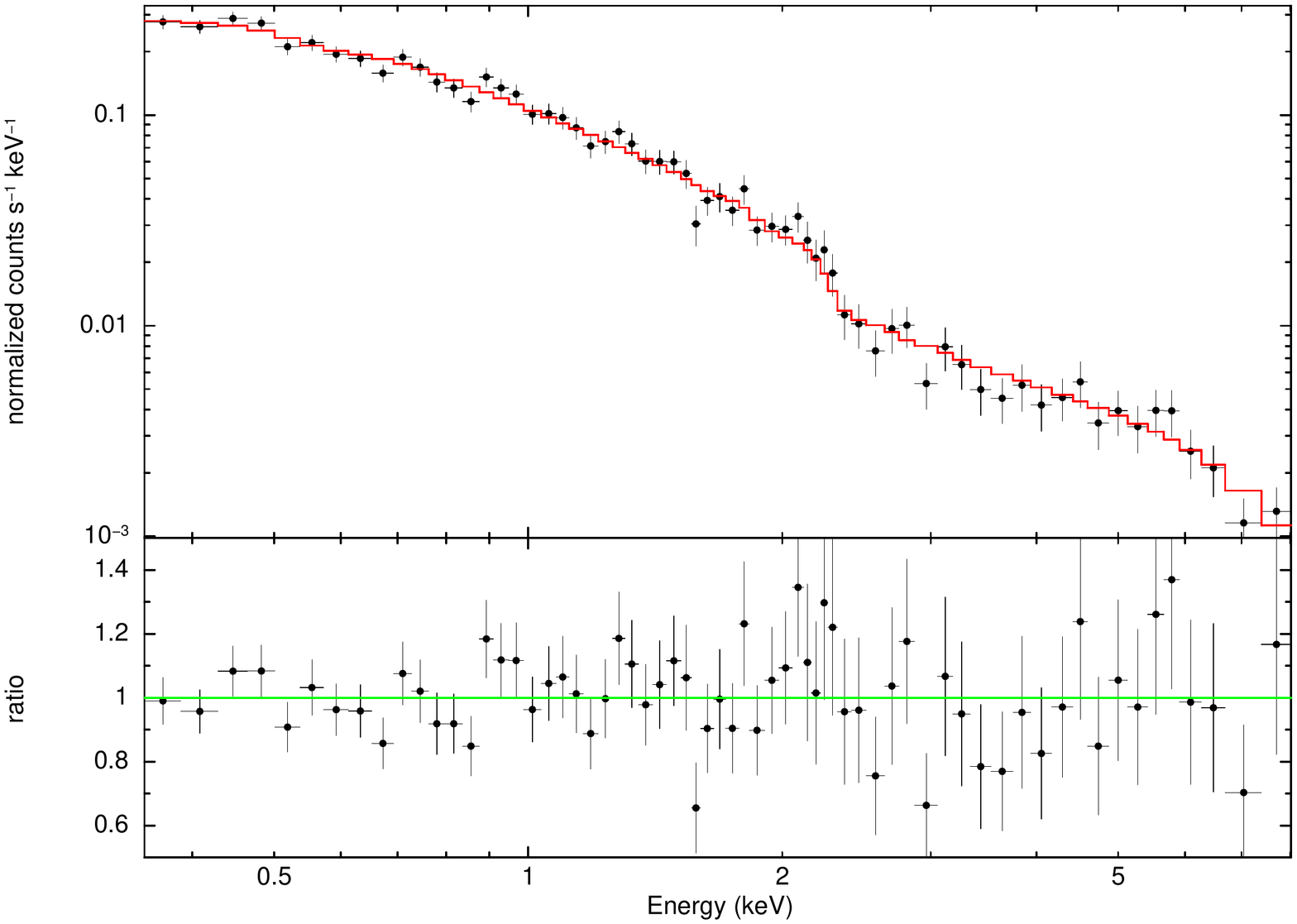}
	\caption{\xmm\ spectrum (black points) of
		J1250+2631 when the model PL+Reflection (red solid line) is applied. The lower panel
		shows the data-to-model ratios.}
	\label{fig:spectra}
\end{figure}

\begin{figure*}
	\centering
	\includegraphics[trim={1.3cm 0 1.5cm 0}, clip=true,scale=0.37]{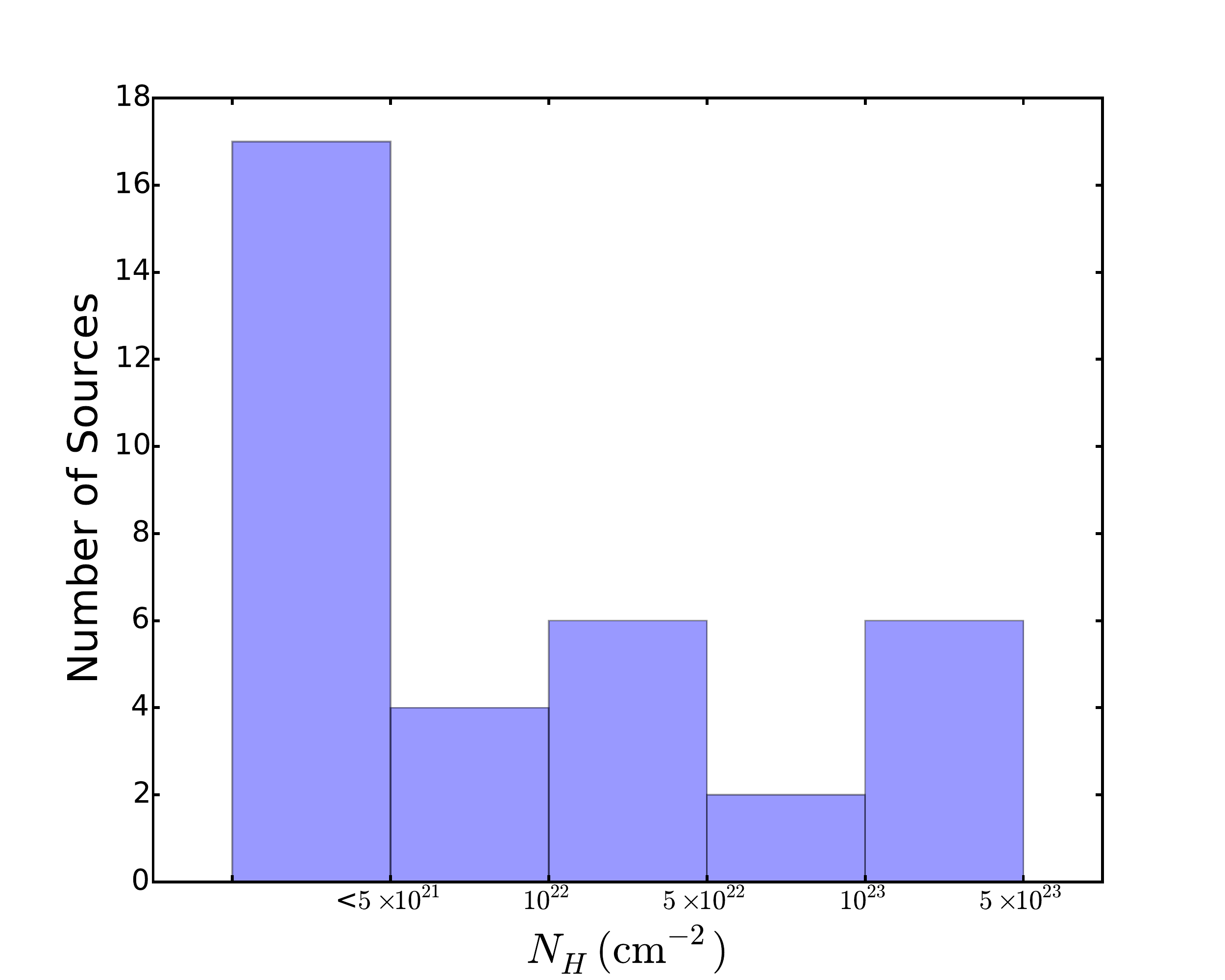}
	\includegraphics[trim={0 0 1.9cm 0}, clip=true,scale=0.37]{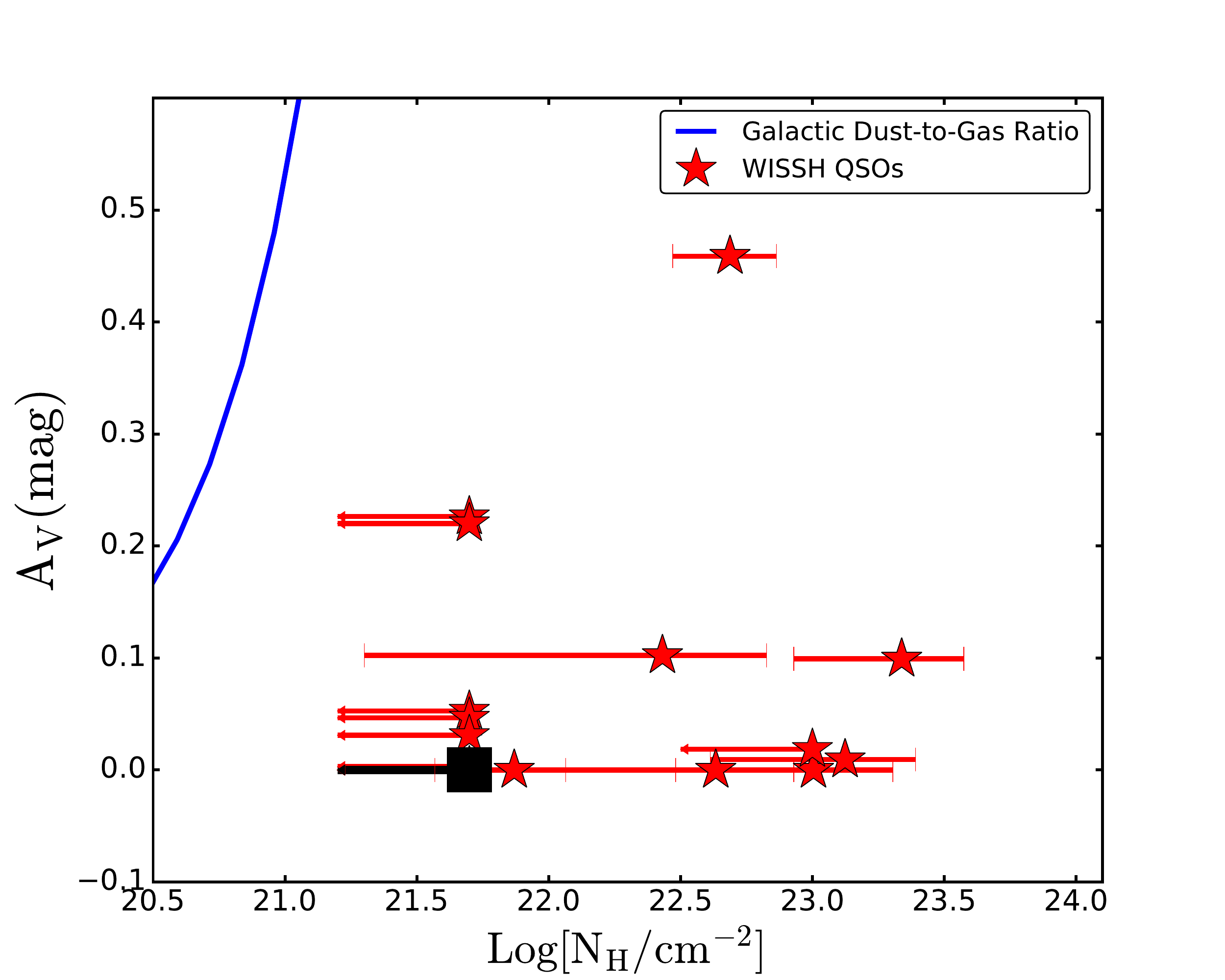}
	\caption{{\it Left Panel}: The distribution of absorption column densities for the 35 X-ray detected quasars, derived by HR or spectral analysis. {\it Right Panel}: ${\rm A_V}$ magnitudes versus X-ray absorption column densities derived by spectral analysis. The black square represents values of \nh = $5 \times 10^{21} \, {\rm cm}^{-2}$ and $A_{\rm V} = 0$ for 8 sources (see Table \ref{tab:data}). The blue solid line represents the relation obtained by assuming a Galactic dust-to-gas ratio, i.e. ${\rm A_V}=10^{(\log {N_H} -21.278)}$ mag (\citealt{maiolino01}).}
	\label{fig:histonh}
\end{figure*}

\begin{table*}
	\caption{X-ray properties of the WISSH quasars. Columns give the following information:(1) SDSS name, (2) X-ray photon index, (3) absorption \nh\ (in units of $10^{22} \, {\rm cm^{-2}}$), (4) 2-10 keV fluxes (in units of $10^{-14} \, {\rm erg \, cm^{-2} \, s^{-1}}$), (5) 2-10 keV unabsorbed luminosities (in units of Log[$L/{\rm erg \, s^{-1}}$]), (6) Hard X-ray fluxes and luminosities were derived by using the \nh\ value from HR and $\Gamma$ = 1.8 or from  
	spectral analysis (models: PL = Power-law, APL = Absorbed Power-law), (7) X-ray-to-optical flux ratios (X/O) in units of $10^{-2}$, (8) \aox values.}            
	\label{tab:data}      
	\centering          
	\begin{tabular}{c c c c c c c c}     
		\toprule\toprule   
		SDSS & $\Gamma$ & \nh & $f_{2-10}$ & Log$L_{2-10}$ & Model & X/O &\aox \\ 
		(1)   & (2) & (3) & (4) & (5) & (6) & (7) & (8)\\
		\midrule                    
		J0045+1438        & 1.8$^f$ 	       &  $\leq 22(4)^b$ 	    & 0.69 & 44.24	 & APL$^*$ & 0.3 & -2.21    \\
		J0209-0005        & $1.25_{-0.15}^{+0.15}$ &  $\leq 1.6$		    & 5.25 & 45.16	& PL  & 2.3 & -1.98   \\
		J0735+2659        & $1.54_{-0.13}^{+0.13}$ &  $\leq 3.6$		    & 7.08 & 45.11	& PL  & 1.4  & -1.89   \\
		J0745+4734        & $1.83_{-0.18}^{+0.18}$ &  $\leq 3.5$		    & 30.9 & 46.37	& PL & 7.2  & -1.54  \\
		J0747+2739        & 1.8$^f$ 	      &  $\leq 30(4)^b$		    & 2.13 & 45.43	& APL$^*$ & 2.2  & -1.72    \\
		J0801+5210        & $1.81_{-0.13}^{+0.14}$  &  $\leq 0.8$	    & 2.66 & 45.25	& PL  & 1.0  &    -1.95    \\
		J0900+4215        & $1.8_{-0.15}^{+0.16}$    &  $\leq 2.3$		& 13.5 & 46.00    & PL  & 4.5 &  -1.66   \\
		J0904+1309        & $2.05_{-0.04}^{+0.04}$ &  $\leq 0.8$		    & 9.04 & 45.89	& PL   & 4.2 & -1.66  \\
		J0947+1421        & 1.8$^f$ 		 &  $\leq 16(0.5)^b$	    & 1.61 & 45.01   &   APL$^*$ & 0.7 &  -1.94  \\
		J1014+4300        & $1.74_{-0.30}^{+0.32}$ &  $\leq 3.3$		    & 3.55 & 45.43	& PL&  0.9  & -1.91  \\
		J1027+3543        & 1.8$^f$	      &  $\leq 14 (1)^b$		    & 9.16 & 45.79	& APL$^*$  & 2.8 & -1.74       \\
		J1057+4555        & 1.8$^f$ 	      &  $\leq 17 (4)^b$		    & 4.62 & 45.77	 & APL$^*$ & 2.7  & -1.70   \\
		J1106+6400        & $2.04_{-0.15}^{+0.15}$ &  $\leq 0.7$		    & 12.3 & 45.69	& PL & 2.2  & -1.70  \\
		J1110+4831        & $1.97_{-0.08}^{+0.08}$ &  $\leq 0.8$		    & 3.17 & 45.36	& PL  & 0.9 & -1.85 \\
		J1111+1336        & $1.78_{-0.13}^{+0.13}$ &  $\leq 5.7$		& 2.61 & 45.36   & PL  &  1.4 &   -1.85  \\
		J1159+1337        & 1.8$^f$ 	     &  $\leq 0.5$	    & 0.91$^c$ & 45.04  & APL$^*$& 0.7 &   -1.99          \\
		J1200+3126       & 1.8$^f$ 	      &  $\leq 15(1)^b$			& 5.78 & 45.55 & APL$^*$  & 1.4  & -1.71    \\
		J1201+0116        & 1.8$^f$ 	     & 40$^\dagger$	      & $\leq$ 4.21$^\ddagger$ & $\leq$ 45.54$^\ddagger$ & -  & $\leq$ 2.5 & $\geq$ -1.75 \\
		J1201+1206       & 1.8$^f$ 	      &  $ \leq 44(7)^b$		& 6.60	 & 45.77   & APL$^*$ & 3.9 & -1.68   \\
		J1210+1741       & 1.8$^f$    & 40$^\dagger$               & $\leq 1.72^\ddagger $ & $\leq 45.25^\ddagger$ & - & $\leq$1.5  &$\geq -1.91 $    \\
		J1215-0034        & $1.61_{-0.33}^{+0.34}$ &  $21.8_{-8.3}^{+9.2}$  & 5.25 & 45.33	 & APL & 2.6 &  -1.62 \\
		J1236+6554       & 1.8$^f$ 	      &  $\leq 0.5$ 		 & 2.55$^c$ & 45.33   & APL$^*$& 1.4  & -1.86   \\
		J1245+0105       & 1.8$^f$ 		   &   $30_{-22}^{+20}$		      & 1.84 & 45.03	  &  APL$^*$&  2.3  & -1.80  \\
		J1249-0159       & 1.8$^f$ 	      &  $ \leq 56(1)^b$			& 1.25 & 45.08   & APL$^*$ & 1.1  & -1.89  \\
		J1250+2631        & $2.35_{-0.07}^{+0.07}$ &  $\leq 0.3$		    & 29.9 & 45.94	 & PL$^a$ & 3.0  &  -1.55 \\
		J1328+5818       & 1.8$^f$ 		   &  $30_{-20}^{+40}$		      & 2.19 & 45.22	 & APL$^*$  & 4.2 &  -1.77 \\
		J1333+1649        & $1.8^f$			&  $0.7_{-0.2}^{+0.3}$  & 25.7 & 45.83	 & APL  & 4.6  & -1.62   \\
		J1421+4633        & $1.8^f$		    &  $13.3_{-5.5}^{+6.2}$   & 1.72 & 45.18	& APL & 0.9  & -1.79  \\
		J1422+4417        & 1.8$^f$ & 40$^\dagger$               &  $\leq 1.84^\ddagger$ &  $\leq 45.28^\ddagger$  & -  & $\leq$1.4 &$\geq $ -1.98   \\
		J1426+6025        & $1.79_{-0.13}^{+0.13}$ &  $\leq 0.8$		& 7.76 & 45.72   & PL  & 1.7 &  -1.83 \\
		J1433+0227        & 1.8$^f$    		    & 40$^\dagger$	       & $\leq 0.97^\ddagger$ & $\leq 45.22^\ddagger$ & - & $\leq$1.5 & $\geq$ -1.86  \\
		J1441+0454        & $1.7_{-0.3}^{+0.3}$    &  $2.7_{-1.6}^{+2.2}$     & 2.95 & 44.84	   & APL & 1.4  &-1.85 \\
		J1506+5220        & 1.8$^f$ 	     &  40$^\dagger$		 & $\leq 1.05^\ddagger$ & $\leq 45.14^\ddagger$ & - & $\leq$1.5  &$\geq$ -1.92  \\
		J1513+0855        & $1.71_{-0.11}^{+0.11}$ &  $\leq 13.4$	    & 7.94 & 45.60	 & PL  & 3.9 & -1.85 \\
		J1521+5202        & $1.4_{-0.4}^{+0.4}$   &  $10.1_{-6.2}^{+6.7}$	     & 3.52 & 44.85	 & APL  & 0.4 & -2.02   \\
		J1538+0855       & 1.8$^f$               & 40$^\dagger$  & $\leq 1.75^\ddagger$ & $\leq 45.24^\ddagger$  & - & $\leq$0.8 &$\geq $-1.91  \\ 
		J1549+1245        & $2.13_{-0.19}^{+0.22}$ &  $4.9_{-1.3}^{+1.6}$  & 3.98 & 45.34	& APL & 2.5  & -1.83   \\
		J1621-0042       & 1.8$^f$ 	      &  $10_{-9}^{+20}$		   & 10.7 & 46.04 & APL$^*$ & 6.1 & -1.58  \\
		J1639+2824       & 1.8$^f$ 		  &   $40_{-35}^{+60}$			& 4.08 & 45.67  & APL$^*$   &  2.2 &-1.87\\
		J1701+6412        & $2.18_{-0.22}^{+0.23}$ &  $4.3_{-2.3}^{+2.5}$  & 6.61 & 45.75	& APL   & 1.0 &  -1.67 \\
		J2123-0050        & $1.71_{-0.05}^{+0.06}$ &  $\leq 0.4$		    & 9.16 & 45.43	 & PL  & 2.2 &  -1.80 \\
		
		\bottomrule\bottomrule                 
	\end{tabular}
	\tablefoot{$^*$: Based on HR analysis. $f$ : frozen value. $^a$: Best-fit model consists of a power law plus cold reflection component. $^b$:
	 value shown in parentheses indicates the \nh\ inferred from the HR computed with the measured soft and hard fluxes. See Sect. \ref{subsec:HR} for further details.  $^c$: flux derived by assuming the upper limit value on \nh.
	 $^\dagger$: X-ray undetected sources: \nh $= 4 \times 10^{23} \, {\rm cm^{-2}}$ is assumed. $^\ddagger$: For the calculation of fluxes and luminosities, the upper limit on count rates listed in Table \ref{tab:xdata} is assumed.}    
\end{table*}

\subsubsection{More Complex Spectral Models}    
We will discuss here the peculiar case of J1250$+$2631, which showed additional spectral complexity with respect to PL and APL models.

The \xmm\ spectrum of J1250$+$2631 showed an evident excess at high energies when fitted with a PL model. As the observed spectral range (0.3-10 keV) corresponds to a rest-frame energy range of $\sim$ 0.9-30 keV, we added
a Compton reflection component to account for this excess, by using \texttt{pexrav} (\citealt{pexrav95}). A cut-off energy of $E=300$ keV was fixed, metal abundances were set to solar values and inclination angle of the reflector cos($i$)=0.45. Thus, we were able to fit both the reflected and intrinsic  continuum X-ray emission, whose slope
results to be quite steep ($\Gamma = 2.35_{-0.10}^{+0.12}$).  This model provides the best description of the spectrum of J1250+2631 (see Fig. \ref{fig:spectra}),  with an associated Cstat/d.o.f. = 45/60, compared to the Cstat/d.o.f.  = 69/61 derived by the PL model.
For the same spectrum, \citet{page04} found  $\Gamma = 2.34\pm0.04$ and a reflection parameter $R=2.87 \pm 0.96$, which is consistent with our result, i.e. $R=3.6_{-1.6}^{+2.6}$.
\cite{lanzuisi16} have recently published the analysis of a  deep ($\sim$ 100 ks) {\it NuSTAR} observation of J1250$+$2631 confirming the steep photon index and  the presence of strong reflection in this source. They interpret the large $R$ in terms of variability of the primary continuum, whereby the intense reflection component represents the light echo of a higher continuum level.

\subsection{X-ray Properties of WISSH quasars: Absorption, Fluxes and Luminosities}
\label{subsec:xrayprop}

The left panel of Figure \ref{fig:histonh} shows the distribution of the absorption column densities of X-ray detected WISSH quasars. Sources for which the PL model represents the best fit are included in the bin at \nh $<5 \times 10^{21} \, {\rm cm^{-2}}$. In case of \nh\ derived by HR analysis, we consider the values corresponding to the measured HR (i.e. the values in parentheses in Table \ref{tab:data}, column 3).
 More than half of the sample (i.e., 60\%) exhibits a \nh value $< 10^{22} \, {\rm cm^{-2}}$, i.e. they are unobscured in X-rays. About 23\% of the sample is moderately obscured, with $10^{22} \, {\rm cm^{-2}} \leq $ \nh $ \leq 10^{23} \, {\rm cm^{-2}}$. Accordingly, the bulk ($\sim$ 70\%) of \nh result to be $<$ 5 $\times$ $10^{22} \, {\rm cm^{-2}}$, in agreement with the broad line classification of WISSH quasars.

In the right panel of Fig.  \ref{fig:histonh} we plot the \nh derived by spectral analysis versus the dust reddening values for X-WISSH quasars.
${\rm A_V}$ were calculated from the color excess $E(B - V)$ through the relation ${\rm A_V} = 3.1 \times E(B - V)$. The color excesses were estimated by the SED fitting
(Duras et al. 2017, in prep.). The black square represents values of \nh = $5 \times 10^{21} \, {\rm cm}^{-2}$ and $A_{\rm V} = 0$ found for 8 sources. The blue solid line indicates the relation calculated by assuming a Galactic dust-to-gas standard value (\citealt{maiolino01}).
It is evident that WISSH quasars have low ${\rm A_V}/$\nh ratios, being apparently in contrast
with the postulate of the Unified Model which states that the dusty torus is the only
responsible for the obscuration of both the X-ray  and UV/optical nuclear radiation.
Lower   ${\rm A_V}/$\nh values
compared to
the Galactic one are usually explained  according to a scenario in which our line of sight does not intercept the torus but a
dust-free X-ray absorbing gas, which is likely located within the dust sublimation radius
(i.e., in the BLR, \citealt{bianchi12}). As pointed out by \citet{maiolino01},
another possible explanation may be  related to a different
dust composition in the surrounding region of AGN. Specifically,  larger grains cause less extinction than the grains in the diffuse Galactic medium.

X-WISSH quasars show hard X-ray (i.e., 2-10 keV) fluxes ranging from $\sim 7 \times 10^{-15}$ up to few $10^{-13} \, {\rm erg \, cm^{-2}s^{-1}}$, with the bulk of them (more than $80\%$) exhibiting  $10^{-14}$ $<$ \fhx\ $<$ $10^{-13} \, {\rm erg \, cm^{-2}s^{-1}}$. Concerning the unabsorbed hard X-ray luminosities, the majority of sources have 10$^{45}$   $<$ \lum\ $<$ $10^{46} \, {\rm erg \, s^{-1}}$ (see Table \ref{tab:data}).

\begin{figure}
	\centering
	\includegraphics[trim={0.8cm 0 0 0}, clip=true,scale=0.36]{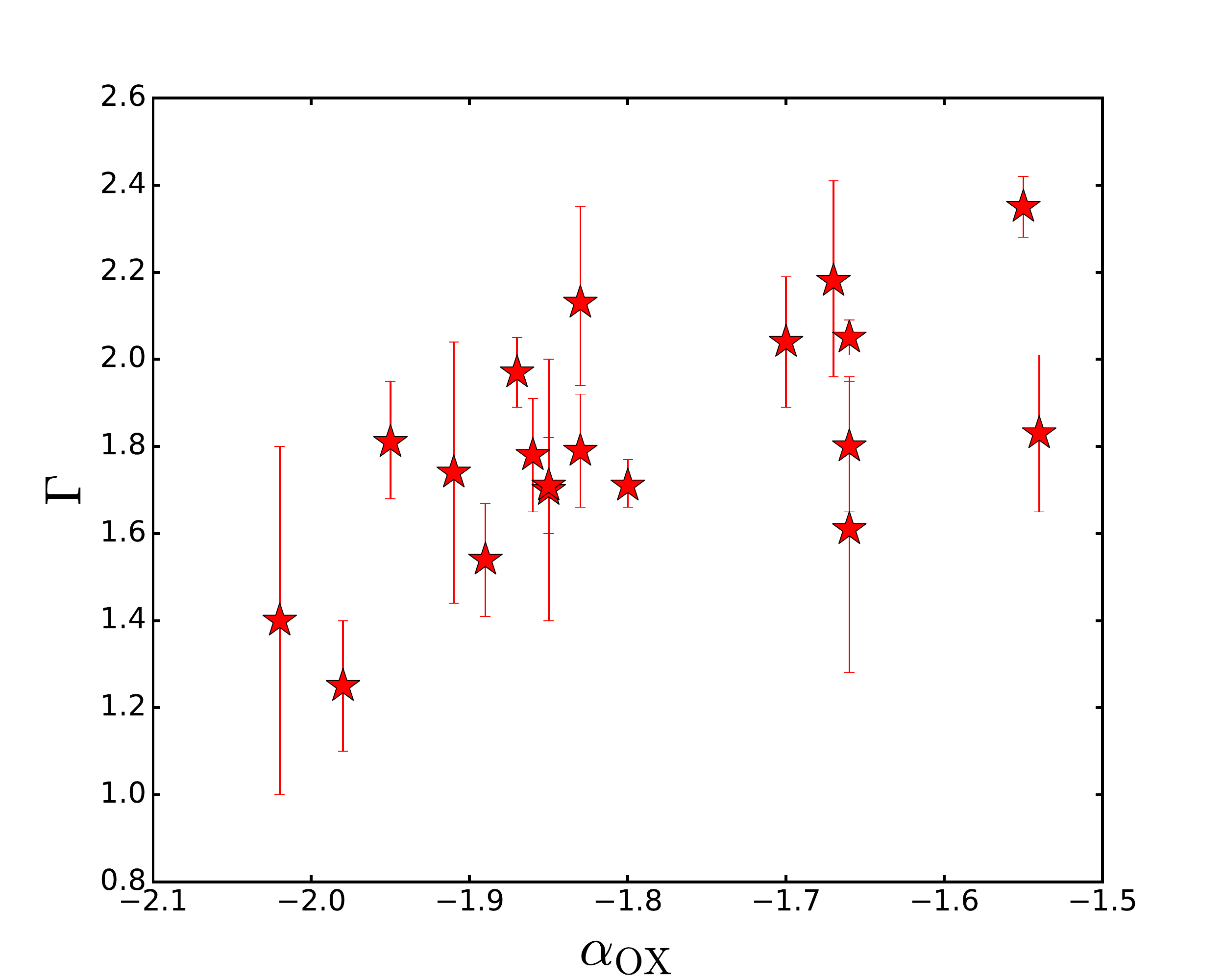}
	\caption{Best fit values of the photon index $\Gamma$ from the spectral analysis as a function of the \aox spectral index for X-WISSH quasars.     Note that the sources for which the $\Gamma$ has been fixed to 1.8 in the spectral analysis are not included (e.g. Table~\ref{tab:data}).}
    \label{fig:gammaaox}
\end{figure}

Fig. \ref{fig:gammaaox} shows the best-fit value of the photon index inferred from our spectra analysis as a function of the   \aox spectral index.
The \aox\ is defined as:
\begin{equation}
	\alpha_{OX} = \frac{\log(L_{2 \, \rm keV}/L_{2500 \AA})}{\log(\nu_{2 \, \rm keV}/\nu_{2500 \AA})},
\end{equation}
i.e., it represents the slope of a power law defined by the rest-frame monochromatic luminosities at 2500\AA \, and 2 keV \citep{avni82}.
For the calculation of the \aox for the WISSH quasars (see Table \ref{tab:data}) we used the rest-frame 2500\AA \, monochromatic luminosities obtained by SED fitting  (Duras et al. 2017, in prep.).
We find that the two sources with the steepest \aox also have the flattest X-ray continuum slope (i.e., X-ray faint compared to optical). However, the limited number of objects and the large uncertainties affecting some $\Gamma$ values does not allow to draw any firm conclusion on a possible trend.

Finally, our analysis of X-WISSH  includes {\it Chandra} observations of
six quasars previously analysed by \citet{just07}
(i.e. J0735+2659, J0900+4215, J1014+4300, J1106+6400, J1236+6554 and J1621-0042).
The X-ray luminosities and spectral properties we derive for these objects are consistent with their results.
Furthermore, our re-analysis of the $\sim$ 37 ks {\it Chandra} observation of J1521+5202 confirms the
results recently published by \citet{luo15}, which also found a slightly flat ($\Gamma$ $\sim$ 1.5) continuum and
an X-ray absorber with a  \nh\ $\sim 10^{23} \, {\rm cm^{-2}}$.

\begin{figure*}
	\centering
	\includegraphics[trim={0 0 2cm 0}, clip=true,scale=0.39]{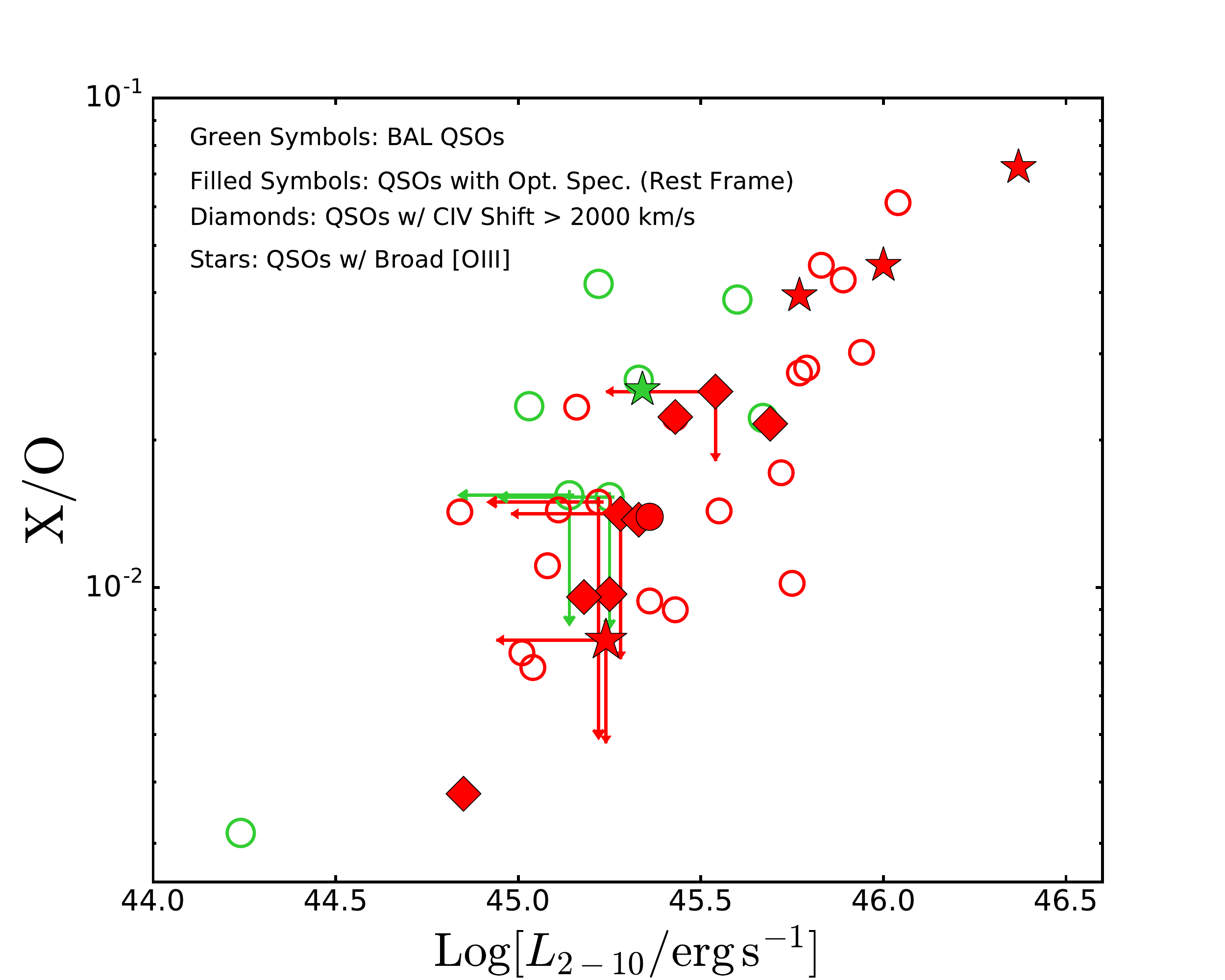}
	\includegraphics[trim={0 0 2cm 0}, clip=true,scale=0.39]{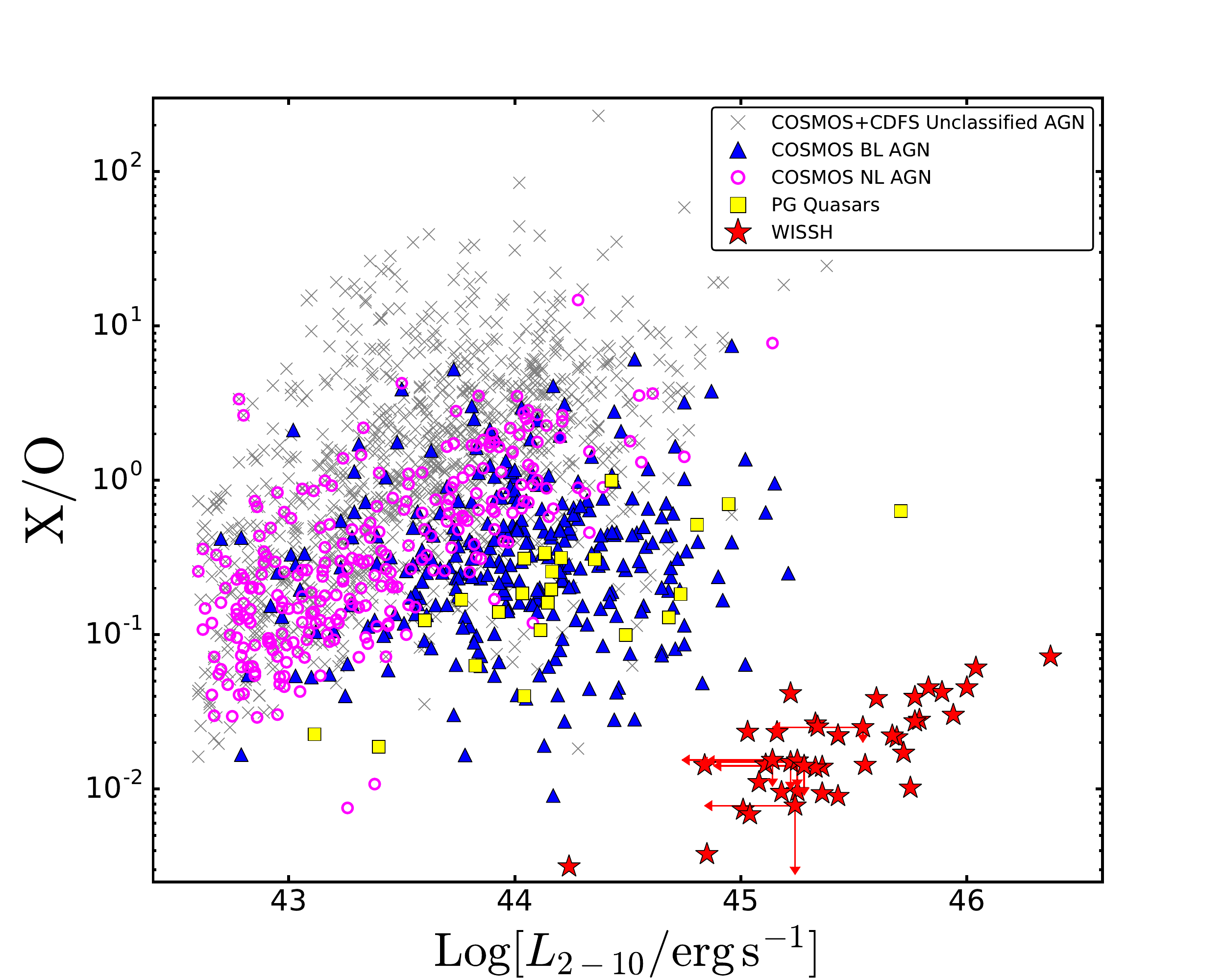}
	\caption{The X/O as a function of unabsorbed 2-10 keV luminosity. In the {\it left panel} X-WISSH quasars are plotted: BAL WISSH quasars
		are represented with green symbols while non-BAL WISSH are represented with red symbols.
		Filled symbols indicate objects which have LBT/LUCI optical spectrum.
		Star and diamond symbols indicate the presence of broad [OIII] emission lines and
		CIV shifts $> 2000 \, {\rm km \, s^{-1}}$, respectively. In the {\it Right panel}
		X-WISSH quasars (red stars) are compared to {\it Chandra}
		COSMOS, CDFS and PG Quasars.
		Blue triangles indicate the X-ray selected broad line COSMOS AGN, while open circles
		indicate X-ray selected narrow-line COSMOS AGN. Unclassified X-ray selected AGN of both COSMOS and CDFS
		are represented with black crosses. Finally, yellow squares indicate PG quasars.}
	\label{fig:XO}
\end{figure*}

\section{X-ray versus Optical and MIR properties}
\label{sec:Xmulti}

\subsection{The X-ray-to-Optical Flux Ratio (X/O)}

The X-ray-to-Optical flux ratio (X/O hereafter) is commonly used to provide a basic classification for X-ray sources without optical identification \citep{fiore03, BH05}.
The bulk of AGN detected in the X-ray surveys exhibit  $0.1<{\rm X/O}<10$, while sources with larger and smaller X/O are typically associated to  obscured AGN and  normal/star-forming galaxies, respectively.
In the calculation of the X/O ratio for the X-WISSH quasars we used  hard X-ray fluxes and  $i$ magnitudes (as done for the COSMOS survey, see \citealt{civano12}), which are publicly available from the SDSS.
The typical magnitudes of our objects range from 16 up to $\sim 18$, with almost 50\% of the sources showing 17 $\lesssim$ $i$ $\lesssim$ 17.6, as listed in Table \ref{tab:xdata}. Given the  redshift of X-WISSH quasars (2 \simlt\ $z$ \simlt\ 4), the  $i$ band  centered at $\sim 7600\AA$ approximately covers the range from $\sim$ 2500\AA\  to 1500\AA. Values of X/O for X-WISSH quasars are reported in column 7 of Table \ref{tab:data}.

In the left panel of Figure \ref{fig:XO} the X/O for X-WISSH quasars is shown as a function of \lum. BAL quasars within our sample (see Table \ref{tab:xdata}) are represented with green symbols and non-BAL WISSH with red symbols. Information about optical rest-frame spectral properties are also reported since 14 out of 41 X-WISSH sources, i.e. $\sim$ 35\% of the sample, have available  LBT/LUCI spectroscopy (\citealt{bischetti16}; Vietri et al. in prep.), and they are represented with filled symbols.
More specifically,  stars  indicate quasars with a broad and blueshifted [OIII] emission line (5 out of 41, $\sim$ 12\%),  while diamonds (8 out of 41, $\sim$ 20\%) represent sources with broad CIV emission line strongly blue-shifted ($> 2000 \, {\rm km \, s^{-1}}$) with respect to the H$\beta$ emission line (indicative of powerful radiatively driven winds in the BLR), and very weak/absent [OIII] emission. Interestingly, there is a hint that quasars showing [OIII] outflows tend to populate the high X/O--high \lum\ region of the plane.

In the right panel of Fig. \ref{fig:XO} we compare the X/O derived for the X-WISSH quasars (red stars), to those of other  AGN samples, i.e. sources
in the the Chandra COSMOS \citep{civano12} and 4Ms Chandra Deep Field South (CDFS) catalog \citep{xue11}. The subsample of CDFS sources used here has been obtained by  matching  the CDFS 4Ms catalog by \citet{xue11} with the GOODS/MUSIC optical multiband catalog \citep{vanzella08}. Only COSMOS and CDFS objects with $L_{2-10} > 3 \times 10^{42} {\rm erg \, s^{-1}}$ have been considered, in order not to be contaminated by non-AGN sources. This  sample of X-ray selected AGN consists of 1658 objects, out of which 296 are Type 1 AGN (blue triangles), 260 are Type 2 AGN (open circles) and the remaining
ones (i.e., 1102) are unclassified AGN (black crosses).
We also plot the X/O values for 23 optically selected quasars from the Palomar-Green (PG) Bright Quasar Survey
of the complete sample by \citet{laor94}.
It is evident from Fig. \ref{fig:XO} that WISSH quasars have very low X-ray-to-optical flux ratios (${\rm X/O}<0.1$) compared to typical AGN values of $0.1<{\rm X/O}<10$ measured  for  the  COSMOS, CDFS and PG sample. Nonetheless, WISSH quasars
exhibit the largest hard X-ray luminosities, reaching extreme values of $L_{2-10}$ $\approx$ 10$^{46} {\rm erg\, s^{-1}}$, and low level of X-ray absorption.

\citet{fiore03} reported the existence of a correlation between X/O and X-ray luminosity for narrow line AGN.
This trend can be also observed in Fig. \ref{fig:XO} where type 2 AGN show increasing  X/O values at increasing \lum, although with a large scatter.  A possible explanation for this
is the largely reduced optical-UV light from these obscured objects which implies that their X/O approximately represents the ratio
between the AGN hard X-ray flux (which is less affected by the absorption than the optical/ultraviolet one) and the host galaxy flux.
Fig. \ref{fig:XO} suggests the presence of  a correlation between the X/O and \lum\  for WISSH quasars.
This is quite unexpected as no correlation between X/O and hard X-ray luminosity
has been reported  for broad-line AGN so far.
However, this trend can be explained in terms of a selection bias due to the 
limited range  of $i$ magnitude of WISSH quasars, with a median value of $i$  $\sim 17.1$, as opposed to their X-ray luminosity, which
ranges from a few $10^{44}$ up to values larger than $10^{46} \, {\rm erg\, s^{-1}}$.


\begin{figure*}
	\centering
	\includegraphics[trim={1.9cm 0 2.5cm 0}, clip=true,scale=0.56]{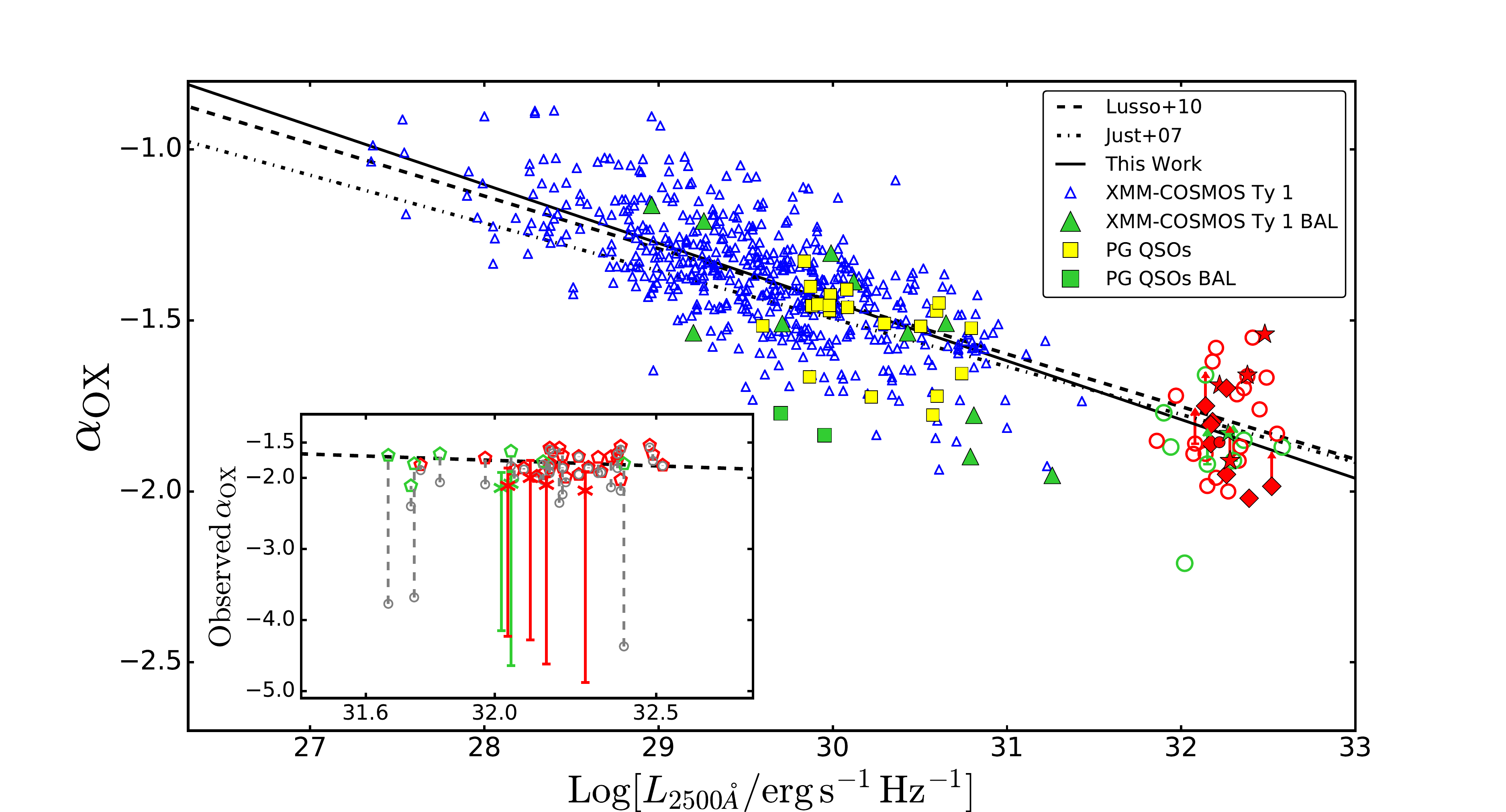}
	\caption{\aox values as a function of extinction-corrected 2500$\AA$ monochromatic luminosities. Symbols for WISSH quasars as in Fig. \ref{fig:XO}. Blue triangles indicate the Broad-line AGN sample from the XMM-Newton-COSMOS survey by \citet{lusso10} while the yellow squares indicate PG quasars. Green symbols represent BAL AGN. The dashed line represents the relation found by \citet{lusso10}, the dash-dot one represents the \citet{just07} relation while the black solid line indicates our linear fit. In the inset on the lower left corner of the figure, open pentagons indicate absorption-corrected \aox values, while gray open circles indicate \aox values when X-ray absorption is not taken into account. Asterisks with error bars indicate \aox values for the undetected X-WISSH sources, calculated with the \nh median value of the sample, i.e. \nh$=8 \times 10^{21}\, {\rm cm^{-2}}$. The error bars were calculated using \nh$=0$ and \nh$=4\times 10^{23}\, {\rm cm^{-2}}$, i.e. the maximum value of the sample.}
	\label{fig:aox}
\end{figure*}

\subsection{The X-ray-to-Optical Index (\aox {\rm vs.} $L_{2500\AA}$)}
\label{subsec:aox}

While the X/O ratio is calculated by observed quantities (i.e., fluxes), the \aox is estimated by considering rest-frame, intrinsic properties, i.e., luminosities.
In unobscured AGN, \aox provides a basic indication of the relative strength between the emission produced in the  accretion disk
(at 2500\AA \, rest-frame) and the one emitted via Compton up-scattering in the hot corona (at 2 keV rest-frame), by connecting two spectral regions encompassing the energy range where the bulk of the AGN radiative output is generated.
The existence of an anticorrelation between \aox and $L_{2500\AA}$ is now well-established \citep{vignali03, steffen06,lusso10} indicating that the relative contribution of the X-ray emission to the total energy output decreases for increasing optical (and bolometric) luminosity.

Fig. \ref{fig:aox} shows the \aox distribution for the X-WISSH quasars as a function of $L_{2500\AA}$, calculated by  using the absorption-corrected X-ray luminosity. This demonstrates that WISSH quasars are in broad agreement with the \aox-$L_{2500\AA}$ anticorrelation, i.e.  when compared with less luminous AGN, the most luminous quasars exhibit a weaker X-ray luminosity.
The \aox values derived for our hyper-luminous objects are plotted with those of  other Type 1 AGN samples at lower \lbol. Blue open triangles represent the Broad-Line AGN sample from the XMM-COSMOS survey by \citet{lusso10} (L10 hereafter),  while PG quasars are indicated with yellow squares as in Fig. \ref{fig:XO}. We have also included information on BAL objects, represented with green symbols. WISSH symbols are as  in Fig. \ref{fig:XO}.
We  also plot  the linear relation found by L10 (dashed line) and the relation by \citet{just07}  (dash-dotted line), as well as the linear relation we derived by  fitting the values of the XMM-COSMOS, PG and X-WISSH samples together. 
The best-fit relation for \aox-$L_{2500 \AA}$, treating $L_{2500 \AA}$ as the independent variable results to be:
\begin{equation}
	\alpha_{\rm OX} = (-0.172\pm0.006)\times \log (L_{2500 \AA}/{\rm erg \, s^{-1} \, Hz^{-1}}) +(3.72\pm0.17).
\label{eq:aoxeq}
\end{equation}
where the errors on the slope and intercept indicate $1\sigma$ significance. The Spearman’s rank test gives a rank coefficient of  $\rho_s =-0.75$, and the probability of deviation from a random distribution is $d_s \sim 10^{-20} $, confirming a very robust anti-correlation between \aox and $L_{2500 \AA}$.

Finally, in the inset of Fig. \ref{fig:aox}, open pentagons  indicate
 absorption-corrected \aox values for X-WISSH quasars, while gray open circles indicate \aox values when X-ray absorption is not taken into account.
Asterisks indicate \aox  for the 6 undetected sources, derived assuming the  median value  of  the \nh\ measured in the X-WISSH sample, i.e. \nh$=8 \times 10^{21}\, {\rm cm^{-2}}$ (the error bars are calculated with \nh$=0$ and \nh$=4\times 10^{23}\, {\rm cm^{-2}}$, i.e. the largest \nh of the sample).  This plot highlights that once corrected for  X-ray absorption, the steeper \aox (typically associated to BAL quasars, marked in green) flatten to values broadly consistent with the L10 relation. A largely suppressed luminosity at 2 keV, and in turn a very steep \aox $< -3$  value, measured in our sources can be  therefore ascribed to obscuration along our line of sight to the AGN.



\begin{figure*}
	\centering
	\includegraphics[trim={0.45cm 0 0 0}, clip=true, scale=0.5]{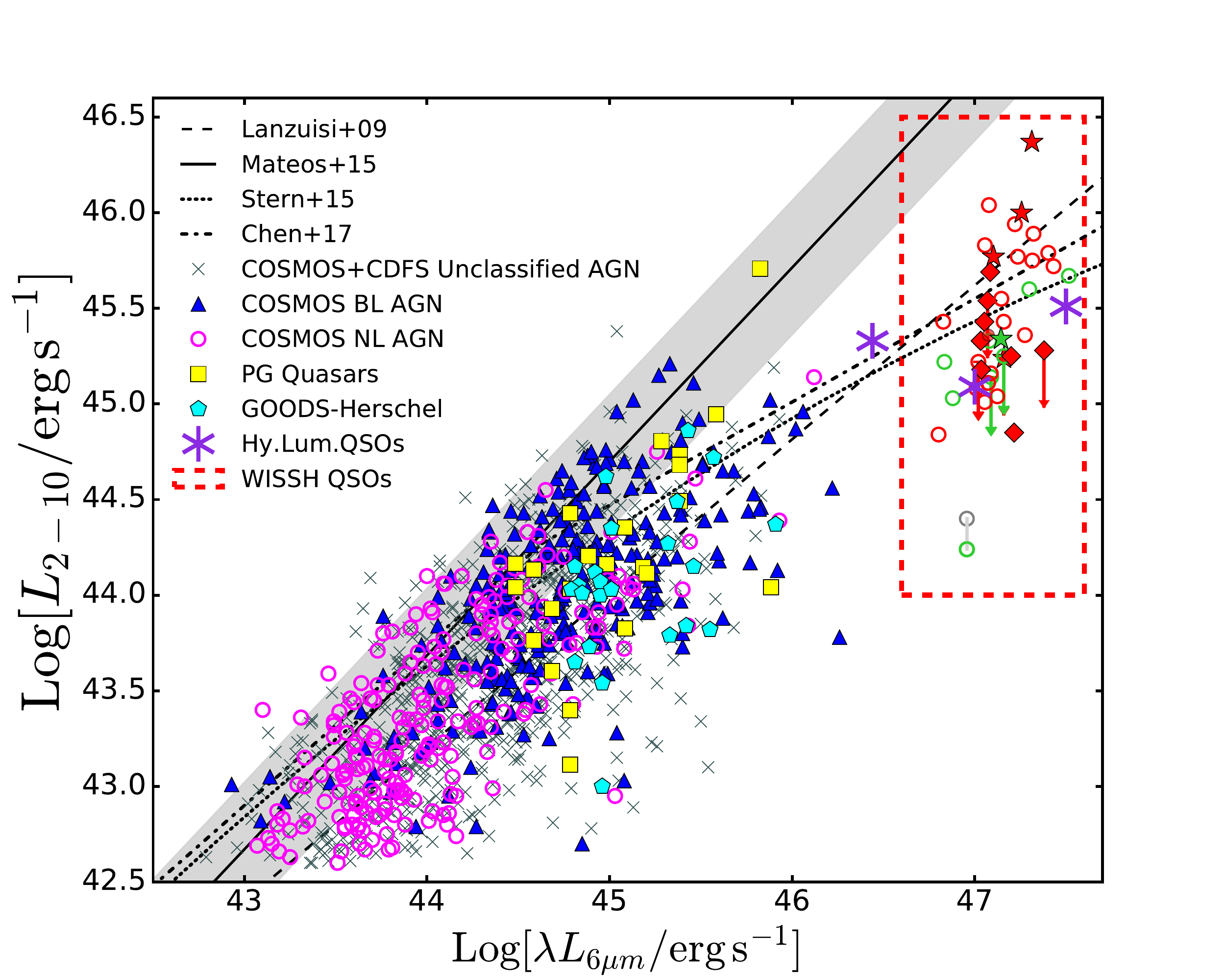}
	\caption{Absorption-corrected 2-10 keV luminosity as a function of MIR luminosity at 6$\mu m$
		(symbols as in Fig. \ref{fig:XO}). Cyan pentagons represent the MIR luminous quasars sample from
		the GOODS-{\it Herschel} fields presented by \citet{delmoro15}. Violet asterisks indicate the hyper-luminous quasars ULAS J1539+0557, ULAS J2315+0143 and 2QZ0028-2830 (see Appendix \ref{sec:app1}). The black solid, dashed, dotted and dash-dotted lines
		represent the   \cite{mateos15}, \cite{lanzuisi09}, \cite{stern15}, \cite{chen17} relations, respectively.
		The grey shaded area indicates the intrinsic scatter about the Mateos relation. The red dashed line box
		indicates the locus of the WISSH quasars. We report two values for the hard X-ray luminosity of J0045+1438 (i.e., the object  with the lowest X-ray luminosity in the X-WISSH sample)  connected by a grey line. We calculated the $L_{2-10}$ by using the HR-based value  of \nh\ (i.e., $4 \times 10^{22} \, {\rm cm^{-2}}$, green open circle) and the upper error (i.e., \nh=$1.5 \times 10^{23} \, {\rm cm^{-2}}$, gray open circle), respectively.
		}
	\label{fig:L6Lx}
\end{figure*}


\subsection{The X-ray to MIR luminosity relation}
\label{subsec:lmir}

WISSH quasars are among the  most luminous AGN in the MIR, being selected to have $z>$1.5 and WISE 22$\mu$m flux density $S_\nu$(22$\mu$m) $>$ 3 mJy (3$\sigma$). This allows us to explore and extend  the correlations involving the MIR luminosity  up to the highest  values.
The X-ray and MIR emission of AGN are believed  to be strongly linked, since
the former  is due to accretion onto the SMBH  while the latter is interpreted in terms of reprocessed nuclear light from the surrounding
absorbing material.
A correlation between these two quantities based on the analysis of
samples of local Seyfert galaxies and X-ray selected AGN has been indeed found (e.g., \citealt{lutz04}; \citealt{mateos15} and references therein).
Furthermore, \citet{fiore09}, \citet{lanzuisi09} discovered that quasars hosted in heavily dust obscured galaxies typically exhibit a lower $L_X/L_{\rm MIR}$ ratio than that reported for ``standard AGN'', suggesting  a more complex scenario for the $L_X-L_{\rm MIR}$ relation in AGN.

As commonly adopted in  recent works, we also use the 6 $\mu m$ luminosity ($\lambda L_{6 \mu m}$) as a proxy of the MIR luminosity, since emission due to AGN heated dust peaks around this wavelength.
$\lambda L_{6 \mu m}$ of WISSH quasars have been estimated by UV-to-MIR SED fitting (Duras et al. 2017, in prep.) and span from $\sim 6 \times 10^{46}$  to $\sim  3 \times 10^{47}\, {\rm erg\,s^{-1}}$. Figure \ref{fig:L6Lx} shows \lum\ as a function of $\lambda L_{6 \mu m}$ for the X-WISSH sources and   a large compilation of AGN samples selected according to different criteria in order to
mitigate any possible bias and obtain a better sampling of the $L_{2-10}-\lambda L_{6 \mu m}$ plane.
By including a total of 1749 AGN, we are able to explore the X--MIR luminosity relation in the ranges  42.5 \simlt\ log (\lum/\ergs) \simlt\ 46.5  and  43.5 \simlt\ log ($\lambda L_{6 \mu m}$/\ergs) \simlt\ 48. More specifically,  we include  AGN from the  \chandra\ COSMOS and  CDFS surveys and the PG quasars (as in Fig. \ref{fig:XO}), and a sample of 24 MIR luminous quasars
($\lambda L_{6 \mu m} \geq 6 \times  10^{44} \, {\rm erg\, s^{-1}}$) at redshifts $z \sim 1-3$, with X-ray luminosity derived by spectral analysis from
\citet{delmoro15} (cyan pentagons).
Finally, we also add the quantities for three hyper luminous quasars (namely ULASJ 1539+0557, ULASJ 2315+0143 and 2QZ0028-2830; indicated as violet asterisks), which show luminosities comparable to those measured for X-WISSH (see Appendix \ref{sec:app1}).
We also plot  the $L_{2-10}-\lambda L_{6 \mu m}$ relations derived by  four previous works, i.e. (i)
the relation by \citet{mateos15} (black solid line), which has been obtained by using a sample of $> 200$ AGN from the Bright Ultra-Hard XMM-Newton Survey (including both broad and narrow line AGN with X-ray luminosities $10^{42} < L_{2-10} < 10^{46}\, {\rm erg\, s^{-1}}$ and redshifts 0.05 < $z$ < 2.8); (ii) the relation reported in
\citet{lanzuisi09} (black dashed line) by using MIR-selected, dust-obscured (with MIR-to-optical flux ratio $F_{24 \mu m}/F_R >2000$) luminous quasars at $z \sim 1-2$ with $L_{2-10} \geq 10^{43} \, {\rm erg\, s^{-1}}$; (iii)  the  polynomial relation derived by  \citet{stern15} (black dotted curve) based on $\sim$200 AGN with luminosities spanning  from the local Seyfert regime to the hyper-luminous quasar one,
which accounts for a progressive decrease of  $L_{2-10}/\lambda L_{6 \mu m}$ by moving towards the highest luminosities; (iv) the bilinear function by \citet{chen17} (black dash-dot curve) which was recently derived by including more than $\sim$ 2500 type 1 AGN from four different X-ray surveys, i.e. Bo\"otes, XMM-COSMOS, XRT-SDSS and XXL-North, covering $L_{2-10}$ in the range 10$^{41}$--10$^{46}$ erg s$^{-1}$ and $\lambda L_{6 \mu m}$ up to $10^{47} \, {\rm erg\, s^{-1}}$.

As expected, X-WISSH  quasars are located far below the
Mateos et al. relation and most of them are located even below the Lanzuisi et al. relation.
Figure \ref{fig:L6Lx} points out the existence of a large scatter in the measurement of \lum\ at a given $\lambda L_{6 \mu m}$, once AGN selected with different criteria have been taken into account. This suggests that using a universal $L_{X} - L_{\rm MIR}$ relation
for evaluating the X-ray luminosity of a source can lead to a heavily over/understimated value.
Still we reinforce  previous results (e.g.,  \citealt{lanzuisi09}; \citealt{stern15}; \citealt{chen17}) indicating that MIR luminous sources clearly offset from the  $L_{X} - L_{\rm MIR}$ relation  derived for local Seyfert/X-ray-selected AGN.
This suggests that in AGN at the highest MIR luminosities, the 
  relative contribution due to X-ray emission to the bolometric luminosity progressively decreases with respect to the MIR one (see Sect. \ref{sec:disc} for a detailed discussion).
We also attempt to interpret the observed trend in the $L_{X} - L_{\rm MIR}$ relation in terms of a luminosity-dependent covering factor ($C_f$) of the torus. We assume $L_{\rm MIR}$ $\propto$ $C_f$ $\times$ $L_{UV}$($\approx$ \lbol)
and the $C_f$ as a function of the luminosity at 5100\AA\  derived by \cite{maiolino07} (see their Eq. 3\footnote{\cite{maiolino07} assume $C_f$ equal to the fraction of obscured AGN.}) for a combined sample of high-z luminous quasars, local quasars and Seyfert galaxies ($L_{5100\AA}$ are estimated from $L_{UV}$
by using the bolometric corrections in \citealt{runnoe12a}). Consistently, we derive \lum\ from $L_{UV}$ according to Eq. (6) in \cite{lusso10},  assuming \lum\ = 1.61 $\times$ $L_{2 \, \rm keV}$ (i.e. adopting $\Gamma$ = 2).
However, the resulting $L_{X} - L_{\rm MIR}$ is unable to reproduce the observed decrease in   $L_{X}/L_{\rm MIR}$ at high luminosities, but instead shows an opposite trend.
This suggests that the observed weakness of the X-ray emission relative to the MIR one in quasars cannot be interpreted to first order as due to a luminosity-dependent  $C_f$.

Alternatively, the observed  $L_{X} - L_{\rm MIR}$ relation  could be caused by the  selection bias due to the inclusion of both X-ray and MIR selected sources (collected from different areas of the sky, i.e. from pencil-beam surveys to all-sky surveys), as well as a redshift bias, according to which the most luminous AGN have been preferentially collected at high $z$.


\begin{figure*}
	\centering
	\includegraphics[trim={0 0 2cm 0}, clip=true,scale=0.37]{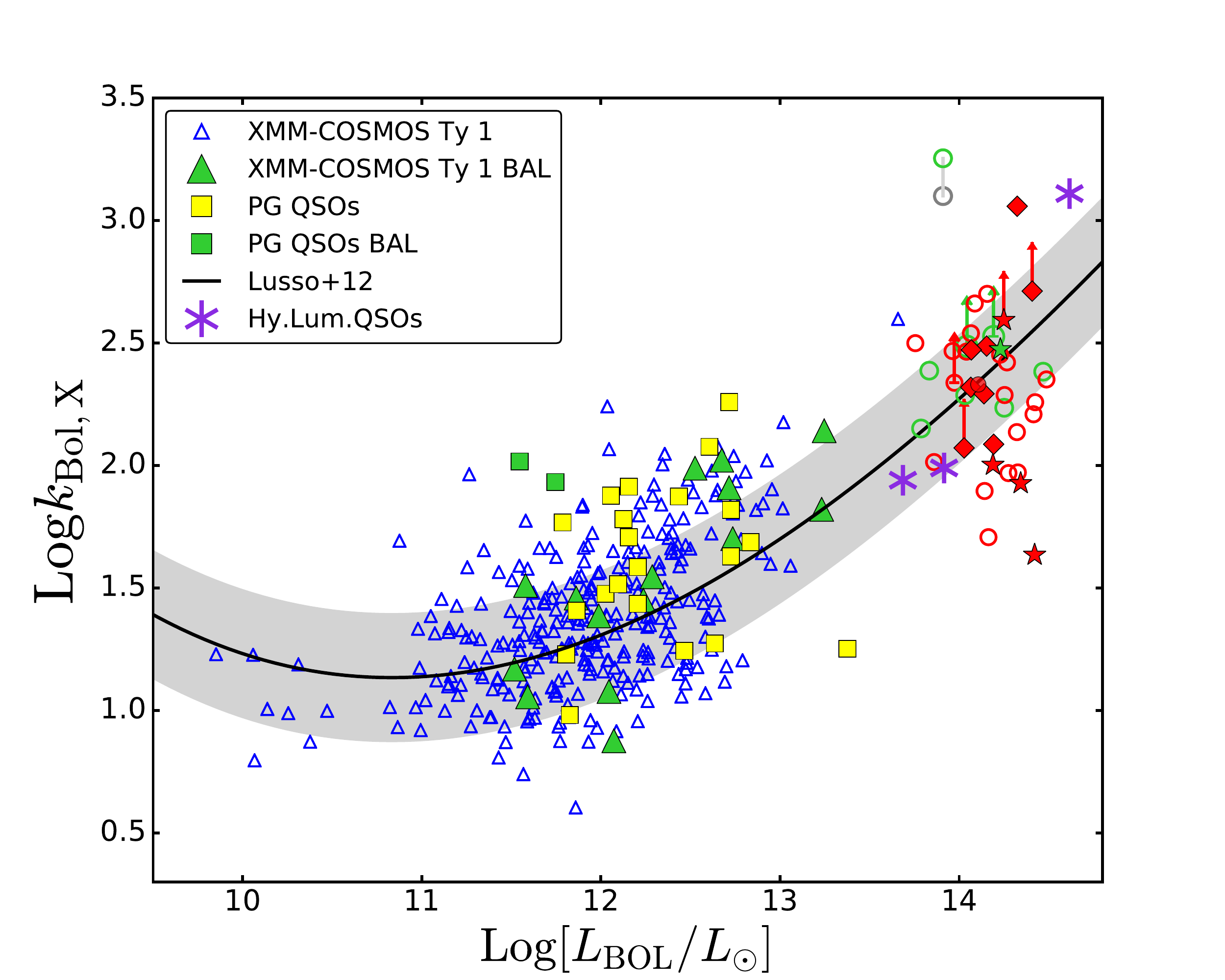}
	\includegraphics[trim={0 0 2cm 0}, clip=true,scale=0.37]{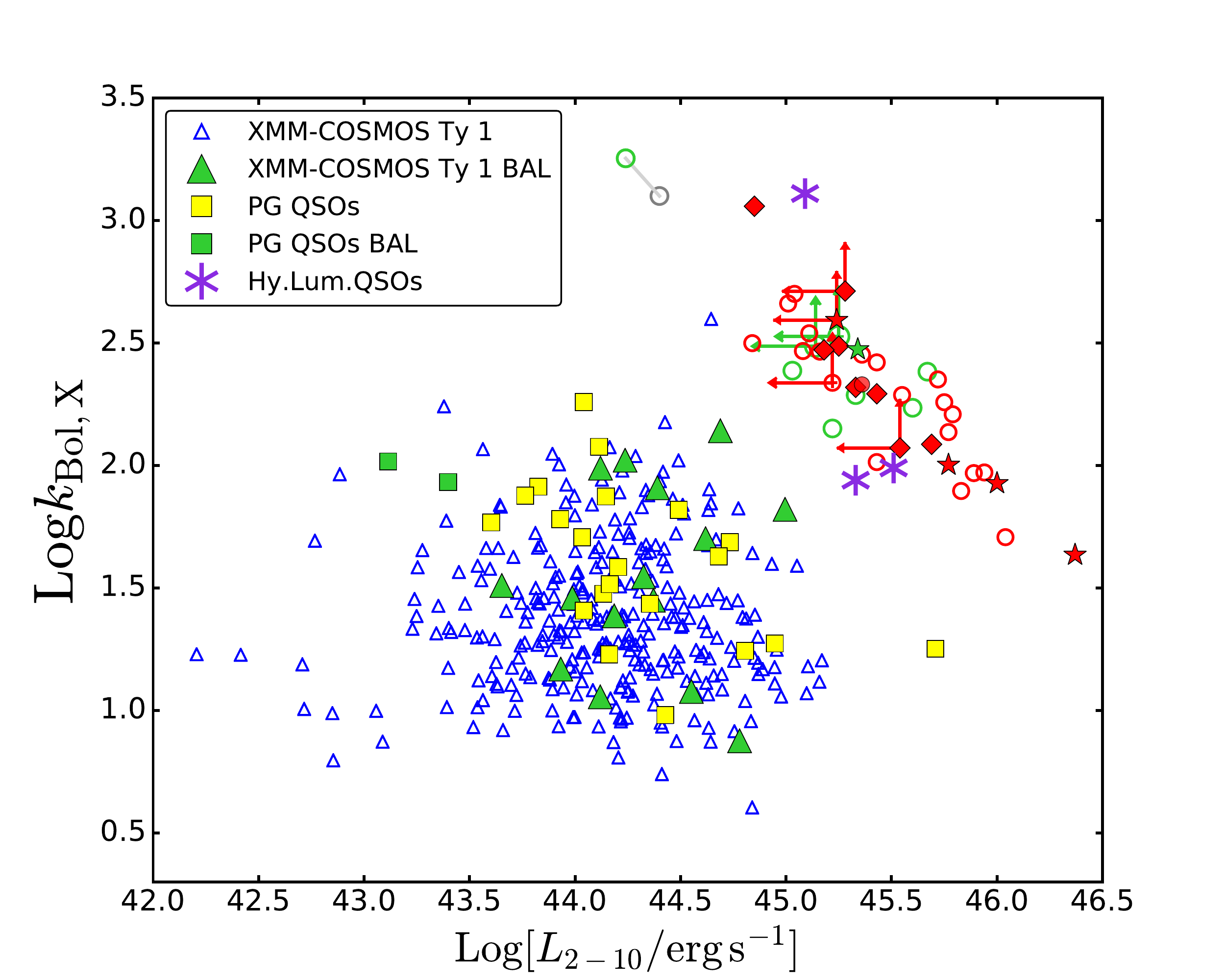}
	\caption{Bolometric correction in the 2-10 keV band as a function of the
		bolometric luminosity {\it (left panel)} and as a function of the 2-10 keV luminosity {\it (right panel)}. Symbols for the WISSH quasars are as in Fig. \ref{fig:XO}. Blue triangles indicate the Broad-line AGN sample from the XMM-Newton-COSMOS survey by \citet{lusso12}. The yellow squares indicate PG quasars while green symbols represent BAL AGN. Violet asterisks indicate the hyper luminous quasars ULASJ 1539+0557, ULASJ 2315+0143 and 2QZ0028-2830. The black solid curve in the {\it left panel} represents the relation found by \cite{lusso12} for type 1 AGN, while the gray shaded area indicates the 1$\sigma$ dispersion on the relation. As in Fig. \ref{fig:L6Lx}, we report two values for the hard X-ray luminosity of J0045+1438 (i.e., the object  with the lowest X-ray luminosity in the X-WISSH sample), connected by a gray line. We calculated the $L_{2-10}$ by using the best fit value  of \nh\ (i.e., $4 \times 10^{22} \, {\rm cm^{-2}}$, green open circle) and the upper error (i.e., \nh=$1.5 \times 10^{23} \, {\rm cm^{-2}}$, gray open circle), respectively.}
	\label{fig:lbol}
\end{figure*}

Given its importance for shedding light on the properties of accretion, reprocessing of the nuclear light and geometry of the circumnuclear medium in AGN, the $L_{X}/L_{\rm MIR}$ ratio certainly deserves further and deeper investigation in the future. For instance, it would be very interesting to populate the right top corner of the $L_{2-10}-\lambda L_{6 \mu m}$ diagram, by selecting AGN
at the highest X-ray luminosities. 
Indeed, the recent work by \citet{chen17} shows that the heterogeneous  flux limits of the different X-ray surveys
contribute significantly to the difference in the $L_X/L_{\rm MIR}$ ratio behaviour between low and high-luminosity AGN.
The planned $eROSITA$ All-Sky X-ray survey \citep{merloni12} will be therefore crucial to pick up a large sample of X-ray selected hyper-luminous quasars and provide unprecedented constraints on the behavior of the $L_{X} - L_{\rm MIR}$ relation  at $L_{2-10} >10^ {45} \, {\rm erg\, s^{-1}}$.

\section{X-ray Bolometric Corrections and Eddington Ratios}
\label{sec:bolo}

Thanks to their spectacular luminosities, WISSH quasars offer the opportunity to investigate
the X-ray bolometric correction (defined as $k_{\rm Bol,X} = L_{\rm Bol}/L_{2-10}$) up to the highest luminosity values (i.e. \lum\ \simgt\   10$^{45}$ and $L_{\rm Bol} > 10^{47} \, {\rm erg \, s^{-1}}$).
Bolometric luminosities for X-WISSH quasars have been estimated by SED fitting (Duras et al. 2017, in prep.) and they span from $L_{\rm Bol} \sim 2 \times 10^{47}$ to $> 10^{48}  \, {\rm erg \, s^{-1}}$.

The left panel of Figure \ref{fig:lbol} shows the  $k_{\rm Bol,X}$  for the X-WISSH quasars as a function of \lbol, while in the right panel $k_{\rm Bol, X}$ is plotted versus the 2-10 keV luminosity.  Our sources (symbols as in Fig. \ref{fig:XO}) are compared to Type 1 AGN from the XMM-Newton-COSMOS survey (presented in \citealt{lusso12}, L12 hereafter, indicated as open triangles), PG quasars (yellow squares) and the hyper-luminous quasars ULASJ 1539+0557, ULASJ 2315+0143 and 2QZ0028-2830, represented with violet asterisks (see Appendix \ref{sec:app1}). As in the previous figures, BAL quasars are indicated by green symbols.
The black solid line represents the $k_{\rm Bol, X}-L_{\rm Bol}$ relation obtained by L12 for Type 1 objects in COSMOS. This curve also provides a good description for our hyper-luminous quasars (the scatter around the L12 relation is $\sigma=0.82$ dex).
Combining all the objects from these samples, the Spearman's rank  correlation coefficient is $\rho_s=0.7$ (and $d_s <10^{-20}$), indicating a strong correlation between $k_{\rm Bol, X}$ and \lbol. 
The ratio of the X-ray luminosity over the bolometric luminosity provides the relative contribution of the corona and accretion disk to the total radiation output from the AGN. Therefore, the positive trend observed between $k_{\rm Bol, X}$ and \lbol\ lends support to the scenario whereby the corona radiative power becomes progressively  weaker with respect to the optical/UV disk emission at increasing bolometric luminosities.
The right panel of Fig. \ref{fig:lbol} reinforces our conclusions on the extreme nature of  WISSH quasars which exhibit both  exceptional X-ray luminosities and large bolometric corrections.


\begin{figure*}
	\centering
	\includegraphics[trim={0 0 2cm 0}, clip=true,scale=0.37]{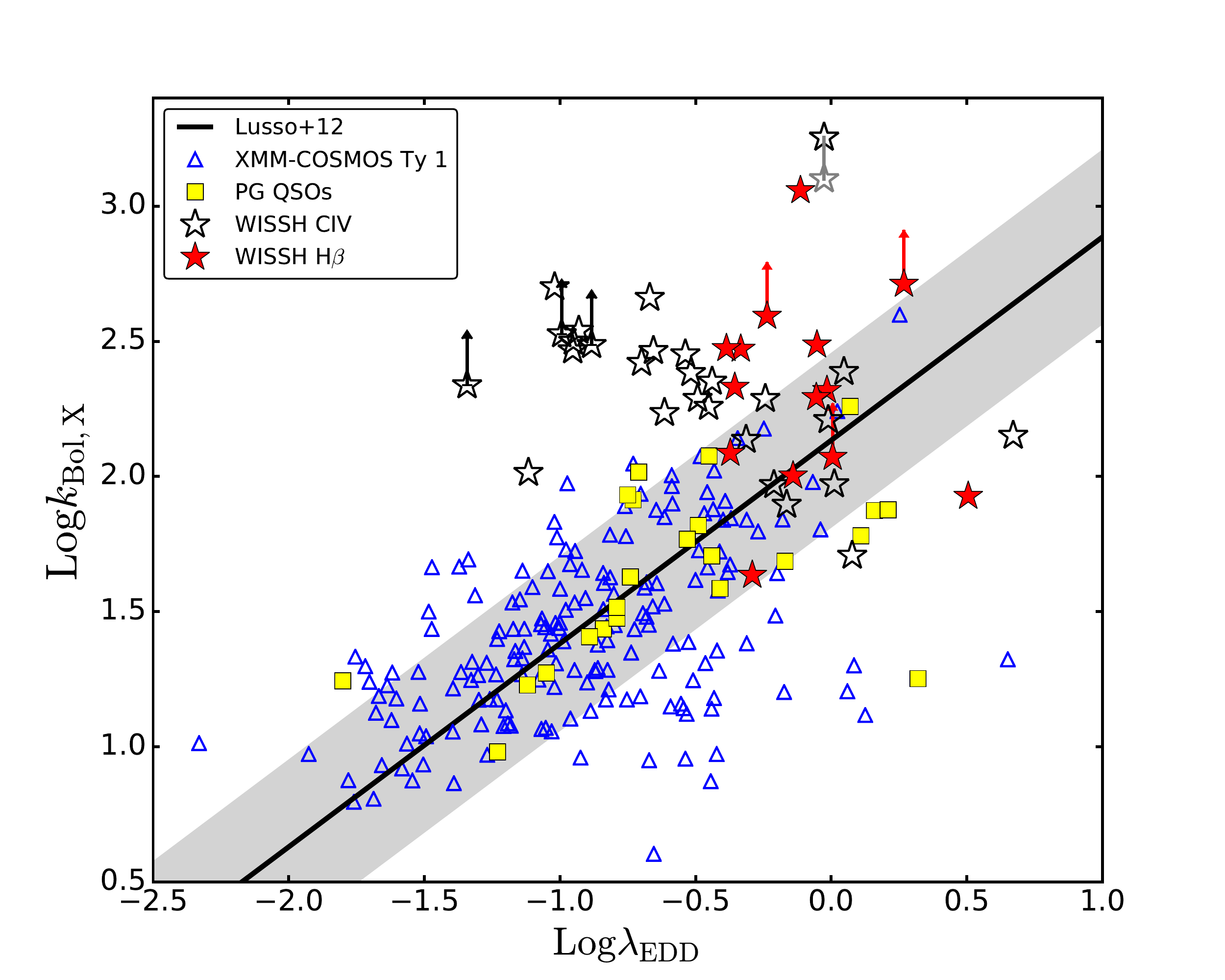}
	\includegraphics[trim={0 0 2cm 0}, clip=true,scale=0.37]{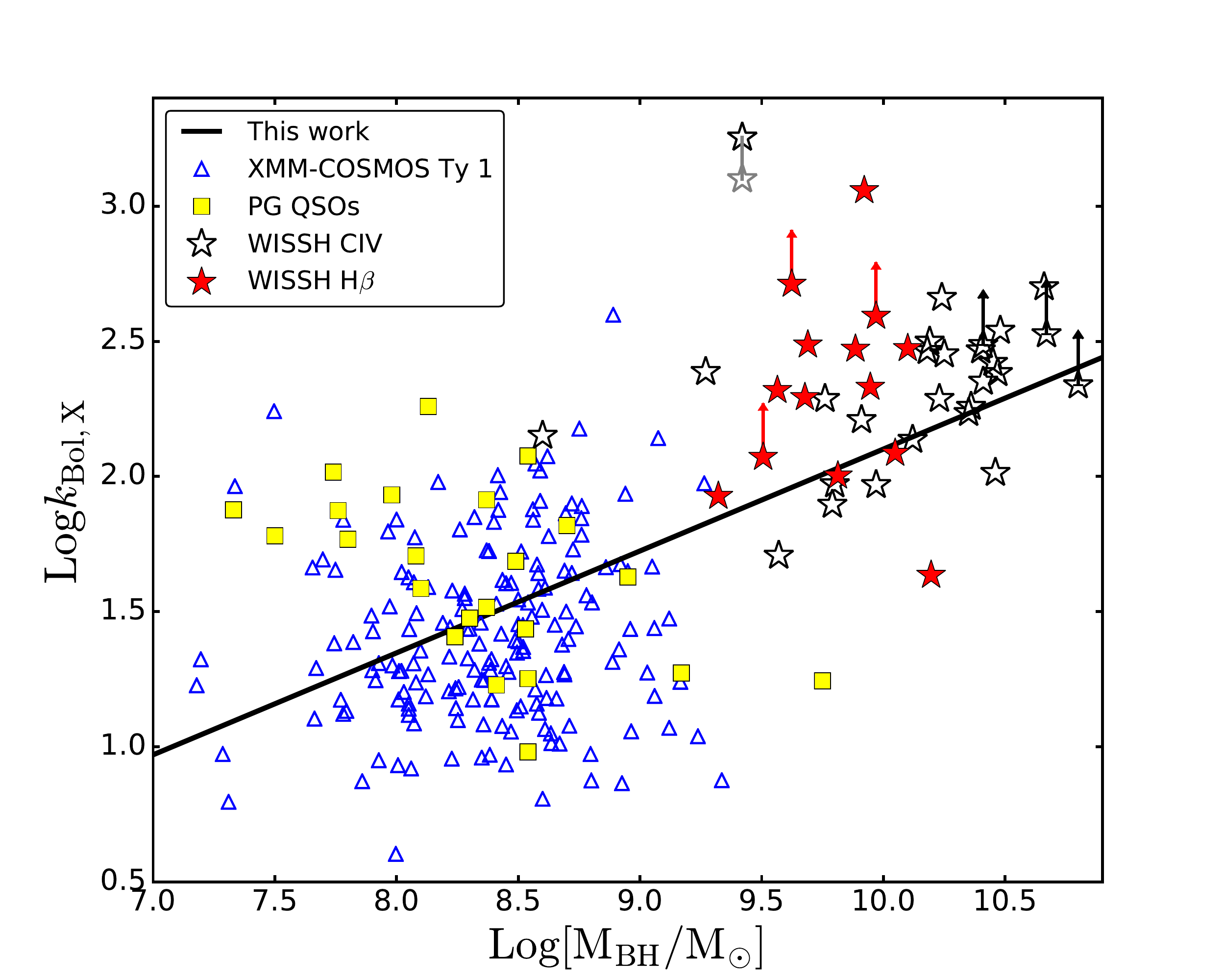}
	\caption{ Bolometric correction in the 2-10 keV band as a function of the
		Eddington ratio $\lambda_{\rm EDD}$ ({\it left panel}) and  as a function of the black hole mass ({\it right panel}). Open stars represent WISSH quasars with CIV-based SMBH masses (\citealt{weedman12}), while red filled stars represent $H\beta$-based SMBH masses (\citealt{bischetti16}, Vietri et al. in prep.). Blue empty triangles indicate the Broad-line AGN sample from the XMM-Newton-COSMOS survey by \citet{lusso12}. The yellow squares indicate PG quasars. The black solid line in the left panel represents the relation found by Lusso et al. 2012 for type 1 AGN, while the gray shaded area indicates the 1$\sigma$ dispersion on the relation. The black solid line in the right panel indicates the linear fit found in this work. As in Fig. \ref{fig:L6Lx}, we report two values for the hard X-ray luminosity of J0045+1438, connected by a gray line. We calculated the $L_{2-10}$ by using the best fit value  of \nh\ (i.e., $4 \times 10^{22} \, {\rm cm^{-2}}$, black open star) and the upper error (i.e., \nh=$1.5 \times 10^{23} \, {\rm cm^{-2}}$, gray open star), respectively.}
	\label{fig:eddratio}
\end{figure*}

The multi-band coverage of WISSH allows us to derive the SMBH mass ($M_{\rm BH}$) for each quasar. Accordingly, we are able to  study the behaviour of  $k_{\rm Bol,X}$ as a function of $M_{\rm BH}$ and Eddington ratio ($\lambda_{\rm EDD} = L_{\rm Bol}/L_{\rm EDD}$).
The SMBH mass for 14 out of 41 objects ($\sim 35\%$ of X-WISSH) are estimated from H$\beta$ lines (\citealt{bischetti16}; Vietri et al. in prep.), while the remaining ones are calculated from CIV lines (\citealt{weedman12}).  
In the left panel of Figure \ref{fig:eddratio} we present the hard X-ray bolometric correction versus $\lambda_{\rm EDD}$), while in the right panel $k_{\rm Bol, X}$ is shown as a function of the $M_{\rm BH}$. Filled (open) stars represent WISSH quasars with $H\beta$ (CIV)-based SMBH masses.
The $M_{\rm BH}$ for COSMOS and PG quasars are computed using the width of the  H$\beta$ (or Mg II) emission line.
Several authors claimed for a positive trend  between $k_{\rm Bol, X}$ and $\lambda_{\rm EDD}$ \citep{vasufabian07,lusso12}, even if affected by large scatter.
For the sources considered here,  the total scatter with respect to the L12 relation is $\sigma=0.43$ dex, with a larger scatter observed at the highest values of $k_{\rm Bol, X}$ and $\lambda_{\rm EDD}$. The Spearman's rank test gives a rank coefficient of  $\rho_s =0.56$, and the probability of deviation from a random distribution is $d_s \sim 10^{-21}$, confirming a likely and robust correlation between the two quantities.
Additionally, by fitting $M_{\rm BH}$ vs. $k_{\rm Bol, X}$,  we found the following linear relation (see right panel of Fig. \ref{fig:eddratio}):
\begin{equation}
	\log k_{\rm Bol,X} =(0.377\pm0.032)\times \log (M_{\rm BH}/M_{\odot}) -(1.67\pm0.28),
\end{equation}
with a large scatter ($\sigma \simeq 0.4$ dex) and $\rho_s = 0.4$. Furthermore, the probability value is very low ($d_s < 10^{-10}$). These values mean that the existence of a correlation between $M_{\rm BH}$ and $k_{\rm Bol, X}$ is significantly likely. The inclusion of WISSH quasars suggests that the largest values of $k_{\rm Bol, X}$ may be observed in objects at the massive end of the $M_{\rm BH}$ distribution.


\begin{figure*}
	\centering
	\includegraphics[scale=0.45]{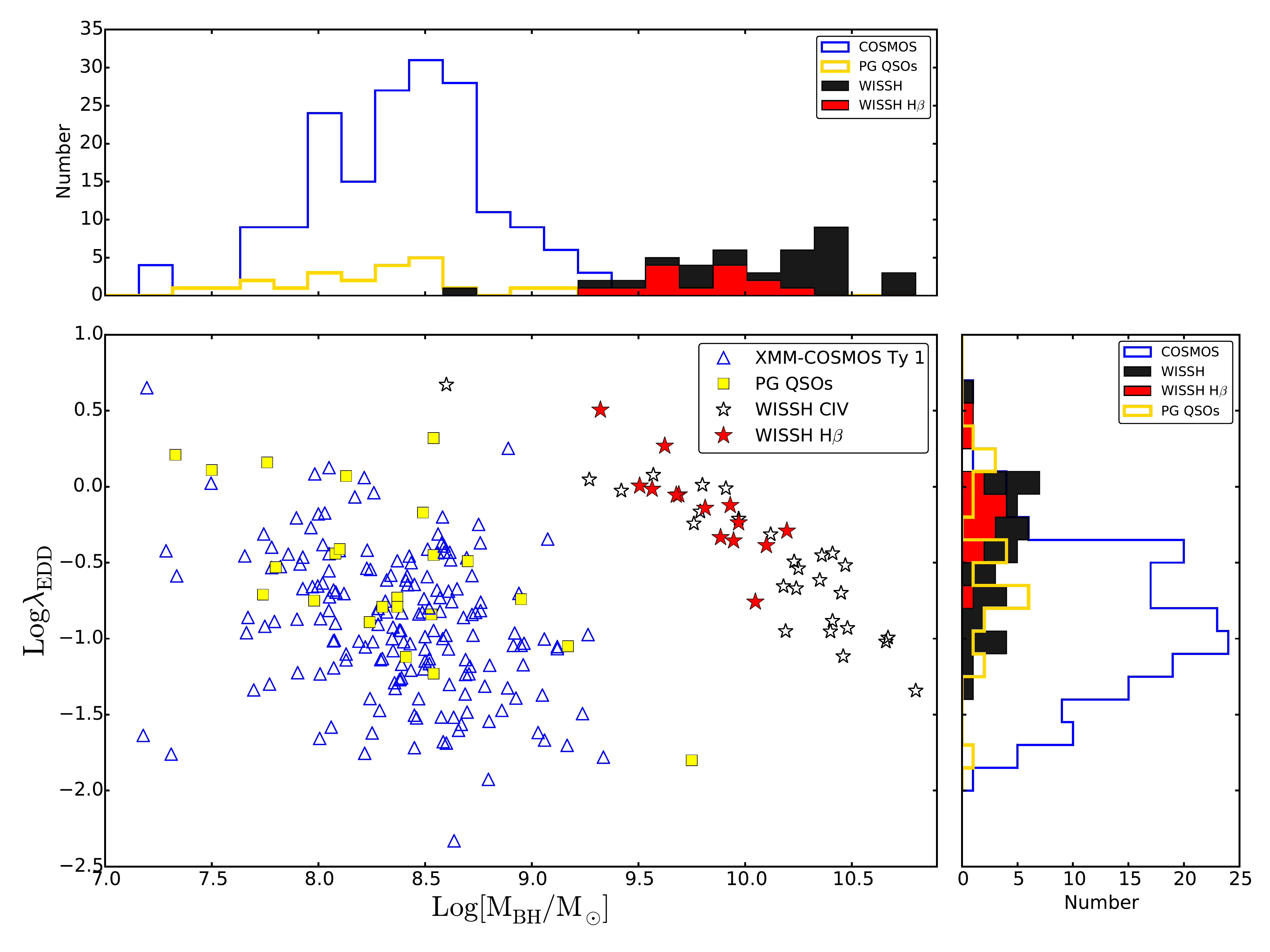}
	\caption{Eddington ratios as a function of black hole masses. The distribution of the masses is on the top, while the distribution of Eddington ratios is shown on the right (Symbols as in Fig. \ref{fig:eddratio}).}
	\label{fig:histos}
\end{figure*}

The WISSH  survey offers the  opportunity of  sampling extreme values 
 of  \lbol,  $k_{\rm Bol, X}$, $M_{\rm BH}$ and $\lambda_{\rm EDD}$  
 poorly sampled so far, i.e.,
$\gtrsim$  10$^{14}$ $L_\odot$, 10$^{2}$, 10$^{10}$ $M_\odot$ and 0.5, respectively, as
 highlighted by Fig. \ref{fig:histos} which  shows the distribution of $\lambda_{\rm EDD}$  as a function of the SMBH mass for the AGN samples considered in  Fig. \ref{fig:eddratio}.

A clear difference  in  the distribution of the $M_{\rm BH}$ is  evident when comparing sources from X-WISSH to those from COSMOS and PG survey, which can be considered as  representative of  the most studied class of broad-line AGN  in X-rays.
We performed a Kolmogorov-Smirnov (KS) test in order to understand if they are consistent with having been sampled from the same parent distribution and  
we derived a {\it p-value}  $<10^{-9}$  for both comparisons.
Accordingly, WISSH sources truly represent an extreme class of AGN showing a $M_{\rm BH}$ distribution centered around 10$^{10}$ $M_\odot$.
When comparing $\log \lambda_{\rm EDD}$ of X-WISSH to COSMOS, the {\it p-value} (i.e. the probability that they belong to the same parent distribution) is $\sim 10^{-6}$, while this probability increases to $\sim 10\%$ if the $\lambda_{\rm EDD}$ of X-WISSH and  PG quasars are considered.
However, it is worth noting that the  H$\beta$-based SMBH mass derived for randomly-selected WISSH quasars with optical rest-frame spectroscopy seems to be typically smaller than those estimated from the CIV emission line. \cite{coatman17} reported a similar trend by comparing  H$\alpha$-based with CIV-based $M_{\rm BH}$, finding that  CIV-based $M_{\rm BH}$ can be overestimated by up to a factor of $\sim$5, especially in case of highly-blueshifted ($v_b >$ 1200 km ${\rm s^{-1}}$) CIV emission lines. The latter are detected in the majority of WISSH quasars (see Sect. \ref{sec:disc}; Vietri et al. in prep.), suggesting that the real distribution of the $\lambda_{\rm EDD}$ of WISSH quasars may be centered at much larger values. If we restrict the comparison to objects with more reliable H$\beta$-based masses, the probability that they belong to the same parent distribution decreases to 0.04\%, which indicates that these two distributions are indeed different, thus confirming the extreme nature of WISSH sources.\\

 
\subsection{X-ray photon index versus $M_{\rm BH}$ and $\lambda_{\rm EDD}$}
\label{sgammabh}

\begin{figure*}[h!]
	\centering
	\includegraphics[trim={0.45cm 0 0 0}, clip=true, scale=0.36]{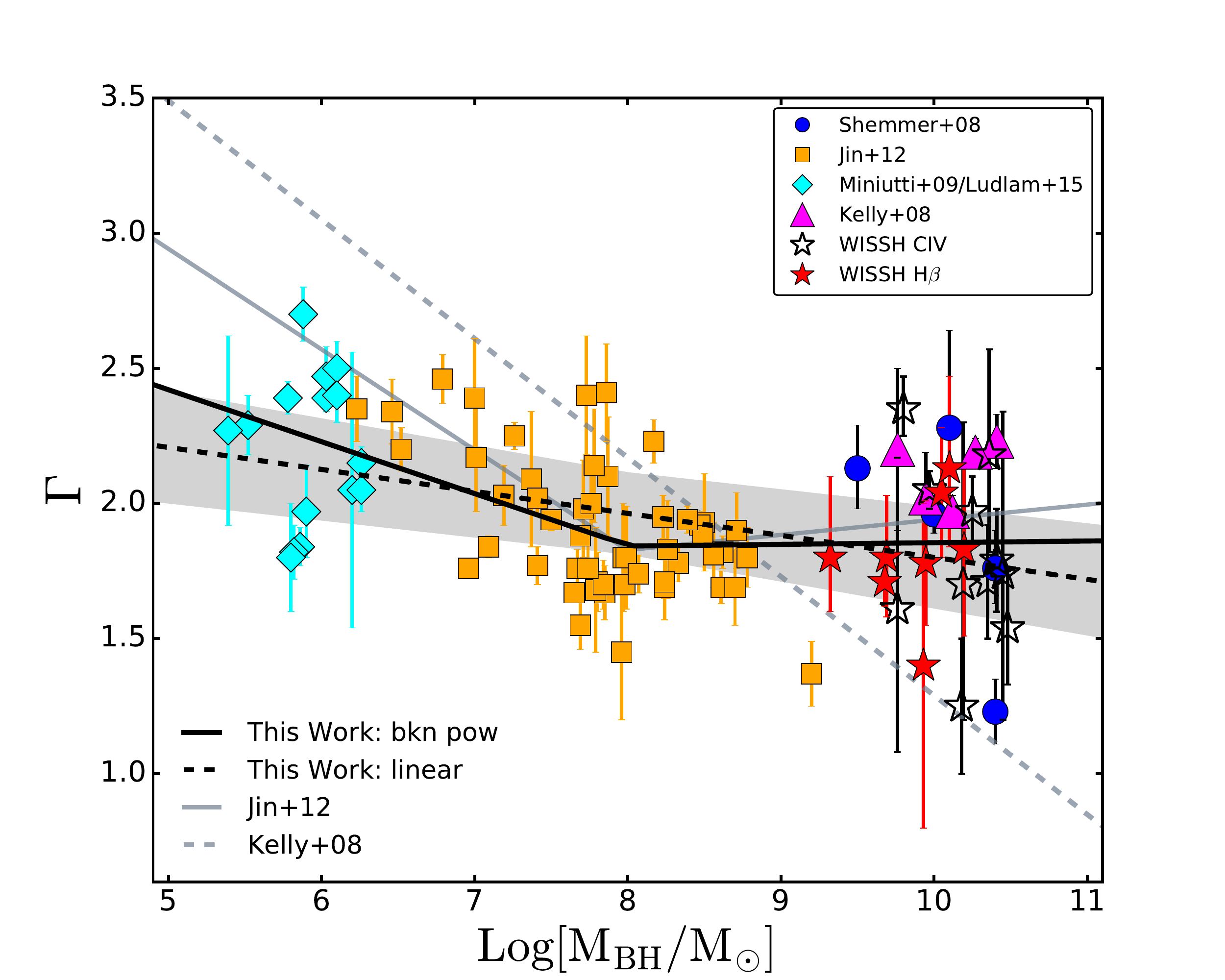}
	\includegraphics[trim={0.45cm 0 0 0}, clip=true, scale=0.36]{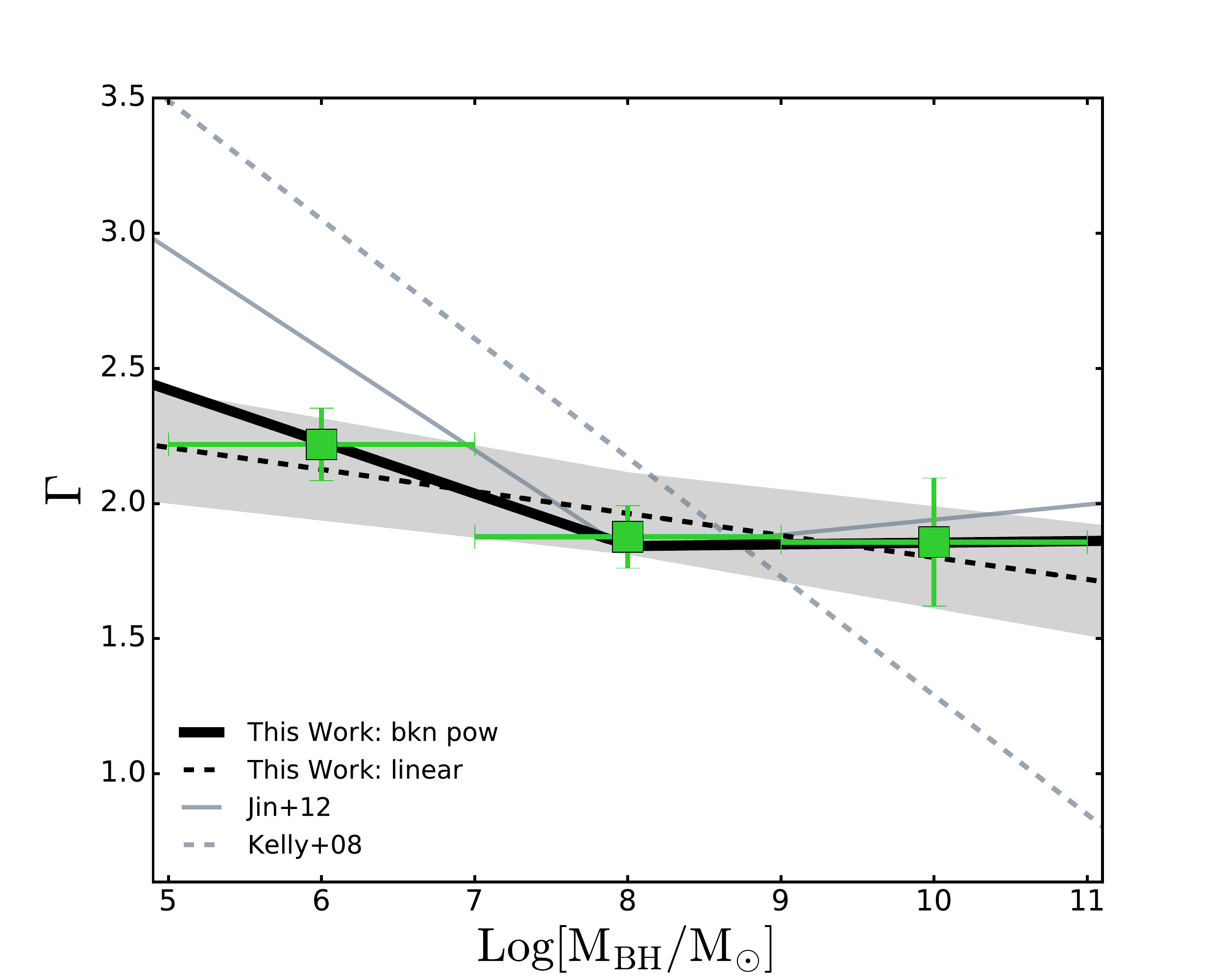}
	\caption{{\it Left Panel}: X-ray photon index as a function of $M_{\rm BH}$ for the X-WISSH quasars with X-ray spectroscopy (filled stars indicate objects with a H$\beta$-based SMBH mass \citealt{bischetti16}, Vietri et al. in prep.), while open stars indicate sources with a CIV-based SMBH mass (\citealt{weedman12}). We also report data for other samples as follows: cyan diamonds indicate intermediate   $M_{\rm BH}$ AGN by \citealt{miniutti09} and \cite{ludlam15}; orange squares mark the AGN from \cite{jin12a}; magenta triangles and blue circles represent high-$z$ quasars from \cite{kelly08} and \cite{shemmer08}, respectively. The black  dashed  and solid line represent the  linear  (the gray shaded area indicates the combined uncertainties in slope and normalization) and broken power-law relations found in this work. The gray  dashed line and the solid line mark the linear and broken power-law  found by \cite{jin12a} and \cite{kelly08}, respectively. Errors on $\Gamma$ are consistently  reported at 90\% c.l. for all the sources. {\it Right Panel}:  binned $\Gamma$ over the following  log($M_{\rm BH}$/$M_{\odot}$)  ranges: 5--7, 7--9 and 9--11.
	}
	\label{fig:gammamh}
\end{figure*}

Previous works focusing on the relation between the photon index $\Gamma$  and $M_{\rm BH}$ found different behaviors. Specifically, \cite{kelly08} reported a strong monotonic decreasing trend for  H$\beta$-based $M_{\rm BH}$ of low-$z$ AGN, while \cite{jin12a} claimed for the existence of two regimes, i.e. one decreasing up to  log($M_{\rm BH}/M_{\odot}$) = 8, and another slightly increasing at larger masses.
In order to further investigate this relation, we extended the range of $M_{\rm BH}$ considered by \cite{jin12a} both at  \simlt 10$^{6}$ and
\simgt 10$^{9}$ by including the \cite{miniutti09} and \cite{ludlam15} samples of intermediate mass SMBHs, and the WISSH quasars, respectively.
Fig. \ref{fig:gammamh} shows the photon index $\Gamma$  as a function of $M_{\rm BH}$ for these samples. 
Notice that we tried to be as less biased as possible by the soft excess contamination. Indeed, we considered the slope of the primary hard X-ray power law derived by a fit with an additional  spectral component to account for soft excess (i.e. Table 3 in  \citealt{miniutti09} and \citealt{ludlam15}) or those derived from a fit limited to the 2-10 keV band (Table 3 in \citealt{jin12b}). 
In case of high-$M_{\rm BH}$ objects, which in this case are all at high $z$\footnote{In addition to WISSH quasars with $\Gamma$ derived via X-ray spectroscopy, we have considered the 5 quasars at $z$ \simgt\ 1.5  listed in Table 2 in \cite{kelly08}, and the 5 quasars listed in Table 2 in \cite{shemmer08}.},   we adopted the photon index resulting from a power-law fit since the soft excess is outside the observed band.  
According to Fig. \ref{fig:gammamh}, a slightly decreasing trend with $M_{\rm BH}$, possibly flattening at large  values is visible.
We have therefore fitted the data with both a (i) linear ($\chi^2$/dof = 341/93) and a (ii) broken power law ( $\chi^2$/dof = 304/91) relation finding the following best-fit parametrizations:\\

\begin{equation}
(i)\hspace{0.5cm} \Gamma = (-0.08\pm0.02)\log(M_{\rm BH}/M_{\odot}) + (2.61\pm0.15) 
\end{equation}

\begin{equation}
(ii)\hspace{0.5cm} \Gamma = (-0.19\pm0.04)\log(M_{\rm BH}/M_{\odot}) + (3.38\pm0.28) 
\end{equation}
\noindent for $\log (M_{\rm BH}/M_{\odot}) \leq (8.01\pm0.48)$;

\begin{equation}
\Gamma =  (0.006\pm0.045)\log(M_{\rm BH}/M_{\odot}) + (1.79\pm0.28)
\end{equation}
\noindent for $\log (M_{\rm BH}/M_{\odot}) > (8.01\pm0.48)$. \\

\begin{figure}
	\centering
	\includegraphics[trim={0.45cm 0 0 0}, clip=true, scale=0.36]{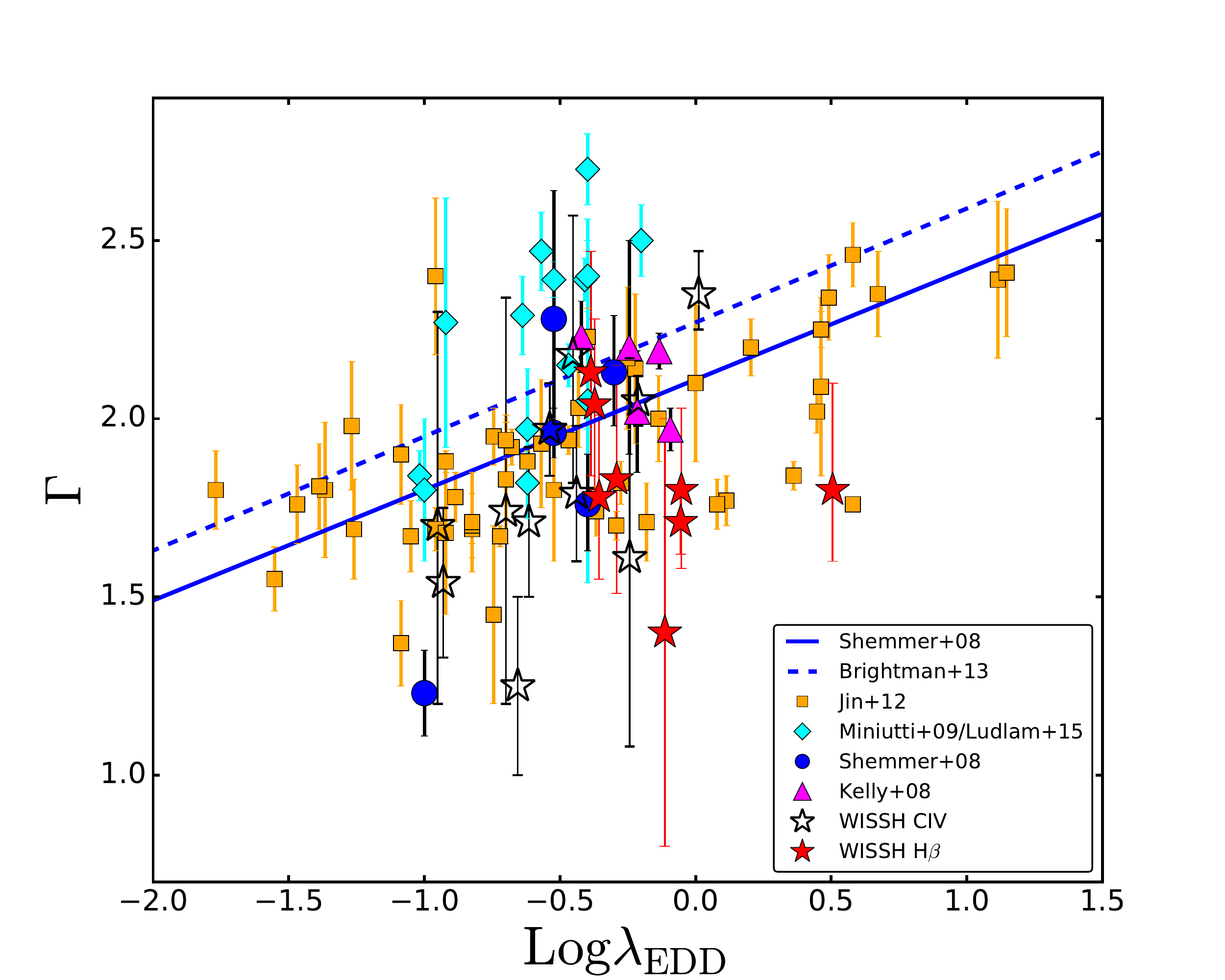}
	\caption{X-ray photon index as a function of $\lambda_{\rm EDD}$ for the X-WISSH quasars with X-ray spectroscopy. Symbol as in Fig. \ref{fig:gammamh}. The blue solid line represents the relation found by \citet{shemmer08}, while the blue dashed line represents the relation by \citet{brightman13}.
	Errors on $\Gamma$ are consistently  reported at 90\% c.l. for all the sources. }
	\label{fig:gammaedd}
\end{figure}

It can be seen that our broken power-law relation implies flatter $\Gamma$ than that derived by \cite{jin12a} over the entire $10^5-10^{10}$  masses. This is likely due to the increased statistics at both extremes of $M_{\rm BH}$ range, which led to a better sampling of the overall $\Gamma$  distribution. 
Notice that unlike \cite{jin12a} we left the break in  $M_{\rm BH}$  free to vary. We find, however, that our measured break is consistent with that assumed by \cite{jin12a}.
The much steeper relation reported by  \cite{kelly08} can be explained in terms of the presence of an unfitted soft excess component in their power-law fit to the 0.3-7 keV spectra of low-$z$ AGN.
Our results suggest that the power-law slope depends weakly on $M_{\rm BH}$. 
The modest increase measured at small masses might be ascribed to an enhancement of the  emission from the accretion disk leading to a stronger X-ray corona cooling, in case of highly-accreting sources.
This effect is more noticeable in low-$z$, low-$M_{\rm BH}$ sources (such as Narrow Emission line Seyfert galaxies, e.g. \citealt{kura00}) by a stronger soft excess which manifests itself at progressively harder energies. However, in order to obtain more comprehensive and less biased view of the $\Gamma$--$M_{\rm BH}$
relation two major improvements are at least necessary: (i) a significant increase of studied AGN with  $M_{\rm BH}$ $\ll$ 10$^{7}$ $M_{\odot}$ to overcome the observed large scatter in $\Gamma$;
(ii) a homogeneous spectral analysis with an accurate modeling of the soft excess component of X-ray data  in the widest possible energy range (i.e. by combining \xmm\ and {\it Chandra} with simultaneous {\it NuSTAR} observations).

Fig. \ref{fig:gammaedd} shows $\Gamma$ as a function of $\log \lambda_{\rm EDD}$ for the same sources of 
Fig. \ref{fig:gammamh}.
 \citet{shemmer08} and \citet{brightman13} reported the existence of a correlation between these two quantities (shown in Fig. \ref{fig:gammaedd} as a blue solid and dashed line, respectively). The former is based on 10 highly luminous radio-quiet  AGN at $z = 1.3-3.2$ plus 25 less luminous PG quasars at $z < 0.5$, while the latter was obtained by including 69 X-ray selected, broad-line AGN up 
to $z \sim  2$ in COSMOS and Extended CDFS, with $10^{42.5} < L_{2-10} < 10^{45.5} \, {\rm erg \, s^{-1}}$.
If on one hand the data appear to be broadly consistent with the reported correlations, on the other hand a large scatter is evident, especially for -1 $\leq$ $\log \lambda_{\rm EDD}$ $\leq$ 0. Furthermore, as expected, most of the steepest $\Gamma$ are those from low-$M_{\rm BH}$ sources and these may both drive the correlation at high $\lambda_{\rm EDD}$ and contribute to the large scatter. Accordingly, before drawing a firm conclusion on the
existence of a strong $\Gamma$--$\log \lambda_{\rm EDD}$ correlation, the same improvements required for the
$\Gamma$--$M_{\rm BH}$ should be taken in to account.

\section{Conclusions and Future Work}
\label{sec:disc}

\subsection{Relative X-ray weakness of hyper-luminous quasars compared to less luminous AGN}
\label{subsec:xweakness}

We have investigated the X-ray properties of 41 sources from the sample of WISSH quasars, which includes among the most luminous AGN known. They exhibit \lum\ $\approx$ 10$^{45-46}$ \ergs, i.e. well above the typical
X-ray luminosity range of AGN extensively investigated in previous spectroscopic studies.
Furthermore, the MIR selection enables us to extend and complete the study of the nuclear properties of hyper-luminous quasars, which have been mainly based on X-ray and/or optically selected AGN samples so far.
As expected from their optical classification of broad-line objects, we found that most of them show a low level of X-ray absorption along the line of sight to the nucleus, with a fraction of $\sim$80\% exhibiting intrinsic obscuration \nh $\leq 5 \times 10^{22} \, {\rm cm^{-2}}$.
We found that WISSH quasars allow to sample different range of the parameters space with respect to the bulk of the AGN population at lower luminosities when their X-ray output is compared to emission properties in other bands.
The X/O colors  derived for WISSH quasars are very low, i.e. 0.01 $<$ X/O $<$ 0.1, and cannot be ascribed to an obscured X-ray emission, since,  as expected from their optical classification of broad-line objects, we found that most of them show a low level of X-ray absorption along the line of sight to the nucleus (see Sect. \ref{subsec:xrayprop}). The largest X/O values observed for less-luminous sources in the CDFS and COSMOS survey (0.1 $<$ X/O $<$ 10)  can be partly explained by the X-ray selection. However,
our results point out that luminous AGN can even  reach X/O values as low as those so far reported for  star-forming and quiescent galaxies (e.g., \citealt{barger03}).
We also found that WISSH quasars follow the well-known \aox-$L_{2500\AA}$ anti-correlation \citep{vignali03} by populating the bright end of the
distribution with the steepest \aox\ (see Eq. \ref{eq:aoxeq} and Fig. \ref{fig:aox}). The combination of variability and  non-simultaneity of UV and X-ray observations contributes to the observed large scatter in this relation, however it is not the main cause of dispersion as
results based on simultaneous data clearly indicate (see \citealt{vagnetti10}).
Steep \aox\  and, therefore, very low X/O clearly imply that, with respect to less-luminous AGN, the X-ray emission of hyper-luminous quasars is relatively weaker as compared to the  optical/UV one.
The results of our X-ray analysis therefore lend support to previous works  that reported different SED for quasars with different luminosity. \citet{krawiz13} analyzed the SEDs of  119,652 broad-line quasars detected
in the SDSS with 0.064 $< z <$ 5.46 and found that  the mean SED of high-luminosity sources, relative to the SED of low-luminosity sources, is characterized by a softer (i.e. redder) far-UV spectral slope, a bluer optical continuum and a stronger hot dust emission peaking at 2--4 $\mu$m.
Remarkably, the formation of radiation line driven winds from the accretion disk are favored in AGN with strong UV luminosity relative to the X-ray one (i.e. {\it X-ray weak}). The efficiency of the line driving mechanism in accelerating the wind is indeed largely enhanced by the abundance of UV photons, while the soft UV-to-X-ray ionizing continuum avoid  the  overionization of the gas in the disk atmosphere (\citealt{leighly04, proga03, proga07} and references therein).

\begin{figure}
	\centering
	\includegraphics[trim={0.25cm 0 0 0}, clip=true, scale=0.39]{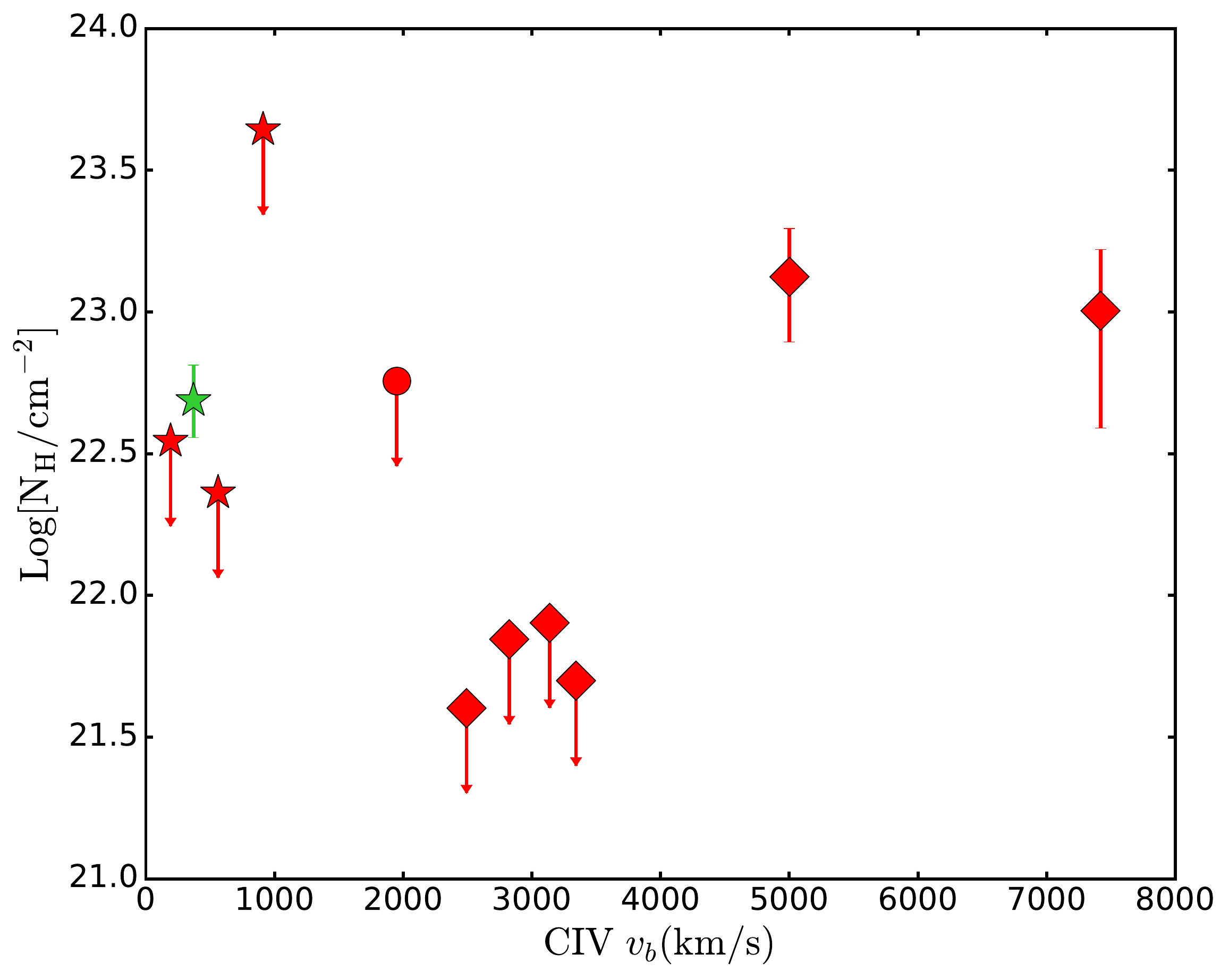}
	\caption{\nh\ derived by spectral or HR analysis for the detected X-WISSH quasars as a function of CIV blueshift ($v_b$). Symbols as in Fig. \ref{fig:XO}.}
	\label{fig:Nhshift}
\end{figure}

This scenario is particularly relevant for high-ionization lines which originate closer to the primary continuum source than low-ionization lines. It also provides an explanation for the larger blueshifts ($v_b$)  of the CIV with respect to low-ionization emission lines such as MgII and H$\beta$ found at higher luminosities in the "wind-dominated" quasar population (e.g. \citealt{richards11, marziani15}), and the  observed anti-correlation between $v_b$ and the EW of the CIV emission line \citep{korista98, krawiz13}. This SED-based scenario for "wind-dominated" quasars rules out orientation as a major cause of these blueshifts.
Such accretion disk winds have been also revealed in the majority of WISSH quasars with available rest-frame optical spectroscopy, showing $v_b$ up to $\sim$ 7000 km s$^{-1}$ (Vietri et al. in prep.). Fig. \ref{fig:Nhshift} shows the \nh\ derived by our spectral X-ray analysis plotted versus  $v_b$. \citet{gallagher05} claimed that large $v_b$ ($>$ 1100 km s$^{-1}$) quasars  exhibit  X-ray absorption at levels of \nh $> 10^{22}$ \cm2, as expected in case of quasars observed at larger inclination angles, i.e. with a line of sight close to the accretion disk plane. However, we find that most of WISSH quasars with such large   $v_b$ values  have \nh \simlt\  $10^{22}$ \cm2, which is at odds with their claim that the detection of large blueshifts is due to an orientation effect.

Furthermore, a softer ionizing continuum scenario at larger luminosity explains the detection of CIV, NV and  SiIV BALs in quasars, and two anti-correlations reported by \citet{brandt00} and \citet{laor02} between \aox\ and the EW of the CIV absorption line, and between luminosity and  maximum outflow velocity of absorption, respectively.

Remarkably, \citet{proga05} suggests that highly-accreting AGN can launch an UV radiation-driven accretion-disk wind being
effective in weakening/destroying the X-ray corona and, therefore, quenching the X-ray emission.
This explanation is very intriguing as it predicts different properties for the X-ray corona in
powerful AGN which are linked to the simultaneous presence of nuclear  outflows.
In this sense, the extreme X-ray weakness of WISSH quasars exhibiting strongly blueshifted CIV emission lines (see Figs. \ref{fig:XO} and \ref{fig:lbol}) lends support to this scenario, although it is too early to draw robust conclusions due to the small number of sources. 
Hyper-luminous quasars thus represent the  ideal laboratory
to study the the coronal-quenching scenario and the link between the AGN energy output and wind acceleration, 
since they satisfy the ideal conditions to develop powerful radiation driven disk winds as found by \cite{proga07}, i.e.,
extreme \lbol\ and very large $M_{\rm BH}$.


\begin{figure}
	\centering
	\includegraphics[trim={0.25cm 0 0 0}, clip=true, scale=0.39]{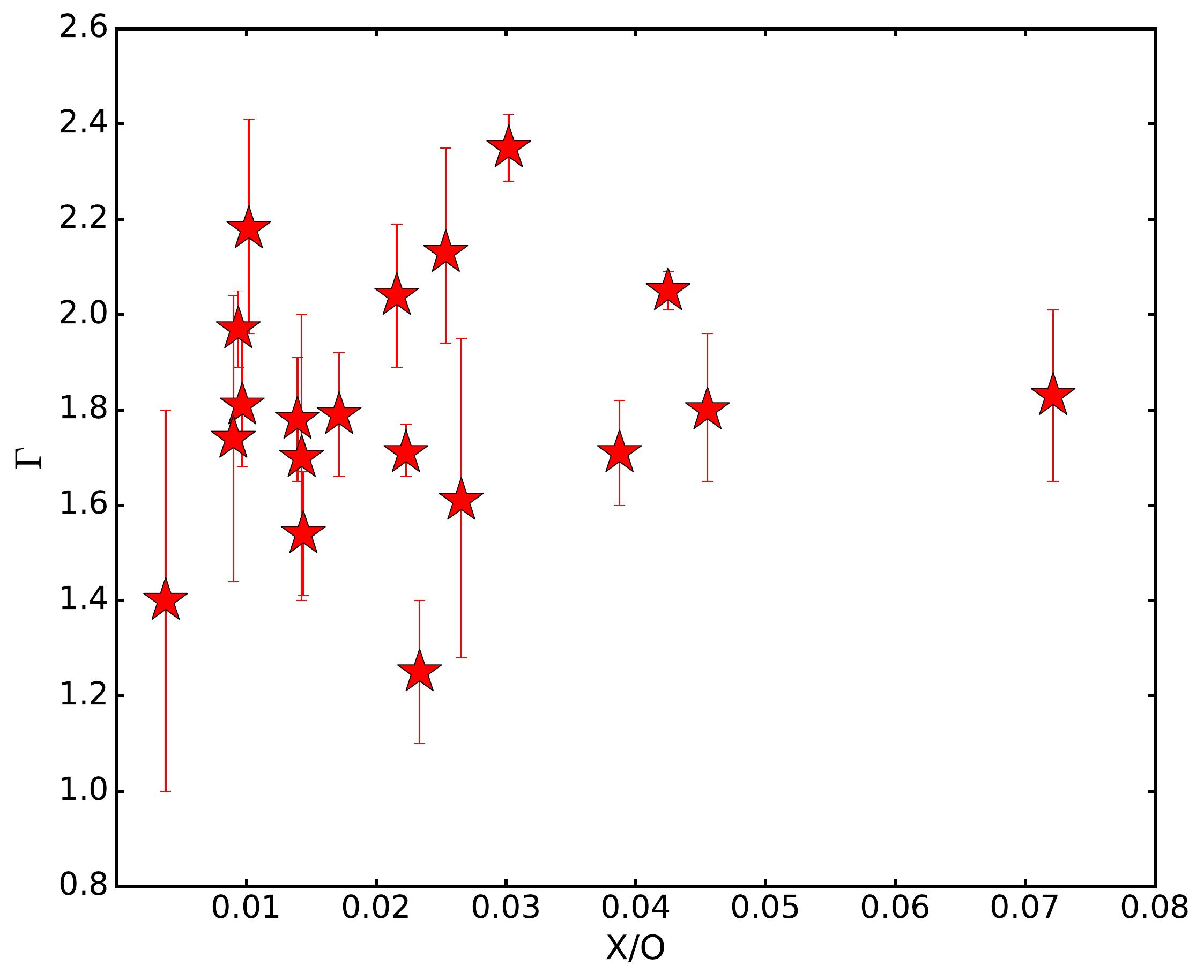}
	\caption{ X-ray photon index as a function of X/O for the X-WISSH quasars with X-ray spectroscopy.}
	\label{fig:XOgamma}
\end{figure}

Alternative explanations for the intrinsic X-ray weakness relative to UV in AGN have been proposed. Specifically, the X-ray under-luminosity could be due to a strong light bending effect which suppresses the X-ray continuum and  produces a reflection-dominated X-ray spectrum \citep{schartel10,miniutti09b}, or a consequence of photon-trapping and advection into the SMBH of X-ray photons in case of high accretion rate regimes (e.g. \citealt{leighly07}).
Furthermore, radiation magneto-hydrodynamic simulations presented by \cite{jiang14} suggest that,  assuming a scenario with the X-ray  corona
heated via  dissipation of turbulence driven by magneto-rotational instability,  X-ray weakness can occur when most of the  energy liberated by accretion is dissipated in the disk due to a large surface density (and possibly associated to a very large accretion rate and luminosity).
 It might also be argued that the weakness observed in the 2-10 keV flux of our hyper-luminous quasars could partly be due to a $\Gamma$ flatter than the canonical range of $\sim$1.8-2, which implies that 
the bulk of the X-ray radiation is  emitted at energies larger than 10 keV. In this case, we should expect that sources with a flattest $\Gamma$ would show the lowest X/O, offering a possible explanation for the X-ray weakness in terms of a flatness of the spectral index.
However, Fig. \ref{fig:XOgamma} does not support this scenario and the Spearman's rank test results into a probability of deviation from a random distribution of $d_s = 0.34$, indicating no significant correlation between X/O and $\Gamma$.

The observed displacement of luminous quasars from the  $L_X-L_{\rm MIR}$ relation inferred for low-luminosity AGN (see Sect. \ref{subsec:lmir} and Fig. \ref{fig:L6Lx}) can be accounted for by the relative X-ray weakness progressively emerging in objects at the bright end of the AGN luminosity function compared to the slowly decreasing bolometric correction in the MIR (\citealt{treister08}; \citealt{runnoe12b}).
 In addition to this effect, the low $L_{2-10}$-to-$\lambda L_{6 \mu m}$ ratios derived for WISSH quasars can be also caused by the strong  AGN-heated, hot dust emission typically observed in the mean SED of hyper-luminous quasars \citep{krawiz13}. This  leads to a luminosity-dependent $L_X-L_{\rm MIR}$ relation, although the observed large scatter  in the $L_{2-10}$-$\lambda L_{6 \mu m}$ plane severely challenges  the idea of the existence of an universal relation.
 Furthermore, an explanation in terms of a luminosity-dependent $C_f$ is not able to reproduce the observed trend at large 
 $L_{\rm MIR}$.

The presence of a relatively weaker X-ray corona in hyper-luminous quasars compared to typical AGN is further supported by the increasing $k_{\rm Bol, X}$ with \lbol\ (see. Fig. \ref{fig:lbol}). WISSH quasars typically exhibit $k_{\rm Bol, X}$ \simgt\ 100, which demonstrates a significantly reduced contribution of the X-ray emission to the radiation output produced by the accretion disk-corona system  in the powerful quasar regime.
A deeper investigation based on a large sample of luminous quasars with  well-determined  X-ray spectral information and $M_{\rm BH}$ is also crucial to establish the possible dependence of  $k_{\rm Bol, X}$ on $\lambda_{EDD}$ and $M_{\rm BH}$, and, if a strong correlation is observed,  to understand which is the dominant parameter among  \lbol, $\lambda_{EDD}$ and  $M_{\rm BH}$. 
 This will also improve our  understanding of the  relation between $\Gamma$ and $M_{\rm BH}$, for which we derived
 a double power-law behaviour flatter than previously found by other works (see Sect. \ref{sgammabh}).  
 The WISSH quasar sample offers a unique data set   to perform such a study as shown in Fig. \ref{fig:histos}.

\begin{figure}
	\centering
	\includegraphics[trim={0.5cm 4cm 0.5cm 4cm}, clip=true, scale=0.4]{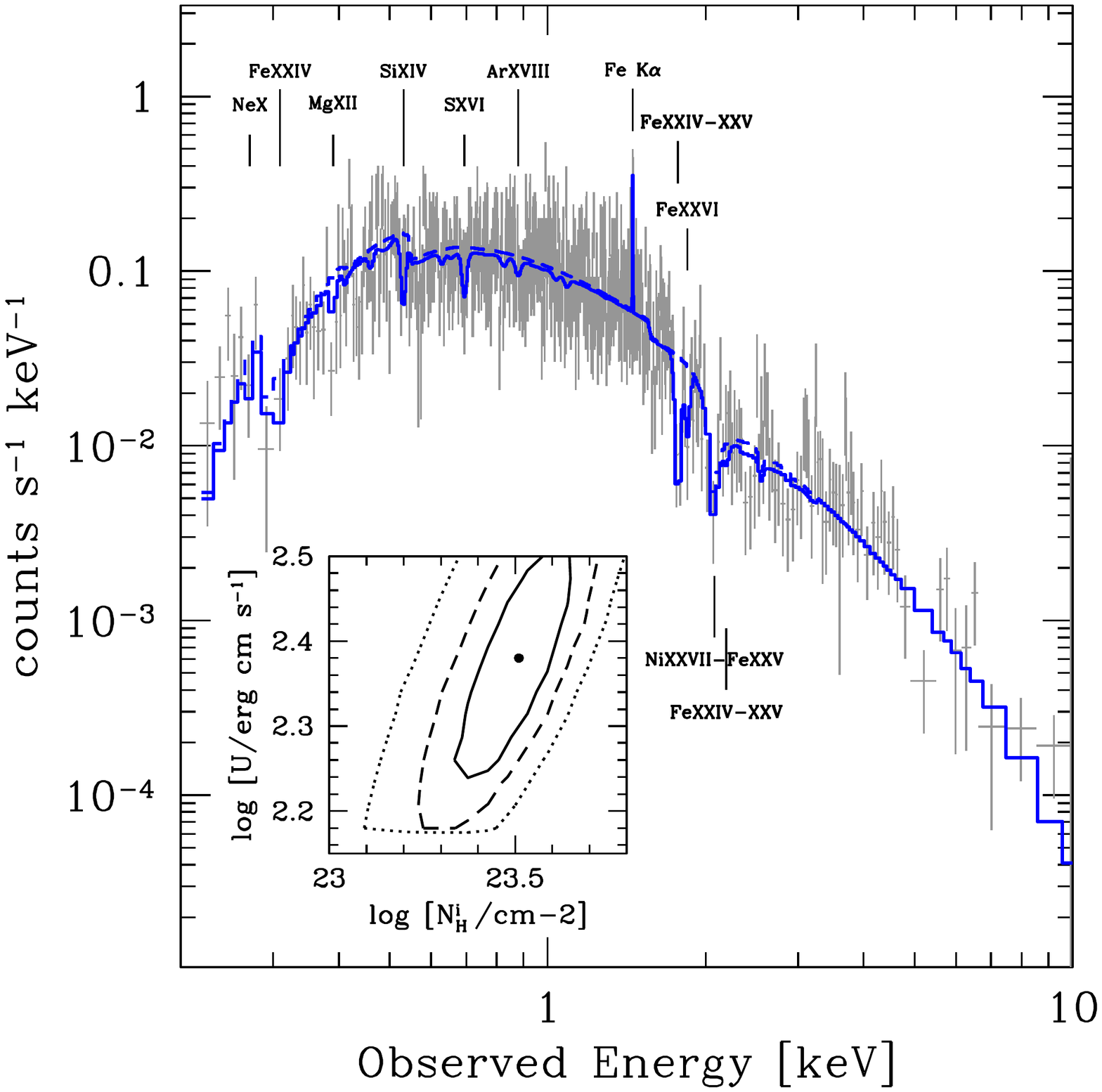}
	\caption{Simulated Athena/X-IFU 15~ks spectrum   for a WISSH-like quasar at $z$ = 3.4 with UFO absorber with velocity 0.15c, $N_{H}^{i}=10^{23.4}$~\cm2; $\log (U) =2.3$  and $v_{turb}=5000$ km s$^{-1}$ (see Sect. \ref{subsec:athena}). The  solid
line indicates	 the best-fit baseline continuum model modified by the UFO and the Fe~K$\alpha$ line (the dotted line
indicates the  baseline continuum component). Absorption lines with observed $EW>1$~keV and the Fe~K$\alpha$ emission line are labelled. The inset shows confidence contours for $U$ and $\log N_{H}^{i}$. Contours are 68, 90\% and 99\% for two interesting parameters.}
	\label{fig:ufospec}
\end{figure}

\subsection{Future perspective with ATHENA}
\label{subsec:athena}

We discuss here the possible development in the study of the X-ray
properties of WISSH quasars in the light of the upcoming ESA's next
X-ray observatory {\it Athena} \citep{nandra13}. The marked improvement in sensitivity
and spectral resolution which will be provided by the X-ray
calorimeter X-IFU on board {\it Athena} will allow us to easily
investigate the global X-ray properties of  WISSH quasars
with a modest amount of time.

We simulated a 0.2-10~keV X-IFU observation for a X-WISSH
source with the lowest flux $f_{2-10} = 7 \times 10^{-15}\,{\rm erg \,
  s^{-1} cm^{-2}}$ and with average redshift $z=3.4$ (median for the
whole WISSH sample), $\Gamma=1.8$ and column density \nh=$8 \times 10^{21} \, {\rm
  cm^{-2}}$ (median for the X-WISSH sample).  With this baseline spectral model,
Athena/X-IFU can easily gather more than 600 counts in only 15~ks
observation. This can give constraints on $\Gamma$ and \nh at $<10$\%
and $\sim50$\% level ($1\sigma$ error range), respectively.

Given the huge luminosity of the WISSH quasars, we expect
them to be likely hosting widespread signs of outflows at all scales.
Several theoretical models \citep{k2010,fgq2012,zub}
 propose that the kpc-scale outflows
routinely observed in the ionized, neutral and molecular phases
are  triggered by relativistic, X-ray winds originating in the nuclear
regions, the so-called Ultra-Fast Outflows
\citep[UFO,][]{t2010}.
Recent observations lend support to this scenario, suggesting an energy-conserving
expansion for  the large-scale outflows \citep{feru15,tombesi15}.

 Our WISSH sample is
characterized by two distinct populations showing in UV and optical mildly
ionized outflows in the nuclear
region as large CIV blueshifts ($\geq2000~\rm{km \, s^{-1}}$; Vietri et
al. in prep.) and at larger galaxy-wide scales as massive [OIII] outflows
($\gtrsim2000~\rm{M_\odot yr^{-1}}$; \citealt{bischetti16}).  We can
envisage a program to follow-up these two categories and investigate
the link between  the UFOs and their outflow manifestations. 

In order to simulate the spectral signatures from relativistic nuclear
winds we used {\sl PHASE}, a self-consistent photoionized absorber code
\citep{k2003}. Under the assumption of ionization balance the code
calculates the line opacity due to highly ionized metal transitions
given (1) the equivalent hydrogen column density of ionized material
($N_{H}^{i}$), (2) ionization parameter $U$\footnote{$U=Q/4\pi r^2cn$,
where $Q$ is the rate of hydrogen ionizing photons from a source at
distance $r$, $n$ is the hydrogen density and c is the speed of
light. The alternative ionization parameter $\xi$ is defined as
$\xi$ = $L$/$nr^{2}$, where $L$ is the isotropic luminosity 
ionizing source in the interval 13.6 eV to 13.6 keV.}, (3) turbulent velocity ($v_{turb}$) and (4) covering factor ($f$) of the
photoionized plasma.

We simulated a Athena/X-IFU 15~ks observation for a baseline model with flux $f_{2-10} = 3 \times
10^{-14}\,{\rm erg \, s^{-1} cm^{-2}}$. This is the lowest flux for
the sub-sample of sources for which we detect optical and UV
signatures from mildly ionized outflows. We modified this spectrum with
{\sl PHASE} assuming as physical parameters for the photoionized
absorber those measured by \citet{g2014} for the smallest
detected feature (in their sequence 2013b) in the local WISSH analog
PDS~456; i.e. one of the brightest $z<0.2$ QSO
(\lbol$=10^{47}$~\ergs), showing recurrent signatures of UFOs \citep{r2003,r2009,r2014}.
The parameters adopted are:
$N_{H}^{i}=10^{23.4} \, {\rm cm ^{-2}}$; log$U$ = 2.3 (which roughly correspond to
the estimated $\log[\xi/{\rm erg \, cm\ s^{-1}}$] = 3.8 in PDS~456) and $v_{turb}=5000$ km s$^{-1}$  (we assumed $f=1$). To
this model we added a narrow Fe~K$\alpha$ line with rest-frame
equivalent width ${\rm EW} = 100$ eV and intrinsic full width at half
maximum of $\sim2000\,\rm km\,s^{-1}$ \citep{shuyaqoobwang10}.

Fig. \ref{fig:ufospec} reports a simulated 15~ks X-IFU spectrum from an
UFO with velocity 0.15c. The 0.2-10~keV collected counts are
$\sim2500$. The solid line shows the best-fit with the UFO signatures
imprinted in the spectrum. The dotted model indicates the same best-fit model
but without the absorption/emission line components. Constraints for $\log U$
and $\log N_{H}^{i}$ are reported in the inset. On average both
parameters can be measured with an accuracy of respectively $\sim5$\%
and $\sim1$\% (90\% c.l. for one interesting parameter). Such a  level of accuracy is possible thanks to
the presence of several highly ionized absorption lines arising in the
0.2-2~keV observed energy band (corresponding to rest-frame 0.7-8~keV)
from elements from OVIII to FeXXV-XXVI. This is clearly seen in
Fig. \ref{fig:ufospec}. The most significant of these lines is the FeXXV at 6.7~keV
which is detected at $\sim10\sigma$ with an observed EW$=800$~eV.  The
turbulent velocity of the UFO can be constrained with an accuracy of
$\sim20$\%.

The EW of the Fe~K$\alpha$ line can be constrained with
an accuracy of $25$\% and its position can be recovered at
levels of 100-150~$\,\rm km\,s^{-1}$, therefore allowing to probe the kinematics of
the reflecting material (i.e. rotation of the accretion disk, bulk motion
of outflowing material) down to such levels. This is crucial to obtain
X-ray based redshift estimates competing with low-dispersion optical 
spectrographs and probe the dynamics of the quasar innermost regions by
comparing it with the conditions of more external and lower ionization
medium.
These constraints allow us to accurately probe UFO properties in a sample of 20
WISSH quasars comprising 10 objects for each of the two WISSH
sub-populations with a relatively modest investment of exposure time (300~ks). 
We can thus investigate the link between SMBH winds and kpc-scale outflows
at extreme luminosity regimes and at the golden epoch of AGN/galaxy
co-evolution ($z \sim$ 2-3). 
Thanks to the enormous leap in sensitivity of Athena
compared to {\it Chandra} and \xmm\  we can study the recurrency of these
phenomena by probing their temporal variability on scales of few hours
and start to investigate their (anti-)correlated variability with the
enormous power radiated by the central engine as recently performed in
a local less luminous AGN by \citet{p2017}. In this way we will
further probe how the integrated energy output injected into the ISM
would affect the subsequent baryon cycle evolution in their host galaxy.


\begin{acknowledgements}
We  thank the anonymous referee for helpful comments that improved the paper.
We also  acknowledge very useful discussions with B. Lusso.
The scientific results reported in this article are based on observations made by the Chandra X-ray Observatory. 
We used observations obtained with XMM-Newton, an ESA science mission with instruments and contributions directly funded by ESA Member States and NASA.
This work was supported by ASI/INAF contract
I/009/10/0 and INAF PRIN 2011, 2012 and 2014.
LZ acknowledges  financial support under ASI/INAF contract  I/037/12/0.
MB and GL acknowledge support from the FP7 Career Integration Grant ``eEASy'': "supermassive black holes through cosmic time: from current surveys to eROSITA-Euclid Synergies" (CIG 321913). GB acknowledges financial support under the INTEGRAL ASI-INAF agreement 2013-025.R1.
This research has made use of the NASA/IPAC Extragalactic Database (NED) which is operated by the Jet Propulsion Laboratory, California Institute of Technology, under contract with the National Aeronautics and Space Administration. 
      
\end{acknowledgements}

%
%


%
%

\bibliographystyle{aa} 
\bibliography{wisshX} 

\begin{thebibliography}{120}
\expandafter\ifx\csname natexlab\endcsname\relax\def\natexlab#1{#1}\fi

\bibitem[{{Avni} \& {Tananbaum}(1982)}]{avni82}
{Avni}, Y. \& {Tananbaum}, H. 1982, ApJl, 262, L17

\bibitem[{{Banerji} {et~al.}(2015){Banerji}, {Alaghband-Zadeh}, {Hewett}, \&
  {McMahon}}]{banerji15}
{Banerji}, M., {Alaghband-Zadeh}, S., {Hewett}, P.~C., \& {McMahon}, R.~G.
  2015, \mnras, 447, 3368

\bibitem[{{Barger} {et~al.}(2003){Barger}, {Cowie}, {Capak}, {Alexander},
  {Bauer}, {Fernandez}, {Brandt}, {Garmire}, \& {Hornschemeier}}]{barger03}
{Barger}, A.~J., {Cowie}, L.~L., {Capak}, P., {et~al.} 2003, AJ, 126, 632

\bibitem[{{Bianchi} {et~al.}(2007){Bianchi}, {Guainazzi}, {Matt}, \& {Fonseca
  Bonilla}}]{bianchi07}
{Bianchi}, S., {Guainazzi}, M., {Matt}, G., \& {Fonseca Bonilla}, N. 2007,
  \aap, 467, L19

\bibitem[{{Bianchi} {et~al.}(2012){Bianchi}, {Maiolino}, \&
  {Risaliti}}]{bianchi12}
{Bianchi}, S., {Maiolino}, R., \& {Risaliti}, G. 2012, Advances in Astronomy,
  2012, 17

\bibitem[{{Bischetti} {et~al.}(2017){Bischetti}, {Piconcelli}, {Vietri},
  {Bongiorno}, {Fiore}, {Sani}, {Marconi}, {Duras}, {Zappacosta}, {Brusa},
  {Comastri}, {Cresci}, {Feruglio}, {Giallongo}, {La Franca}, {Mainieri},
  {Mannucci}, {Martocchia}, {Ricci}, {Schneider}, {Testa}, \&
  {Vignali}}]{bischetti16}
{Bischetti}, M., {Piconcelli}, E., {Vietri}, G., {et~al.} 2017, \aap, 598, A122

\bibitem[{{Brandt} \& {Hasinger}(2005)}]{BH05}
{Brandt}, W.~N. \& {Hasinger}, G. 2005, \araa, 43, 827

\bibitem[{{Brandt} {et~al.}(2000){Brandt}, {Laor}, \& {Wills}}]{brandt00}
{Brandt}, W.~N., {Laor}, A., \& {Wills}, B.~J. 2000, \apj, 528, 637

\bibitem[{{Brightman} {et~al.}(2013){Brightman}, {Silverman}, {Mainieri},
  {Ueda}, {Schramm}, {Matsuoka}, {Nagao}, {Steinhardt}, {Kartaltepe},
  {Sanders}, {Treister}, {Shemmer}, {Brandt}, {Brusa}, {Comastri}, {Ho},
  {Lanzuisi}, {Lusso}, {Nandra}, {Salvato}, {Zamorani}, {Akiyama}, {Alexander},
  {Bongiorno}, {Capak}, {Civano}, {Del Moro}, {Doi}, {Elvis}, {Hasinger},
  {Laird}, {Masters}, {Mignoli}, {Ohta}, {Schawinski}, \&
  {Taniguchi}}]{brightman13}
{Brightman}, M., {Silverman}, J.~D., {Mainieri}, V., {et~al.} 2013, MNRAS, 433,
  2485

\bibitem[{{Bruni} {et~al.}(2012){Bruni}, {Mack}, {Salerno},
  {Montenegro-Montes}, {Carballo}, {Benn}, {Gonz{\'a}lez-Serrano}, {Holt}, \&
  {Jim{\'e}nez-Luj{\'a}n}}]{bruni12}
{Bruni}, G., {Mack}, K.-H., {Salerno}, E., {et~al.} 2012, \aap, 542, A13

\bibitem[{{Carniani} {et~al.}(2015){Carniani}, {Marconi}, {Maiolino},
  {Balmaverde}, {Brusa}, {Cano-D{\'{\i}}az}, {Cicone}, {Comastri}, {Cresci},
  {Fiore}, {Feruglio}, {La Franca}, {Mainieri}, {Mannucci}, {Nagao}, {Netzer},
  {Piconcelli}, {Risaliti}, {Schneider}, \& {Shemmer}}]{carniani15}
{Carniani}, S., {Marconi}, A., {Maiolino}, R., {et~al.} 2015, A\&A, 580, A102

\bibitem[{{Cash}(1979)}]{cash79}
{Cash}, W. 1979, ApJ, 228, 939

\bibitem[{{Chen} {et~al.}(2017){Chen}, {Hickox}, {Goulding}, {Stern}, {Assef},
  {Kochanek}, {Brown}, {Harrison}, {Hainline}, {Alberts}, {Alexander},
  {Brodwin}, {Del Moro}, {Forman}, {Gorjian}, {Jones}, {Murray}, {Pope}, \&
  {Rovilos}}]{chen17}
{Chen}, C.-T.~J., {Hickox}, R.~C., {Goulding}, A.~D., {et~al.} 2017, ArXiv
  e-prints [\eprint[arXiv]{1701.05207}]

\bibitem[{{Cicone} {et~al.}(2014){Cicone}, {Maiolino}, {Sturm},
  {Graci{\'a}-Carpio}, {Feruglio}, {Neri}, {Aalto}, {Davies}, {Fiore},
  {Fischer}, {Garc{\'{\i}}a-Burillo}, {Gonz{\'a}lez-Alfonso},
  {Hailey-Dunsheath}, {Piconcelli}, \& {Veilleux}}]{cicone14}
{Cicone}, C., {Maiolino}, R., {Sturm}, E., {et~al.} 2014, \aap, 562, A21

\bibitem[{{Civano} {et~al.}(2012){Civano}, {Elvis}, {Brusa}, {Comastri},
  {Salvato}, {Zamorani}, {Aldcroft}, {Bongiorno}, {Capak}, {Cappelluti},
  {Cisternas}, {Fiore}, {Fruscione}, {Hao}, {Kartaltepe}, {Koekemoer}, {Gilli},
  {Impey}, {Lanzuisi}, {Lusso}, {Mainieri}, {Miyaji}, {Lilly}, {Masters},
  {Puccetti}, {Schawinski}, {Scoville}, {Silverman}, {Trump}, {Urry},
  {Vignali}, \& {Wright}}]{civano12}
{Civano}, F., {Elvis}, M., {Brusa}, M., {et~al.} 2012, \apjs, 201, 30

\bibitem[{{Coatman} {et~al.}(2017){Coatman}, {Hewett}, {Banerji}, {Richards},
  {Hennawi}, \& {Prochaska}}]{coatman17}
{Coatman}, L., {Hewett}, P.~C., {Banerji}, M., {et~al.} 2017, \mnras, 465, 2120

\bibitem[{{Del Moro} {et~al.}(2016){Del Moro}, {Alexander}, {Bauer}, {Daddi},
  {Kocevski}, {McIntosh}, {Stanley}, {Brandt}, {Elbaz}, {Harrison}, {Luo},
  {Mullaney}, \& {Xue}}]{delmoro15}
{Del Moro}, A., {Alexander}, D.~M., {Bauer}, F.~E., {et~al.} 2016, MNRAS, 456,
  2105

\bibitem[{{Faucher-Gigu{\`e}re} \& {Quataert}(2012)}]{fgq2012}
{Faucher-Gigu{\`e}re}, C.-A. \& {Quataert}, E. 2012, \mnras, 425, 605

\bibitem[{{Ferland} \& {Rees}(1988)}]{ferland88}
{Ferland}, G.~J. \& {Rees}, M.~J. 1988, \apj, 332, 141

\bibitem[{{Feruglio} {et~al.}(2014){Feruglio}, {Bongiorno}, {Fiore}, {Krips},
  {Brusa}, {Daddi}, {Gavignaud}, {Maiolino}, {Piconcelli}, {Sargent},
  {Vignali}, \& {Zappacosta}}]{feruglio14}
{Feruglio}, C., {Bongiorno}, A., {Fiore}, F., {et~al.} 2014, A\&A, 565, A91

\bibitem[{{Feruglio} {et~al.}(2015){Feruglio}, {Fiore}, {Carniani},
  {Piconcelli}, {Zappacosta}, {Bongiorno}, {Cicone}, {Maiolino}, {Marconi},
  {Menci}, {Puccetti}, \& {Veilleux}}]{feru15}
{Feruglio}, C., {Fiore}, F., {Carniani}, S., {et~al.} 2015, \aap, 583, A99

\bibitem[{{Fiore} {et~al.}(2003){Fiore}, {Brusa}, {Cocchia}, {Baldi},
  {Carangelo}, {Ciliegi}, {Comastri}, {La Franca}, {Maiolino}, {Matt},
  {Molendi}, {Mignoli}, {Perola}, {Severgnini}, \& {Vignali}}]{fiore03}
{Fiore}, F., {Brusa}, M., {Cocchia}, F., {et~al.} 2003, A\&A, 409, 79

\bibitem[{{Fiore} {et~al.}(2009){Fiore}, {Puccetti}, {Brusa}, {Salvato},
  {Zamorani}, {Aldcroft}, {Aussel}, {Brunner}, {Capak}, {Cappelluti}, {Civano},
  {Comastri}, {Elvis}, {Feruglio}, {Finoguenov}, {Fruscione}, {Gilli},
  {Hasinger}, {Koekemoer}, {Kartaltepe}, {Ilbert}, {Impey}, {Le Floc'h},
  {Lilly}, {Mainieri}, {Martinez-Sansigre}, {McCracken}, {Menci}, {Merloni},
  {Miyaji}, {Sanders}, {Sargent}, {Schinnerer}, {Scoville}, {Silverman},
  {Smolcic}, {Steffen}, {Santini}, {Taniguchi}, {Thompson}, {Trump}, {Vignali},
  {Urry}, \& {Yan}}]{fiore09}
{Fiore}, F., {Puccetti}, S., {Brusa}, M., {et~al.} 2009, ApJ, 693, 447

\bibitem[{{Gallagher} {et~al.}(2002){Gallagher}, {Brandt}, {Chartas}, \&
  {Garmire}}]{gallagher02}
{Gallagher}, S.~C., {Brandt}, W.~N., {Chartas}, G., \& {Garmire}, G.~P. 2002,
  \apj, 567, 37

\bibitem[{{Gallagher} {et~al.}(2005){Gallagher}, {Richards}, {Hall}, {Brandt},
  {Schneider}, \& {Vanden Berk}}]{gallagher05}
{Gallagher}, S.~C., {Richards}, G.~T., {Hall}, P.~B., {et~al.} 2005, AJ, 129,
  567

\bibitem[{{Gehrels}(1986)}]{gehrels}
{Gehrels}, N. 1986, ApJ, 303, 336

\bibitem[{{Gibson} {et~al.}(2008){Gibson}, {Brandt}, \& {Schneider}}]{gibson08}
{Gibson}, R.~R., {Brandt}, W.~N., \& {Schneider}, D.~P. 2008, \apj, 685, 773

\bibitem[{{Gofford} {et~al.}(2014){Gofford}, {Reeves}, {Braito}, {Nardini},
  {Costa}, {Matzeu}, {O'Brien}, {Ward}, {Turner}, \& {Miller}}]{g2014}
{Gofford}, J., {Reeves}, J.~N., {Braito}, V., {et~al.} 2014, \apj, 784, 77

\bibitem[{{Green} \& {Mathur}(1996)}]{greenmathur96}
{Green}, P.~J. \& {Mathur}, S. 1996, \apj, 462, 637

\bibitem[{{Haardt} \& {Maraschi}(1991)}]{hm91}
{Haardt}, F. \& {Maraschi}, L. 1991, ApJL, 380, L51

\bibitem[{{Haardt} \& {Maraschi}(1993)}]{hm93}
{Haardt}, F. \& {Maraschi}, L. 1993, ApJ, 413, 507

\bibitem[{{Haardt} {et~al.}(1994){Haardt}, {Maraschi}, \&
  {Ghisellini}}]{haardt94}
{Haardt}, F., {Maraschi}, L., \& {Ghisellini}, G. 1994, ApJL, 432, L95

\bibitem[{{Hall} {et~al.}(2002){Hall}, {Anderson}, {Strauss}, {York},
  {Richards}, {Fan}, {Knapp}, {Schneider}, {Vanden Berk}, {Geballe}, {Bauer},
  {Becker}, {Davis}, {Rix}, {Nichol}, {Bahcall}, {Brinkmann}, {Brunner},
  {Connolly}, {Csabai}, {Doi}, {Fukugita}, {Gunn}, {Haiman}, {Harvanek},
  {Heckman}, {Hennessy}, {Inada}, {Ivezi{\'c}}, {Johnston}, {Kleinman},
  {Krolik}, {Krzesinski}, {Kunszt}, {Lamb}, {Long}, {Lupton}, {Miknaitis},
  {Munn}, {Narayanan}, {Neilsen}, {Newman}, {Nitta}, {Okamura}, {Pentericci},
  {Pier}, {Schlegel}, {Snedden}, {Szalay}, {Thakar}, {Tsvetanov}, {White}, \&
  {Zheng}}]{hall02}
{Hall}, P.~B., {Anderson}, S.~F., {Strauss}, M.~A., {et~al.} 2002, \apjs, 141,
  267

\bibitem[{{Imanishi} \& {Terashima}(2004)}]{imanishi04}
{Imanishi}, M. \& {Terashima}, Y. 2004, \aj, 127, 758

\bibitem[{{Jiang} {et~al.}(2014){Jiang}, {Stone}, \& {Davis}}]{jiang14}
{Jiang}, Y.-F., {Stone}, J.~M., \& {Davis}, S.~W. 2014, \apj, 784, 169

\bibitem[{{Jim{\'e}nez-Bail{\'o}n} {et~al.}(2005){Jim{\'e}nez-Bail{\'o}n},
  {Piconcelli}, {Guainazzi}, {Schartel}, {Rodr{\'{\i}}guez-Pascual}, \&
  {Santos-Lle{\'o}}}]{jim05}
{Jim{\'e}nez-Bail{\'o}n}, E., {Piconcelli}, E., {Guainazzi}, M., {et~al.} 2005,
  \aap, 435, 449

\bibitem[{{Jin} {et~al.}(2012{\natexlab{a}}){Jin}, {Ward}, \& {Done}}]{jin12a}
{Jin}, C., {Ward}, M., \& {Done}, C. 2012{\natexlab{a}}, \mnras, 425, 907

\bibitem[{{Jin} {et~al.}(2012{\natexlab{b}}){Jin}, {Ward}, {Done}, \&
  {Gelbord}}]{jin12b}
{Jin}, C., {Ward}, M., {Done}, C., \& {Gelbord}, J. 2012{\natexlab{b}}, \mnras,
  420, 1825

\bibitem[{{Just} {et~al.}(2007){Just}, {Brandt}, {Shemmer}, {Steffen},
  {Schneider}, {Chartas}, \& {Garmire}}]{just07}
{Just}, D., {Brandt}, W., {Shemmer}, O., {et~al.} 2007, ApJ, 665, 1004

\bibitem[{{Kaastra} {et~al.}(2014){Kaastra}, {Ebrero}, {Arav}, {Behar},
  {Bianchi}, {Branduardi-Raymont}, {Cappi}, {Costantini}, {Kriss}, {De Marco},
  {Mehdipour}, {Paltani}, {Petrucci}, {Pinto}, {Ponti}, {Steenbrugge}, \& {de
  Vries}}]{kaastra14}
{Kaastra}, J.~S., {Ebrero}, J., {Arav}, N., {et~al.} 2014, \aap, 570, A73

\bibitem[{{Kalberla} {et~al.}(2005){Kalberla}, {Burton}, {Hartmann}, {Arnal},
  {Bajaja}, {Morras}, \& {P{\"o}ppel}}]{kalberla05}
{Kalberla}, P., {Burton}, W., {Hartmann}, D., {et~al.} 2005, A\&A, 440, 775

\bibitem[{{Kelly} {et~al.}(2008){Kelly}, {Bechtold}, {Trump}, {Vestergaard}, \&
  {Siemiginowska}}]{kelly08}
{Kelly}, B.~C., {Bechtold}, J., {Trump}, J.~R., {Vestergaard}, M., \&
  {Siemiginowska}, A. 2008, \apjs, 176, 355

\bibitem[{{King} \& {Pounds}(2015)}]{kingpounds15}
{King}, A. \& {Pounds}, K. 2015, \araa, 53, 115

\bibitem[{{King}(2010)}]{k2010}
{King}, A.~R. 2010, \mnras, 402, 1516

\bibitem[{{King} \& {Pounds}(2003)}]{kingpounds03}
{King}, A.~R. \& {Pounds}, K.~A. 2003, \mnras, 345, 657

\bibitem[{{Korista} {et~al.}(1998){Korista}, {Baldwin}, \&
  {Ferland}}]{korista98}
{Korista}, K., {Baldwin}, J., \& {Ferland}, G. 1998, ApJ, 507, 24

\bibitem[{{Krawczyk} {et~al.}(2013){Krawczyk}, {Richards}, {Mehta}, {Vogeley},
  {Gallagher}, {Leighly}, {Ross}, \& {Schneider}}]{krawiz13}
{Krawczyk}, C.~M., {Richards}, G.~T., {Mehta}, S.~S., {et~al.} 2013, ApJs, 206,
  4

\bibitem[{{Krongold} {et~al.}(2003){Krongold}, {Nicastro}, {Brickhouse},
  {Elvis}, {Liedahl}, \& {Mathur}}]{k2003}
{Krongold}, Y., {Nicastro}, F., {Brickhouse}, N.~S., {et~al.} 2003, \apj, 597,
  832

\bibitem[{{Kuraszkiewicz} {et~al.}(2000){Kuraszkiewicz}, {Wilkes}, {Czerny}, \&
  {Mathur}}]{kura00}
{Kuraszkiewicz}, J., {Wilkes}, B.~J., {Czerny}, B., \& {Mathur}, S. 2000, \apj,
  542, 692

\bibitem[{{La Franca} {et~al.}(2005){La Franca}, {Fiore}, {Comastri}, {Perola},
  {Sacchi}, {Brusa}, {Cocchia}, {Feruglio}, {Matt}, {Vignali}, {Carangelo},
  {Ciliegi}, {Lamastra}, {Maiolino}, {Mignoli}, {Molendi}, \&
  {Puccetti}}]{lafranca05}
{La Franca}, F., {Fiore}, F., {Comastri}, A., {et~al.} 2005, \apj, 635, 864

\bibitem[{{Lanzuisi} {et~al.}(2016){Lanzuisi}, {Perna}, {Comastri}, {Cappi},
  {Dadina}, {Marinucci}, {Masini}, {Matt}, {Vagnetti}, {Vignali}, {Ballantyne},
  {Bauer}, {Boggs}, {Brandt}, {Brusa}, {Christensen}, {Craig}, {Fabian},
  {Farrah}, {Hailey}, {Harrison}, {Luo}, {Piconcelli}, {Puccetti}, {Ricci},
  {Saez}, {Stern}, {Walton}, \& {Zhang}}]{lanzuisi16}
{Lanzuisi}, G., {Perna}, M., {Comastri}, A., {et~al.} 2016, \aap, 590, A77

\bibitem[{{Lanzuisi} {et~al.}(2009){Lanzuisi}, {Piconcelli}, {Fiore},
  {Feruglio}, {Vignali}, {Salvato}, \& {Gruppioni}}]{lanzuisi09}
{Lanzuisi}, G., {Piconcelli}, E., {Fiore}, F., {et~al.} 2009, A\&A, 498, 67

\bibitem[{{Laor} \& {Brandt}(2002)}]{laor02}
{Laor}, A. \& {Brandt}, W. 2002, ApJ, 569, 641

\bibitem[{{Laor} {et~al.}(1994){Laor}, {Fiore}, {Elvis}, {Wilkes}, \&
  {McDowell}}]{laor94}
{Laor}, A., {Fiore}, F., {Elvis}, M., {Wilkes}, B.~J., \& {McDowell}, J.~C.
  1994, Apj, 435, 611

\bibitem[{{Leighly}(2004)}]{leighly04}
{Leighly}, K.~M. 2004, ApJ, 611, 125

\bibitem[{{Leighly} {et~al.}(2007){Leighly}, {Halpern}, {Jenkins}, {Grupe},
  {Choi}, \& {Prescott}}]{leighly07}
{Leighly}, K.~M., {Halpern}, J.~P., {Jenkins}, E.~B., {et~al.} 2007, \apj, 663,
  103

\bibitem[{{Ludlam} {et~al.}(2015){Ludlam}, {Cackett}, {G{\"u}ltekin}, {Fabian},
  {Gallo}, \& {Miniutti}}]{ludlam15}
{Ludlam}, R.~M., {Cackett}, E.~M., {G{\"u}ltekin}, K., {et~al.} 2015, \mnras,
  447, 2112

\bibitem[{{Luo} {et~al.}(2013){Luo}, {Brandt}, {Alexander}, {Harrison},
  {Stern}, {Bauer}, {Boggs}, {Christensen}, {Comastri}, {Craig}, {Fabian},
  {Farrah}, {Fiore}, {Fuerst}, {Grefenstette}, {Hailey}, {Hickox}, {Madsen},
  {Matt}, {Ogle}, {Risaliti}, {Saez}, {Teng}, {Walton}, \& {Zhang}}]{luo13}
{Luo}, B., {Brandt}, W.~N., {Alexander}, D.~M., {et~al.} 2013, \apj, 772, 153

\bibitem[{{Luo} {et~al.}(2015){Luo}, {Brandt}, {Hall}, {Wu}, {Anderson},
  {Garmire}, {Gibson}, {Plotkin}, {Richards}, {Schneider}, {Shemmer}, \&
  {Shen}}]{luo15}
{Luo}, B., {Brandt}, W.~N., {Hall}, P.~B., {et~al.} 2015, ApJ, 805, 122

\bibitem[{{Lusso} {et~al.}(2012){Lusso}, {Comastri}, {Simmons}, {Mignoli},
  {Zamorani}, {Vignali}, {Brusa}, {Shankar}, {Lutz}, {Trump}, {Maiolino},
  {Gilli}, {Bolzonella}, {Puccetti}, {Salvato}, {Impey}, {Civano}, {Elvis},
  {Mainieri}, {Silverman}, {Koekemoer}, {Bongiorno}, {Merloni}, {Berta}, {Le
  Floc'h}, {Magnelli}, {Pozzi}, \& {Riguccini}}]{lusso12}
{Lusso}, E., {Comastri}, A., {Simmons}, B., {et~al.} 2012, MNRAS, 425, 623

\bibitem[{{Lusso} {et~al.}(2010){Lusso}, {Comastri}, {Vignali}, {Zamorani},
  {Brusa}, {Gilli}, {Iwasawa}, {Salvato}, {Civano}, {Elvis}, {Merloni},
  {Bongiorno}, {Trump}, {Koekemoer}, {Schinnerer}, {Le Floc'h}, {Cappelluti},
  {Jahnke}, {Sargent}, {Silverman}, {Mainieri}, {Fiore}, {Bolzonella}, {Le
  F{\`e}vre}, {Garilli}, {Iovino}, {Kneib}, {Lamareille}, {Lilly}, {Mignoli},
  {Scodeggio}, \& {Vergani}}]{lusso10}
{Lusso}, E., {Comastri}, A., {Vignali}, C., {et~al.} 2010, \aap, 512, A34

\bibitem[{{Lusso} \& {Risaliti}(2016)}]{lussorisaliti16}
{Lusso}, E. \& {Risaliti}, G. 2016, \apj, 819, 154

\bibitem[{{Lutz} {et~al.}(2004){Lutz}, {Maiolino}, {Spoon}, \&
  {Moorwood}}]{lutz04}
{Lutz}, D., {Maiolino}, R., {Spoon}, H.~W.~W., \& {Moorwood}, A.~F.~M. 2004,
  A\&A, 418, 465

\bibitem[{{Magdziarz} \& {Zdziarski}(1995)}]{pexrav95}
{Magdziarz}, P. \& {Zdziarski}, A. 1995, MNRAS, 273, 837

\bibitem[{{Maiolino} {et~al.}(2001){Maiolino}, {Marconi}, {Salvati},
  {Risaliti}, {Severgnini}, {Oliva}, {La Franca}, \& {Vanzi}}]{maiolino01}
{Maiolino}, R., {Marconi}, A., {Salvati}, M., {et~al.} 2001, A\&A, 365, 28

\bibitem[{{Maiolino} {et~al.}(2007){Maiolino}, {Shemmer}, {Imanishi}, {Netzer},
  {Oliva}, {Lutz}, \& {Sturm}}]{maiolino07}
{Maiolino}, R., {Shemmer}, O., {Imanishi}, M., {et~al.} 2007, \aap, 468, 979

\bibitem[{{Marziani} {et~al.}(2016){Marziani}, {Mart{\'{\i}}nez Carballo},
  {Sulentic}, {Del Olmo}, {Stirpe}, \& {Dultzin}}]{marziani15}
{Marziani}, P., {Mart{\'{\i}}nez Carballo}, M.~A., {Sulentic}, J.~W., {et~al.}
  2016, \apss, 361, 29

\bibitem[{{Mateos} {et~al.}(2015){Mateos}, {Carrera}, {Alonso-Herrero},
  {Rovilos}, {Hern{\'a}n-Caballero}, {Barcons}, {Blain}, {Caccianiga}, {Della
  Ceca}, \& {Severgnini}}]{mateos15}
{Mateos}, S., {Carrera}, F., {Alonso-Herrero}, A., {et~al.} 2015, MNRAS, 449,
  1422

\bibitem[{{Mathur} {et~al.}(2000){Mathur}, {Green}, {Arav}, {Brotherton},
  {Crenshaw}, {deKool}, {Elvis}, {Goodrich}, {Hamann}, {Hines}, {Kashyap},
  {Korista}, {Peterson}, {Shields}, {Shlosman}, {van Breugel}, \&
  {Voit}}]{mathur00}
{Mathur}, S., {Green}, P.~J., {Arav}, N., {et~al.} 2000, \apjl, 533, L79

\bibitem[{{Matt} {et~al.}(1991){Matt}, {Perola}, \& {Piro}}]{matt91}
{Matt}, G., {Perola}, G.~C., \& {Piro}, L. 1991, \aap, 247, 25

\bibitem[{{Merloni} {et~al.}(2012){Merloni}, {Predehl}, {Becker},
  {B{\"o}hringer}, {Boller}, {Brunner}, {Brusa}, {Dennerl}, {Freyberg},
  {Friedrich}, {Georgakakis}, {Haberl}, {Hasinger}, {Meidinger}, {Mohr},
  {Nandra}, {Rau}, {Reiprich}, {Robrade}, {Salvato}, {Santangelo}, {Sasaki},
  {Schwope}, {Wilms}, \& {German eROSITA Consortium}}]{merloni12}
{Merloni}, A., {Predehl}, P., {Becker}, W., {et~al.} 2012, ArXiv e-prints
  [\eprint[arXiv]{1209.3114}]

\bibitem[{{Miniutti} {et~al.}(2009{\natexlab{a}}){Miniutti}, {Fabian},
  {Brandt}, {Gallo}, \& {Boller}}]{miniutti09b}
{Miniutti}, G., {Fabian}, A.~C., {Brandt}, W.~N., {Gallo}, L.~C., \& {Boller},
  T. 2009{\natexlab{a}}, \mnras, 396, L85

\bibitem[{{Miniutti} {et~al.}(2009{\natexlab{b}}){Miniutti}, {Ponti}, {Greene},
  {Ho}, {Fabian}, \& {Iwasawa}}]{miniutti09}
{Miniutti}, G., {Ponti}, G., {Greene}, J.~E., {et~al.} 2009{\natexlab{b}},
  \mnras, 394, 443

\bibitem[{{Murray} {et~al.}(1995){Murray}, {Chiang}, {Grossman}, \&
  {Voit}}]{murray95}
{Murray}, N., {Chiang}, J., {Grossman}, S.~A., \& {Voit}, G.~M. 1995, \apj,
  451, 498

\bibitem[{{Nandra} {et~al.}(2013){Nandra}, {Barret}, {Barcons}, {Fabian}, {den
  Herder}, {Piro}, {Watson}, {Adami}, {Aird}, {Afonso}, \& et~al.}]{nandra13}
{Nandra}, K., {Barret}, D., {Barcons}, X., {et~al.} 2013, ArXiv e-prints
  [\eprint[arXiv]{1306.2307}]

\bibitem[{{Nandra} {et~al.}(1997){Nandra}, {George}, {Mushotzky}, {Turner}, \&
  {Yaqoob}}]{nandra97}
{Nandra}, K., {George}, I.~M., {Mushotzky}, R.~F., {Turner}, T.~J., \&
  {Yaqoob}, T. 1997, \apj, 477, 602

\bibitem[{{Nandra} {et~al.}(2000){Nandra}, {Le}, {George}, {Edelson},
  {Mushotzky}, {Peterson}, \& {Turner}}]{nandra00}
{Nandra}, K., {Le}, T., {George}, I.~M., {et~al.} 2000, \apj, 544, 734

\bibitem[{{Nanni} {et~al.}(2017){Nanni}, {Vignali}, {Gilli}, {Moretti}, \&
  {Brandt}}]{nanni17}
{Nanni}, R., {Vignali}, C., {Gilli}, R., {Moretti}, A., \& {Brandt}, W.~N.
  2017, ArXiv e-prints [\eprint[arXiv]{1704.08693}]

\bibitem[{{Page} {et~al.}(2004){Page}, {Reeves}, {O'Brien}, {Turner}, \&
  {Worrall}}]{page04}
{Page}, K.~L., {Reeves}, J.~N., {O'Brien}, P.~T., {Turner}, M.~J.~L., \&
  {Worrall}, D.~M. 2004, \mnras, 353, 133

\bibitem[{{Parker} {et~al.}(2017){Parker}, {Pinto}, {Fabian}, {Lohfink},
  {Buisson}, {Alston}, {Kara}, {Cackett}, {Chiang}, {Dauser}, {De Marco},
  {Gallo}, {Garcia}, {Harrison}, {King}, {Middleton}, {Miller}, {Miniutti},
  {Reynolds}, {Uttley}, {Vasudevan}, {Walton}, {Wilkins}, \& {Zoghbi}}]{p2017}
{Parker}, M.~L., {Pinto}, C., {Fabian}, A.~C., {et~al.} 2017, \nat, 543, 83

\bibitem[{{Petrucci} {et~al.}(2000){Petrucci}, {Haardt}, {Maraschi}, {Grandi},
  {Matt}, {Nicastro}, {Piro}, {Perola}, \& {De Rosa}}]{petrucci00}
{Petrucci}, P.~O., {Haardt}, F., {Maraschi}, L., {et~al.} 2000, ApJ, 540, 131

\bibitem[{{Petrucci} {et~al.}(2013){Petrucci}, {Paltani}, {Malzac}, {Kaastra},
  {Cappi}, {Ponti}, {De Marco}, {Kriss}, {Steenbrugge}, {Bianchi},
  {Branduardi-Raymont}, {Mehdipour}, {Costantini}, {Dadina}, \&
  {Lubi{\'n}ski}}]{petrucci13}
{Petrucci}, P.-O., {Paltani}, S., {Malzac}, J., {et~al.} 2013, \aap, 549, A73

\bibitem[{{Piconcelli} {et~al.}(2005){Piconcelli}, {Jimenez-Bail{\'o}n},
  {Guainazzi}, {Schartel}, {Rodr{\'{\i}}guez-Pascual}, \&
  {Santos-Lle{\'o}}}]{picoPG05}
{Piconcelli}, E., {Jimenez-Bail{\'o}n}, E., {Guainazzi}, M., {et~al.} 2005,
  A\&A, 432, 15

\bibitem[{{Piconcelli} {et~al.}(2015){Piconcelli}, {Vignali}, {Bianchi},
  {Zappacosta}, {Fritz}, {Lanzuisi}, {Miniutti}, {Bongiorno}, {Feruglio},
  {Fiore}, \& {Maiolino}}]{pico15}
{Piconcelli}, E., {Vignali}, C., {Bianchi}, S., {et~al.} 2015, \aap, 574, L9

\bibitem[{{Proga}(2003)}]{proga03}
{Proga}, D. 2003, ApJ, 585, 406

\bibitem[{{Proga}(2005)}]{proga05}
{Proga}, D. 2005, \apjl, 630, L9

\bibitem[{{Proga}(2007)}]{proga07}
{Proga}, D. 2007, in Astronomical Society of the Pacific Conference Series,
  Vol. 373, The Central Engine of Active Galactic Nuclei, ed. L.~C. {Ho} \&
  J.-W. {Wang}, 267

\bibitem[{{Reeves} {et~al.}(2014){Reeves}, {Braito}, {Gofford}, {Sim}, {Behar},
  {Costa}, {Kaspi}, {Matzeu}, {Miller}, {O'Brien}, {Turner}, \& {Ward}}]{r2014}
{Reeves}, J.~N., {Braito}, V., {Gofford}, J., {et~al.} 2014, \apj, 780, 45

\bibitem[{{Reeves} {et~al.}(2009{\natexlab{a}}){Reeves}, {O'Brien}, {Braito},
  {Behar}, {Miller}, {Turner}, {Fabian}, {Kaspi}, {Mushotzky}, \&
  {Ward}}]{reeves09}
{Reeves}, J.~N., {O'Brien}, P.~T., {Braito}, V., {et~al.} 2009{\natexlab{a}},
  \apj, 701, 493

\bibitem[{{Reeves} {et~al.}(2009{\natexlab{b}}){Reeves}, {O'Brien}, {Braito},
  {Behar}, {Miller}, {Turner}, {Fabian}, {Kaspi}, {Mushotzky}, \&
  {Ward}}]{r2009}
{Reeves}, J.~N., {O'Brien}, P.~T., {Braito}, V., {et~al.} 2009{\natexlab{b}},
  \apj, 701, 493

\bibitem[{{Reeves} {et~al.}(2003){Reeves}, {O'Brien}, \& {Ward}}]{r2003}
{Reeves}, J.~N., {O'Brien}, P.~T., \& {Ward}, M.~J. 2003, \apjl, 593, L65

\bibitem[{{Reeves} \& {Turner}(2000)}]{reevesturner00}
{Reeves}, J.~N. \& {Turner}, M.~J.~L. 2000, \mnras, 316, 234

\bibitem[{{Reis} \& {Miller}(2013)}]{reis13}
{Reis}, R.~C. \& {Miller}, J.~M. 2013, ApJL, 769, L7

\bibitem[{{Reynolds}(1997)}]{reynolds97}
{Reynolds}, C.~S. 1997, \mnras, 286, 513

\bibitem[{{Richards} {et~al.}(2011){Richards}, {Kruczek}, {Gallagher}, {Hall},
  {Hewett}, {Leighly}, {Deo}, {Kratzer}, \& {Shen}}]{richards11}
{Richards}, G.~T., {Kruczek}, N.~E., {Gallagher}, S.~C., {et~al.} 2011, AJ,
  141, 167

\bibitem[{{Runnoe} {et~al.}(2012{\natexlab{a}}){Runnoe}, {Brotherton}, \&
  {Shang}}]{runnoe12a}
{Runnoe}, J.~C., {Brotherton}, M.~S., \& {Shang}, Z. 2012{\natexlab{a}},
  \mnras, 422, 478

\bibitem[{{Runnoe} {et~al.}(2012{\natexlab{b}}){Runnoe}, {Brotherton}, \&
  {Shang}}]{runnoe12b}
{Runnoe}, J.~C., {Brotherton}, M.~S., \& {Shang}, Z. 2012{\natexlab{b}},
  \mnras, 426, 2677

\bibitem[{{Sabra} \& {Hamann}(2001)}]{sabra01}
{Sabra}, B.~M. \& {Hamann}, F. 2001, \apj, 563, 555

\bibitem[{{Schartel} {et~al.}(2010){Schartel}, {Rodr{\'{\i}}guez-Pascual},
  {Santos-Lle{\'o}}, {Jim{\'e}nez-Bail{\'o}n}, {Ballo}, \&
  {Piconcelli}}]{schartel10}
{Schartel}, N., {Rodr{\'{\i}}guez-Pascual}, P.~M., {Santos-Lle{\'o}}, M.,
  {et~al.} 2010, \aap, 512, A75

\bibitem[{{Shemmer} {et~al.}(2008){Shemmer}, {Brandt}, {Netzer}, {Maiolino}, \&
  {Kaspi}}]{shemmer08}
{Shemmer}, O., {Brandt}, W.~N., {Netzer}, H., {Maiolino}, R., \& {Kaspi}, S.
  2008, ApJ, 682, 81

\bibitem[{{Shu} {et~al.}(2010){Shu}, {Yaqoob}, \& {Wang}}]{shuyaqoobwang10}
{Shu}, X.~W., {Yaqoob}, T., \& {Wang}, J.~X. 2010, \apjs, 187, 581

\bibitem[{{Steffen} {et~al.}(2006){Steffen}, {Strateva}, {Brandt}, {Alexander},
  {Koekemoer}, {Lehmer}, {Schneider}, \& {Vignali}}]{steffen06}
{Steffen}, A.~T., {Strateva}, I., {Brandt}, W.~N., {et~al.} 2006, \aj, 131,
  2826

\bibitem[{{Stern}(2015)}]{stern15}
{Stern}, D. 2015, ApJ, 807, 129

\bibitem[{{Sturm} {et~al.}(2011){Sturm}, {Gonz{\'a}lez-Alfonso}, {Veilleux},
  {Fischer}, {Graci{\'a}-Carpio}, {Hailey-Dunsheath}, {Contursi}, {Poglitsch},
  {Sternberg}, {Davies}, {Genzel}, {Lutz}, {Tacconi}, {Verma}, {Maiolino}, \&
  {de Jong}}]{sturm11}
{Sturm}, E., {Gonz{\'a}lez-Alfonso}, E., {Veilleux}, S., {et~al.} 2011, \apjl,
  733, L16

\bibitem[{{Teng} {et~al.}(2014){Teng}, {Brandt}, {Harrison}, {Luo},
  {Alexander}, {Bauer}, {Boggs}, {Christensen}, {Comastri}, {Craig}, {Fabian},
  {Farrah}, {Fiore}, {Gandhi}, {Grefenstette}, {Hailey}, {Hickox}, {Madsen},
  {Ptak}, {Rigby}, {Risaliti}, {Saez}, {Stern}, {Veilleux}, {Walton}, {Wik}, \&
  {Zhang}}]{teng14}
{Teng}, S.~H., {Brandt}, W.~N., {Harrison}, F.~A., {et~al.} 2014, \apj, 785, 19

\bibitem[{{Tombesi} {et~al.}(2010){Tombesi}, {Cappi}, {Reeves}, {Palumbo},
  {Yaqoob}, {Braito}, \& {Dadina}}]{t2010}
{Tombesi}, F., {Cappi}, M., {Reeves}, J.~N., {et~al.} 2010, \aap, 521, A57

\bibitem[{{Tombesi} {et~al.}(2015){Tombesi}, {Mel{\'e}ndez}, {Veilleux},
  {Reeves}, {Gonz{\'a}lez-Alfonso}, \& {Reynolds}}]{tombesi15}
{Tombesi}, F., {Mel{\'e}ndez}, M., {Veilleux}, S., {et~al.} 2015, \nat, 519,
  436

\bibitem[{{Treister} {et~al.}(2008){Treister}, {Krolik}, \&
  {Dullemond}}]{treister08}
{Treister}, E., {Krolik}, J.~H., \& {Dullemond}, C. 2008, \apj, 679, 140

\bibitem[{{Turner} \& {Miller}(2009)}]{turnermiller09}
{Turner}, T.~J. \& {Miller}, L. 2009, \aapr, 17, 47

\bibitem[{{Vagnetti} {et~al.}(2010){Vagnetti}, {Turriziani}, {Trevese}, \&
  {Antonucci}}]{vagnetti10}
{Vagnetti}, F., {Turriziani}, S., {Trevese}, D., \& {Antonucci}, M. 2010, A\&A,
  519, A17

\bibitem[{{Vanzella} {et~al.}(2008){Vanzella}, {Cristiani}, {Dickinson},
  {Giavalisco}, {Kuntschner}, {Haase}, {Nonino}, {Rosati}, {Cesarsky},
  {Ferguson}, {Fosbury}, {Grazian}, {Moustakas}, {Rettura}, {Popesso},
  {Renzini}, {Stern}, \& {GOODS Team}}]{vanzella08}
{Vanzella}, E., {Cristiani}, S., {Dickinson}, M., {et~al.} 2008, A\&A, 478, 83

\bibitem[{{Vasudevan} \& {Fabian}(2007)}]{vasufabian07}
{Vasudevan}, R.~V. \& {Fabian}, A.~C. 2007, MNRAS, 381, 1235

\bibitem[{{Vignali} {et~al.}(2010){Vignali}, {Alexander}, {Gilli}, \&
  {Pozzi}}]{vignali10}
{Vignali}, C., {Alexander}, D.~M., {Gilli}, R., \& {Pozzi}, F. 2010, \mnras,
  404, 48

\bibitem[{{Vignali} {et~al.}(2003){Vignali}, {Brandt}, \&
  {Schneider}}]{vignali03}
{Vignali}, C., {Brandt}, W.~N., \& {Schneider}, D.~P. 2003, AJ, 125, 433

\bibitem[{{Weedman} {et~al.}(2012){Weedman}, {Sargsyan}, {Lebouteiller},
  {Houck}, \& {Barry}}]{weedman12}
{Weedman}, D., {Sargsyan}, L., {Lebouteiller}, V., {Houck}, J., \& {Barry}, D.
  2012, ApJ, 761, 184

\bibitem[{{Wu} {et~al.}(2012){Wu}, {Brandt}, {Anderson}, {Diamond-Stanic},
  {Hall}, {Plotkin}, {Schneider}, \& {Shemmer}}]{wu12}
{Wu}, J., {Brandt}, W.~N., {Anderson}, S.~F., {et~al.} 2012, \apj, 747, 10

\bibitem[{{Xue} {et~al.}(2011){Xue}, {Luo}, {Brandt}, {Bauer}, {Lehmer},
  {Broos}, {Schneider}, {Alexander}, {Brusa}, {Comastri}, {Fabian}, {Gilli},
  {Hasinger}, {Hornschemeier}, {Koekemoer}, {Liu}, {Mainieri}, {Paolillo},
  {Rafferty}, {Rosati}, {Shemmer}, {Silverman}, {Smail}, {Tozzi}, \&
  {Vignali}}]{xue11}
{Xue}, Y., {Luo}, B., {Brandt}, W., {et~al.} 2011, ApJs, 195, 10

\bibitem[{{Young} {et~al.}(2009){Young}, {Elvis}, \& {Risaliti}}]{young09}
{Young}, M., {Elvis}, M., \& {Risaliti}, G. 2009, \apjs, 183, 17

\bibitem[{{Zdziarski} {et~al.}(1996){Zdziarski}, {Johnson}, \&
  {Magdziarz}}]{zdz96}
{Zdziarski}, A.~A., {Johnson}, W.~N., \& {Magdziarz}, P. 1996, MNRAS, 283, 193

\bibitem[{{Zubovas} \& {King}(2012)}]{zub}
{Zubovas}, K. \& {King}, A. 2012, \apjl, 745, L34

\end{thebibliography}


\begin{appendix}
	
	\section{X-ray spectral properties of additional hyper-luminous quasars}
	\label{sec:app1}
	
	We analysed the \xmm\ observations of three additional quasars which have redshift and bolometric luminosities comparable with those of WISSH objects, namely ULAS J1539+0557 ($z$ = 2.658; \citealt{feruglio14}), ULAS J2315+0143 ($z$ = 2.56;  \citealt{banerji15}), 2QZ 0028-2830 ($z$ = 2.4; \citealt{shemmer08}).
	The X-ray data reduction and analysis were carried out following the same procedure outlined in Sect. \ref{subsec:dr} and Sect. \ref{subsec:xspec}, respectively. We can summarize the main results  from the spectral analysis as follows:
	\begin{itemize}
		\item \textbf{ULAS J1539+0557}.
		This reddened quasar has a bolometric luminosity of ${\rm Log}[L_{\rm Bol}/{\rm erg\, s^{-1}}] = 48.2$ and a MIR luminosity ${\rm Log}[\lambda L_{6 \mu m}/{\rm erg\, s^{-1}}] = 47$, respectively \citep{feruglio14}.
		It was observed by \xmm\ for $\sim 44.9$ ks on 2015-02-16.
		The  application of the PL model to the X-ray spectrum results into a very flat power law with $\Gamma \sim 1$. The addition of a  rest-frame absorption component to the fitting model yields a significant improvement the quality of the fit (from  Cstat/dof = 19/15 to 12/14). The best fit value of the column density is \nh = $4.0_{-1.7}^{+2.6} \times 10^{22}$ \cm2. The photon index derived by the APL model  remains slightly flat ($\Gamma = 1.5_{-0.2}^{+0.2}$), but fixing $\Gamma$ to the canonical value $\Gamma=1.8$ results into a worsening of the fit.
		Using the APL model, we measured a hard X-ray flux of $f_{2-10} \sim 4\times 10^{-14} \, {\rm erg \, cm^{-2}s^{-1}}$ and a luminosity of ${\rm Log}[L_{2-10}/{\rm erg\, s^{-1}}] = 45.1$.
		
		\vspace{0.3cm}
		\item \textbf{ULAS J2315+0143}. This reddened quasar  has a  bolometric luminosity of  ${\rm Log}[L/{\rm erg\, s^{-1}}] = 47.5$  \citep{banerji15} and was targeted by \xmm\  on 2014-12-16 for $\sim 55.9$ ks. The X-ray spectrum is consistent with with a power law ($\Gamma =1.64_{-0.09}^{+0.09}$) absorbed
		by a modest \nh\ of $0.7_{-0.4}^{+0.4} \times 10^{22} \, {\rm cm^{-2}}$. Fixing $\Gamma$ to the canonical value of 1.8 does not significantly improve the quality of the fit. The hard X-ray flux and luminosity derived by the \xmm\ spectrum fitted by the APL model are  $f_{2-10} \sim 10^{-13} \, {\rm erg \, cm^{-2}s^{-1}}$ and ${\rm Log}[L_{2-10}/{\rm erg\, s^{-1}}] = 45.5$, respectively.
		
		\vspace{0.3cm}
		\item \textbf{2QZ 0028-2830}. This quasar was observed by \xmm\  on 2009-12-07 for $\sim 22.2$ ks. The PL model provides the best description for the X-ray spectrum. The resulting photon index is  $\Gamma = 1.7_{-0.1}^{+0.1}$. The addition of an intrinsic absorber is not required by the fit, and only an upper limit on \nh\ can be inferred, i.e. \nh$\leq 7.6 \times 10^{21}\, {\rm cm^{-2}}$.
		The hard X-ray flux of  2QZ0028$-$2830  is $f_{2-10} \sim 7 \times 10^{-14} \, {\rm erg \, cm^{-2}s^{-1}}$, which corresponds to a ${\rm Log}[L_{2-10}/{\rm erg\, s^{-1}}] \sim 45.3$.
		This quasar exhibit a bolometric luminosity of ${\rm Log}[L_{\rm Bol}/{\rm erg\, s^{-1}}] = 47.3$ \citep{carniani15} and MIR luminosity of ${\rm Log}[\lambda L_{6 \mu m}/{\rm erg\, s^{-1}}] = 46.4$ derived  by  {\it WISE} photometric data.
	\end{itemize}
	Multiwavelength information about ULAS J1539+0557, ULAS J2315+0143 and 2QZ 0028-2830 are listed in Table \ref{tab:addQSO}.

	\begin{table*}
		\caption{Summary of the multiwavelength properties of the three additional hyper-luminous quasars considered in this work.
			Columns give the following information: (1) quasar ID, (2) redshift, (3) X-ray photon index,
			 (4) absorption column densities \nh (in units of $10^{22} \, {\rm cm^{-2}}$), (5) 2-10 keV fluxes (in units of $10^{-14} \, {\rm erg \, cm^2 \, s^{-1}}$), (6) 2-10 keV unabsorbed luminosities, (7) 6 $\mu m$ luminosities, and (8) bolometric luminosities (in units of Log[$L/{\rm erg \, s^{-1}}$]).} 
		\label{tab:addQSO}
		\centering
		\begin{tabular}{c c c c c c c c}     
			\toprule\toprule
			
			Name & z & $\Gamma$ & \nh & $f_{2-10}$ & Log$L_{2-10}$ & Log$\lambda L_{6 \mu m}$ & Log$L_{\rm Bol}$\\
			(1)   & (2) & (3) & (4) & (5) & (6) & (7) & (8)\\
			\midrule
			ULASJ1539+0557 & 2.658  & $1.5_{-0.2}^{+0.2}$ & $4.0_{-1.7}^{+2.6}$ & 4.0 & 45.1 & 47.0  & 48.2\\
			ULASJ2315+0143 & 2.56   & $1.64_{-0.09}^{+0.09}$ & $0.7_{-0.4}^{+0.4}$ & 9.2& 45.5 & 47.5  & 47.5\\
			2QZ0028-2830   & 2.4    & $1.7_{-0.1}^{+0.1}$ & $\leq 0.76$            & 6.8 & 45.3 & 46.4  & 47.3\\  
			\bottomrule\bottomrule
		\end{tabular}
	\end{table*}

\end{appendix}

\end{document}